\documentclass[iopjournal,twocolumn,showkeys,superscriptaddress]{revtex4-2}
\usepackage{graphicx}
\usepackage{amsmath,amssymb}
\usepackage{color}
\usepackage{bm}
\usepackage{physics}
\usepackage{float}
\usepackage{soul}
\usepackage[dvipsnames]{xcolor}
 \usepackage{multirow}

\usepackage[
    colorlinks=true,
    linkcolor=black,
    citecolor=black,
    urlcolor=blue,
    pdfborder={0 0 0}
]{hyperref}
\begin{document}

\title{Empirical approaches to Fr\"ohlich excitonic polarons in polar semiconductors}

\author{Jacky Even}
\email{jacky.even@insa-rennes.fr}
\affiliation{Univ Rennes, INSA Rennes, CNRS, Institut FOTON, UMR 6082, Rennes, France}
\author{Simon Thebaud}
\affiliation{Univ Rennes, INSA Rennes, CNRS, Institut FOTON, UMR 6082, Rennes, France}
\author{Aseem Rajan Kshirsagar}
\affiliation{Univ Rennes, ENSCR, CNRS, ISCR-UMR 6226, Rennes F-35000, France}
\author{Zeli Xu}
\affiliation{Univ Rennes, INSA Rennes, CNRS, Institut FOTON, UMR 6082, Rennes, France}
\author{Laurent Pedesseau}
\affiliation{Univ Rennes, INSA Rennes, CNRS, Institut FOTON, UMR 6082, Rennes, France}
\author{Marios Zacharias}
\affiliation{Univ Rennes, INSA Rennes, CNRS, Institut FOTON, UMR 6082, Rennes, France}
\author{Claudine Katan}
\email{claudine.katan@univ-rennes.fr}
\affiliation{Univ Rennes, ENSCR, CNRS, ISCR-UMR 6226, Rennes F-35000, France}
\begin{abstract}

The interaction between charge carriers with opposite charge, i.e. excitonic correlation between electron and hole, and the charge carriers' interactions with underlying crystal lattice govern the opto-electronic response of semiconductors. The latter is dominated in ionic crystals by the Fr\"{o}hlich interaction corresponding to the long-range electric field generated by polar phonons at the zone-centre. Theoretical description of complex interplay arising from competition and interference between these two interactions, excitonic and polaronic, has been a formidable challenge.  At the same time, the emergence of optoelectronics based on ionic semiconductors such as metal-halide perovskites has made it highly relevant to revisit theoretical models developed in the past, and push them further for application to real ionic semiconductors of interest. \\
To this end, the present paper reviews the physics of Fr\"ohlich excitonic polarons from the point of view of empirical approaches and supplements it with a few original developments to the state-of-the-art. At first, we review the models for excitonic polarons in ionic semiconductors, initially built by Toyozawa and Hermanson in analogy with Lee-Low-Pines (LLP) model for free polarons, and later extended by Pollman and B\"uttner (PB). These models have been applied in past to the simpler cases of weakly interacting polarons (e.g. GaAs). We further consider their applicability to ionic solids such as TlCl or 3D lead halide perovskites, where the
electron-hole correlations are relatively stronger. In these compounds, electrons and holes have almost equal effective masses, which allows us to derive a new analytical expression of the approximated version of the PB effective interaction potential. The refined Kane approach to PB's model is shown to (i) bridge the regime between weakly interacting polarons and excitonic polarons with strong electron-hole correlations and (ii) recover the LLP model for free polarons in the limit of vanishing correlation. Various theoretical developments carried out in this paper also include extension of Kane and PB's semi-empirical models to incorporate Fr\"ohlich-like interaction with multiple polar phonons, essential for reconciliation of experimental observations in the case of multi-atom systems. In the end, we discuss the relation of the empirical approaches with emerging \textit{ab initio} methods and provide an outlook of their application to lower-dimensional systems. \\  {\scriptsize Dedicated to the memory of Prof. Jozef T. Devreese [1937-2023]} 

\end{abstract}

\maketitle

\tableofcontents

\section{Introduction}

\begin{figure}[htb]
\includegraphics[width=0.45\textwidth]{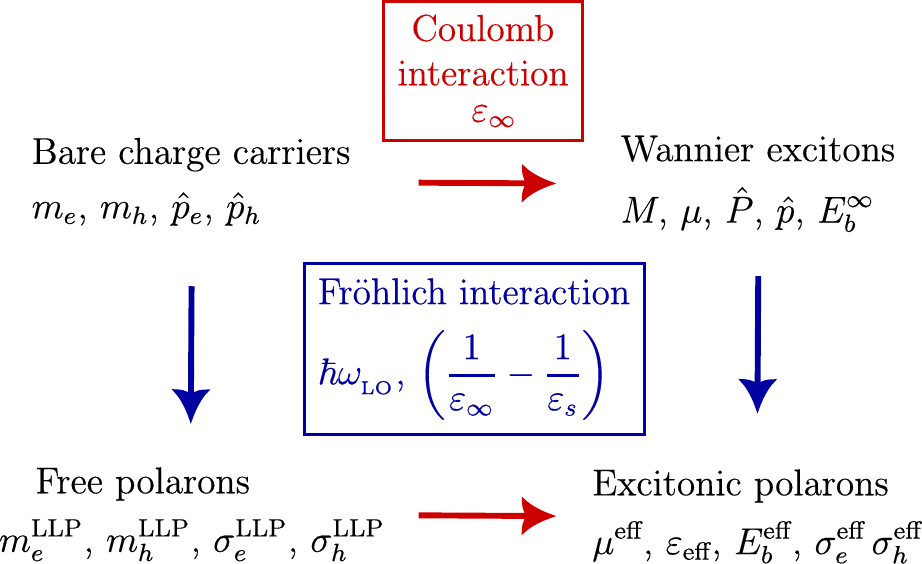}
\caption{Various levels of theory analyzed in this work,
with corresponding material parameters and physical observables.
Continuum-based model parameters for uncorrelated charge
carriers and the 1S exciton ground state as well as acronyms used in this review are further introduced in Tab.~\ref{tab:table2}.}\label{fig3}
\end{figure}

Optoelectronic devices based on ionic semiconductors are at the heart of a prominent class of disruptive energy and quantum technologies that includes next-generation solar cells, nanocrystals for quantum optics, solar-to-hydrogen converters and low-threshold lasers~\cite{Snaith2018-kw,Kovalenko2017-zl,Blancon2018,Fehr2023-gq,Deschler2014-ft}
(see Appendix \ref{AppA}). In these materials, the physical description of the 
electron-hole pairs generated or recombined to absorb or emit light is dominated by two
 phenomena: (1) the interaction between the charge carriers, and (2) the interaction between the charge carriers and ionic lattice, or electron-phonon interaction . On  one hand, the former phenomenon of  Coulombic attraction between the electron and the hole can bind them together, creating  a hydrogen atom-like state called an
exciton. In ionic , or even in covalent semiconductors, the spatial extent of  such
excitons typically spans  several unit cells of the crystal lattice; these excitons described using a Hydrogen-like model are termed as Wannier excitons ~\cite{Wannier_1937,Knox_1983,Dresselhaus_1956} (see Fig.~\ref{fig3}). In this simple picture, the
typical electron-hole distance is represented by the exciton Bohr radius
$a_{B} = \frac{4\pi\varepsilon_{0}\varepsilon_{\infty}\hslash^{2}}{\mu e^{2}}$,
where $\mu$ is a reduced effective mass and the Coulomb attraction is
assumed to be screened by the macroscopic high-frequency dielectric constant
$\varepsilon_{\infty}$.

\begin{figure}[htb]
\includegraphics[]{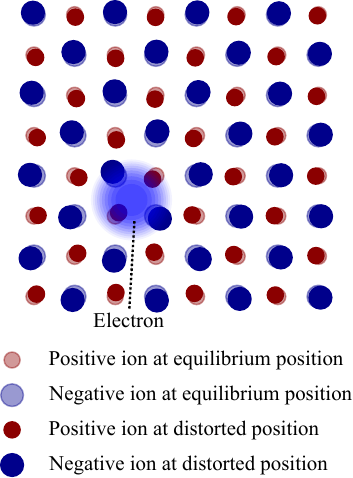}
\caption{ Artist view of the polar lattice distortion
for a negative Fr\"ohlich polaron. The electron is represented by a large shallow sphere, while lattice anions or cations are represented by smaller negatively or positively charged spheres. }\label{fig1}
\end{figure}

\begin{table*}[]
\def\arraystretch{1.25}
\centering
\begin{tabular}{c|c|c|c|c|c|c}
\hline  
       & \multicolumn{2}{c|}{\textbf{Uncorrelated e-h}}                                           & \multicolumn{4}{c}{\textbf{1S ground state exciton}}                                                                                         \\ \hline 
\textbf{Continuum-based models }                    & Self-         & Reduced                                            & Self-         & Reduced                                            & Mutual  & Effective dielectric  \\
        & energies & mass & energies & mass & interaction & constant \\ \hline 
Wannier (high frequency)                   & 0                     & $\mu$                                                  & 0                     & $\mu$                                                  & Coulomb            & $\varepsilon_{\infty}$        \\
Wannier (low frequency)                    & 0                     & $\mu$                                                  & 0                     & $\mu$                                                  & Coulomb            & $\varepsilon_{s}$             \\
Haken                                      & $\sigma_{e(h)}^{\text{LLP}}$ & $\mu^{\text{LLP}}$                                            & $\sigma_{e(h)}^{\text{LLP}}$ & $\mu^{\text{LLP}}$                                            & Hak                & $\varepsilon_{\text{eff}}$           \\
Bajaj                                      & $\sigma_{e(h)}^{\text{LLP}}$ & $\mu^{\text{LLP}}$ & $\sigma_{e(h)}^{\text{LLP}}$ & $\mu^{\text{LLP}}$ & Baj                & $\varepsilon_{\text{eff}}$           \\
Pollman-Buttner, weak coupling             & $\sigma_{e(h)}^{\text{LLP}}$ & $\mu$                                                  & $\sigma_{e(h)}^{\text{LLP}}$ & $\mu$                                                  & PB$_{\text{app}}$           & $\varepsilon_{\text{eff}}$           \\
Pollman-Buttner, weak coupling, modified   & $\sigma_{e(h)}^{\text{LLP}}$ & $\mu^{\text{LLP}}$ & $\sigma_{e(h)}^{\text{LLP}}$ & $\mu^{\text{LLP}}$ & PBK$_{\text{app}}$           & $\varepsilon_{\text{eff}}$           \\
Adamowski-Bednarek-Suffczynski             & $\sigma_{e(h)}^{\text{LLP}}$ & $\mu$                                                  & $\sigma_{e(h)}^{\text{eff}}$ & $\mu$                                                  & ABS                & $\varepsilon_{\text{eff}}$           \\
Adamowski-Bednarek-Suffczynski, simplified & $\sigma_{e(h)}^{\text{LLP}}$ & $\mu$                                                  & $\sigma_{e(h)}^{\text{eff}}$ & $\mu$                                                  & ABS$_{\text{app}}$         & $\varepsilon_{\text{eff}}$           \\
Pollman-Buttner                            & $\sigma_{e(h)}^{\text{LLP}}$ & $\mu$                                                  & $\sigma_{e(h)}^{\text{eff}}$ & $\mu$                                                  & PB                 & $\varepsilon_{\text{eff}}$           \\
Kane                                       & $\sigma_{e(h)}^{\text{LLP}}$ & $\mu^{\text{LLP}}$ & $\sigma_{e(h)}^{\text{eff}}$ & $\mu^{\text{eff}}$ & PBK                & $\varepsilon_{\text{eff}}$           \\
Iadonisi                                   & $\sigma_{e(h)}^{\text{LLP}}$ & $\mu^{\text{LLP}}$ & $\sigma_{e(h)}^{\text{LLP}}$ & $\mu^{\text{eff}}$ & Iad                & $\varepsilon_{\text{eff}}$          \\ \hline  
\end{tabular}%
\caption{Continuum-based model parameters used
for uncorrelated charge carriers and the 1S exciton ground state. The acronyms used in the text to identify each model are listed in the “mutual interaction” column.}
\label{tab:table2}
\end{table*}

On the other hand, the coupling of electrons or holes to the polar lattice of ionic 
semiconductors (see Fig.~\ref{fig1}) is
generally dominated by the long-range electric field created by
zone-centre longitudinal optic (LO) phonons. The original description of
this phenomenon was introduced by Fr\"ohlich as early as
1937,~\cite{Frohlich1937} hence its designation as the Fr\"ohlich
electron-phonon (e-ph) interaction. In the case of a diatomic lattice with the
approximation of a single undispersed optical phonon branch, Fr\"ohlich
managed to relate in an elegant way the microscopic e-ph coupling
Hamiltonian to the difference between the macroscopic high frequency and
static dielectric constants, $\varepsilon_{\infty}$ and
$\varepsilon_{s}$, respectively (see Fig.~\ref{fig3}). 
This is intuitive since $\varepsilon_{s}$
includes the additional contribution of ionic lattice motion to the dielectric response , which is absent at higher frequencies. The strength of the e-ph coupling in such a case is
usually represented by  a dimensionless constant~\cite{Frohlich_1950}
\begin{equation} \label{eq1}
\alpha_{e} = \frac{e^{2}}{4\pi\varepsilon_{0}\hslash}\left( \frac{m_{e}}{2\hbar \omega_{\text{LO}}} \right)^{1/2}\frac{1}{\varepsilon^{*}}
\end{equation}
\noindent
where $m_{e}$ and $e$ are the electron effective mass and charge,
the polarizability of the ionic lattice is represented by
$\frac{1}{\varepsilon^{*}} = \left( \frac{1}{\varepsilon_{\infty}} - \frac{1}{\varepsilon_{s}} \right)$
and $\hbar \omega_{\text{LO}}$ is the energy of the characteristic polar
LO phonon at the $\Gamma$ point of the Brillouin zone (BZ) in the diatomic lattice.
$\alpha_{e} \ll 1\ $defines the weak coupling regime, relevant for
classical III-V semiconductors such as GaAs, InP or InAs, while the
intermediate coupling regime   corresponds roughly to
$\alpha_{e}\sim 3$, and is relevant for soft ionic semiconductors such as metal-halide perovskites. In the intermediate and strong coupling regimes,
the physical picture is that of an electron surrounded -- ``dressed'' --
at all times by a lattice distortion cloud~\cite{Devreese2016,Franchini_2021}. Such a state is termed as a polaron (see Fig.~\ref{fig1}) whose characteristic size is given by the so called ``free polaron
radius''

\begin{equation} \label{eq2}
R_{e} = \left( \frac{\hslash}{2{m_{e}\omega}_{\text{LO}}} \right)^{1/2}.
\end{equation}
\noindent
The radius is larger for particles with small masses and soft lattices.
The dimensionless constant can be rewritten as
$\alpha_{e} = \frac{e^{2}}{8\pi\varepsilon_{0}\hbar \omega_{\text{LO}}R_{e}}\frac{1}{\varepsilon^{*}}$
evidencing that $\alpha_{e}$ includes $R_{e}$ as well as the lattice
polarizability through $\frac{1}{\varepsilon^{*}}$. In terms of their impact on physical properties, polaronic effects are important for free charge carriers and may deeply affect charge transport and optical absorption~\cite{Devreese2016}.

 The strength of polaronic effects were historically demonstrated to be connected to the ionicity of conventional semiconductors. This can be traced back to the dependence of $\alpha_{e}$ on the lattice polarisability. Recently, an additional effect has sparked interest amid the boom in new classes of semiconductors such as halide perovskites, where the softness of the lattice may also play an important role. This can be qualitatively understood considering the inverse dependence of $\alpha_{e}$ on the phonon frequency.

Electron-hole pairs that are the subject of both strong excitonic and polaronic
effects are host to a complex interplay  at the characteristic
lengthscales: $a_{B}$ and $R_{e}$,  as the correlations between charge
carriers may also impact lattice ion displacements. Such scenarios have been described using the concept of excitonic polaron (also sometimes referred to as polaronic
excitons or excitons-polarons).   Ample and 
longstanding theoretical literature has been devoted to the description
of such excitonic polarons  and of their optical properties
(absorption, photoluminescence, \ldots), using a variety of approaches
such as variational methods, path-integral techniques, and more recently
atomistic density-functional simulations~\cite{Antonius_2022,Adamska_2021,Dai_2024,Bai_2024}. This body comprised of seminal works is, however, difficult to parse due to the multiplicity of
methods, notations and unit conventions, and remains largely  concerned with
simple model systems akin to Fr\"ohlich's original case study. At this juncture, the problem of excitonic polarons is more relevant than ever
due to the rise of halide perovskite materials, complex crystals
characterized by strong ionicity, structural disorder and multiple optic
phonon branches (see Appendix \ref{AppA}). Consequently, several groups have 
recently attempted to describe  excitonic polaron effects in strongly polar materials through
first-principles methodologies~\cite{Antonius_2022,Adamska_2021,Dai_2024}. Thus, a systematic overview of theoretical approaches for description of excitonic polarons  in real materials has become  pertinent at this juncture to  draw insights from
these theories .

In this review, we present a complete non-atomistic and empirical
theoretical framework  which treats the physics of free polarons (Fig.~\ref{fig1}) and excitonic polarons (Fig.~\ref{fig4}(a-b))  on an equal footing (Fig.~\ref{fig3}), in the regimes from weak to intermediate strength of e-ph interaction. 
It can provide insights into
physical properties and can readily be connected to experimental data from
various semiconductors without the need to resort to complex and
computationally expensive first-principles or machine
learning/artificial intelligence driven approaches. We will base our
description of excitonic polarons~\cite{Toyozawa_1968,Hermanson_1970} on the
Pollman-B\"uttner's (PB) model,~\cite{Pollmann_1975,Pollmann_1977} later refined by
Kane~\cite{Kane_1978} (herein quoted as the Pollman-B\"uttner-Kane
(PBK) model). As for the strength of the exciton-phonon interaction, it 
can be qualitatively divided into three regimes: weak, intermediate and strong coupling. 
However, in his 1978 paper Kane pointed out~\cite{Kane_1978}: \textit{`Pollmann and Büttner 
… introduced a method patterned after the intermediate coupling regime of LLP [Lee-Low-Pines] designed to 
treat the electron polaron … '}. Therefore, we stress right from this introduction that none of the models related to PB 
is designed to be relevant for the strong coupling regime. Besides, some recent papers 
report on the use of PB's model, whereas in practice an approximate version is used instead, which is valid only in the weak coupling regime. For instance, the one employed for metal halide perovskites~\cite{Men_ndez_Proupin_2015,Lizaraga2025,Muhammad2025,Park_2022} has historically been introduced by Pollman-Büttner for the weak coupling regime (herein named PB$_{\text{app}}$), neglecting  electron-hole correlations in the interaction potential (vide infra). PB$_{\text{app}}$ is thus irrelevant for the intermediate coupling regime. 
Here, we will investigate in details the analytic derivations of various empirical 
models to carefully assess their ranges of validity. (See Table \ref{tab:table2} for the zoology of the models.) 

It seems natural to draw on the theories developed for polarons, which are
of a wide variety. The path integral method is known to provide an accurate description of
free polarons from the weak to the strong coupling regime,~\cite{Feynman_1955} but the 
associated literature on excitonic polarons is scarce and the numerical results need to be  
compared with other approaches~\cite{Adamowski_1981,Park_2022}. Interestingly, the 
transformation introduced initially by Toyozawa and Hermanson (TH) for excitonic polarons~\cite{Toyozawa_1968,Hermanson_1970} affords a proper connection to 
the Lee-Low-Pines (LLP) theory~\cite{Lee_1952,Lee_1953} developed for free polarons. Even 
so other frameworks exist, e.g. the alternative variational approaches for free polarons 
by Haga~\cite{Haga_1955} and excitonic polarons by Aldrich~\cite{Aldrich_1977}, 
{\it our aim is to use the same theoretical framework for the various interaction regimes 
(Fig.~\ref{fig3} and Tab.~\ref{tab:table2})} and therefore we will  specifically develop upon the 
work by LLP, TH and PB. We do not include comparison to work based on quasiparticle path 
integral molecular dynamics,~\cite{Park_2022}, since the original work does not provide a 
consistent set of parameters. More, their comparison is limited to PB$_{\text{app}}$, suitable for 
weak-coupling, while PB or PBK models are the most relevant, across the weak to intermediate coupling regimes. 

\begin{figure}[htb]
\includegraphics[width=0.48\textwidth]{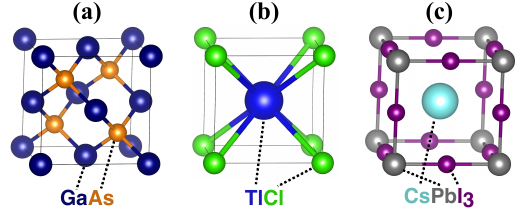}
\caption{ Schematic representation of (a) the zinc-blende
structure of GaAs (b) the CsCl structure of TlCl (c) the perovskite
structure of CsPbI\textsubscript{3}. \label{fig2}
}
\end{figure}

This review not only provides a critical comparison of various 
approximate models available in the literature, but also includes 
several original analytical expressions, in particular 
for the specific case of $m_{h} \approx m_{e}$.  This analytical description includes an
effective interaction mediated by the lattice, self-energies, center of mass motion, virtual phonon populations, Huang-Rhys factors for phonon sidebands and total energy of the excitonic polarons. We further extend PB's framework to include the multimode character of the lattice dynamics, particularly suitable for metal halide perovskites, using Toyozawa's generalization of Fr\"ohlich semi-empirical approach to polar optical e-ph coupling 
Hamiltonian.~\cite{Devreese1972,Yu_2016} This enables us to apply the framework to various semiconductors exhibiting the classical zinc-blende (GaAs, Fig.~\ref{fig2}a), CsCl (TlCl, Fig.~\ref{fig2}b) or perovskite (CsPbI\textsubscript{3}, Fig.~\ref{fig2}c) crystallographic structures, covering a wide range of e-ph coupling strengths. Among these materials, we will put special emphasis on halide perovskites (Fig.~\ref{fig2}c) which are most relevant to emerging optoelectronic technologies. A brief summary on their basic physical properties can be found in Appendix~\ref{AppA}.
On the other hand, TlCl (Fig.~\ref{fig2}b) offers an ideal playground to
test the accuracy of the exciton-phonon coupling models as this material
shares common features with more complex halide perovskite materials
(Tab.~\ref{tab:Table1}), namely: 1) almost equal electron and hole masses, 2) strong
ionicity, 3) intermediate coupling regime, but 4) a simpler situation
where a single polar optical phonon is a valid approximation.

Besides, the present work also complements the recent theoretical developments providing an
atomistic description of free and excitonic polarons~\cite{Lafuente_Bartolome_2022,Dai_2024,Alvertis2024,Jin_2024,Bai_2024,Dai_2024,Sio_2019}.
Some of these recent developments might be already relevant to the weak to
intermediate coupling regime discussed in this work, but thorough
numerical demonstrations are needed.~\cite{Alvertis2024} The PB
framework has the advantage of being numerically very efficient and already extended to 
more complex problems including bipolarons,~\cite{Bassani_1991} 
biexcitons,~\cite{Mokross_1979} but
also 2D quantum wells or 0D nanostructures, with inclusion of both dielectric and quantum 
confinement effects for excitons and phonons.~\cite{Zheng_1998,Oshiro_1999,Beril_1992,Licari_1977}

The review is structured as follows. Sec.~\ref{SecII} presents
the convention used for units as well as materials parameters, before introducing 
the Wannier 
model for excitons. These initial considerations end with a brief discussion on how the
polar nature of the lattice can be accounted for based on an \emph{ad 
hoc} effective dielectric constant. Sec.~\ref{SecIII} focuses on a semi-empirical model
for a free polaron (a single electron or hole moving in a polar
semiconductor, Fig.~\ref{fig1})), including the variational solution 
of Lee, Low and Pines, shedding light on the significance of the so-called phonon cloud 
and its relation to the free polaron radius (Fig.~\ref{fig1}). 
In  Sec.~\ref{SecIV}, we extend this model to excitonic polarons and go through the
variational solution derived by Pollman and B\"uttner, later refined by
Kane. This leads to a clarification of the interplay between
electron-hole correlations and polaronic regimes, and their impact on
the effective mass renormalization and electronic band gap
renormalization.  Section~\ref{SecV} is devoted to a discussion of effective
electron-hole interaction potentials, which can be derived from a full
excitonic polaron model and offer a simple way to account  for the
coupling to the lattice. In Sec.~\ref{SecVII} we extend empirical
approaches to more realistic situations in which multiple phonon modes
are coupled to excitons. This makes it possible to explicitly account 
for the complexity of the halide perovskite lattice dynamics due to the
multimode character of the lattice. Sec.~\ref{SecVI} discusses possible links between
empirical models and complementary first-principles approaches based on
density-functional theory. Sec.~\ref{SecVIII} includes the determination 
of physical observables: population of phonons, intermediate dielectric constants 
needed for realistic description of the excitons,~\cite{Even_2014} 
origin of phonon sidebands observed close to the photoluminescence peak. 
The final section~\ref{SecIX} is devoted to conclusions and future directions.  Appendix~\ref{AppA} provides a brief perspective on the physics of halide perovskites. Appendices~\ref{AppB}--\ref{AppK}, gather analytical derivations and theoretical approximations supporting discussions of the main text. Those include both original expressions as well as preexisting ones, with the aim of reducing the perpetuation of errors in the literature.

\section{Initial considerations} \label{SecII}

\subsection{Unit conventions and material parameters}

Throughout this review, we use the international system of units (SI) with usual 
meanings of $\varepsilon_{0}$ and $\hslash$.
Material parameters used for excitonic polaron models, with a single
phonon mode, are summarized in Tab.~\ref{tab:Table1}. They include
the relative effective masses for the electron and the hole, the high
frequency and static relative dielectric constants, as well as the effective
longitudinal optical phonon energies. All material parameters for GaAs,
CdS, TlCl (\ref{fig2}) are taken from ref.~\cite{Kane_1978} , with a slight
modification for TlCl, where the average of $m_{e} = 0.37$ and
$m_{h} = 0.36$ is considered. For halide perovskites, effective
longitudinal optical phonon energies $\hbar \omega_{\text{LO}}$ (values
given with an asterisk *) are deduced within the PBK model so as to
retrieve the experimental exciton binding energy of 16/25/64 meV for
MAPbI\textsubscript{3}~\cite{Miyata_2015}/MAPbBr\textsubscript{3}~\cite{Galkowski_2016}/CsPbCl\textsubscript{3}.~\cite{Baranowski_2020}
The static and high frequency relative dielectric constants of
MAPbI\textsubscript{3} and MAPbBr\textsubscript{3} are taken from 
ref.~\cite{Sendner_2016}, $\varepsilon_{\infty} = 3.7$ for
CsPbCl\textsubscript{3} is estimated from the experimental
optical LO and TO phonon frequencies~\cite{Wakamura_2001} using
Cochran-Cowley relation~\cite{Cochran_1962}, which is a generalization
of Lyddane-Sachs-Teller equation~\cite{Lyddane_1941} (Eq.~\eqref{eq13}). The
relative effective masses are deduced from the experimentally determined
reduced mass,~\cite{Galkowski_2016,Miyata_2015,Sendner_2016} assuming $m_{e} = m_{h}$. Data
for LiF are taken from Dai and coworkers.~\cite{Dai_2024} Computed
free polaron radii (Eq.~\eqref{eq2}) and $\alpha$ values (Eq.~\eqref{eq1}) are also given.

\begin{table*}[htb]
\def\arraystretch{1.5}
\centering

\begin{tabular}{c||c|c|c|c|c|c|c|c|c|c|c}
\hline \hline 
\multicolumn{2}{c|}{} &  \multicolumn{5}{c|}{\textbf{parameters}}   & \multicolumn{5}{c}{\textbf{free polaron model}}    \\ \hline \hline

\textbf{Semiconductor} & \textbf{Structure type} & \textbf{Lattice} & $\bm{m_e}$ & $\bm{m_h}$ & $\bm{\varepsilon_{\infty}}$ & $\bm{\varepsilon_s}$ & $\bm{\hbar\omega_{\text{\textbf{LO}}}}$ & $\bm{R_e}$ & $\bm{R_h}$ & $\bm{\alpha_e}$ & $\bm{\alpha_h}$ \\

& & \textbf{constant (\AA{})} &  &  &  & & \textbf{(meV)} & \textbf{(nm)} & \textbf{(nm)} &  &  \\ \hline \hline

\textbf{GaAs}          & Zinc blende             & 5.7                                                            & 0.0685         & 0.500          & 11.1                            & 13.1                     & 36.8                                     & 3.8                 & 1.4                 & 0.07                & 0.19                \\
\textbf{CdS}           & Zinc blende             & 5.4                                                            & 0.185          & 0.700          & 5.3                             & 8.6                      & 36.8                                     & 2.4                 & 1.0                 & 0.61                & 1.48                \\
\textbf{TlCl}          & Caesium Chloride        & 3.8                                                            & 0.365          & 0.365          & 5.1                             & 37.6                     & 21.5                                     & 2.2                 & 2.2                 & 2.58                & 2.58                \\
\textbf{MAPbI$_3$}     & Perovskite              & 6.3                                                            & 0.208          & 0.208          & 5.0                             & 34.8                     & 8.2*                                     & 4.7                 & 4.7                 & 3.18                & 3.18                \\
\textbf{MAPbBr$_3$}    & Perovskite              & 5.9                                                            & 0.234          & 0.234          & 4.3                             & 25.5                     & 13.1*                                    & 3.5                 & 3.5                 & 3.01                & 3.01                \\
\textbf{CsPbCl$_3$}    & Perovskite              & 5.5                                                            & 0.404          & 0.404          & 3.7                             & 26.0                     & 25.6*                                    & 1.9                 & 1.9                 & 3.40                & 3.40                \\
\textbf{LiF}           & Sodium Chloride         & 4.0                                                            & 0.880          & 4.40           & 2.0                             & 10.6                     & 77                                       & 0.75                & 0.33                & 4.94                & 11.0        \\ \hline \hline         
\end{tabular}
\caption{ Ground state parameters for excitonic polaron
calculations with a single polar optical phonon and corresponding data}
\label{tab:Table1}
\end{table*}
 
\subsection{The Wannier exciton model}
It is convenient to start with the classic Wannier Hamiltonian based on
effective mass description of the interacting electron and hole
(exciton denoted by $X$):~\cite{Wannier_1937,Dresselhaus_1956,Knox_1983}
\begin{equation} \label{eq3}
{\widehat{H}}_{X} = E_{g} + \frac{{{\widehat{p}}_{e}}^{2}}{2m_{e}} + \frac{{{\widehat{p}}_{h}}^{2}}{2m_{h}} - \frac{e^{2}}{4\pi\varepsilon_{0}\varepsilon_{\text{eff}}\left| {\overrightarrow{r}}_{e} - {\overrightarrow{r}}_{h} \right|},
\end{equation}
where ${\overrightarrow{r}}_{e}$ (${\overrightarrow{r}}_{h}$)
denotes the electron (hole) coordinates, ${\widehat{p}}_{e}$
(${\widehat{p}}_{h}$) its momentum, $m_{e}$ ($m_{h}$) its
effective mass, $\varepsilon_{\text{eff}}$ is a relative effective
dielectric constant and $E_{g}$ corresponds to the fundamental band
gap of the semiconductor. To account for the polar coupling of charge
carriers and match the experimental result for the exciton binding
energy, a relative effective dielectric constant is sometimes taken
arbitrarily in the range: $\varepsilon_\infty < \varepsilon_{\rm eff} < \varepsilon_{\rm s}$. At this stage, several points
can be stressed. This Hamiltonian is formulated in real space, which
allows an attractive analogy with the text-book quantum mechanical
problem of the hydrogen atom. It approximates the more rigorous
Bethe-Salpeter equation (BSE) for the exciton in a periodic solid, which
is expressed in reciprocal space. In the Wannier Hamiltonian framework,
the description of a non-interacting electron (hole) is implicitly
approximated by a wavefunction of the form:
\begin{eqnarray} \label{eq4}
\varphi_{{\overrightarrow{k}}_{e}}\left( {\overrightarrow{r}}_{e} \right) &\approx& u_{{\overrightarrow{k}}_{e} = 0}\left( {\overrightarrow{r}}_{e} \right)\frac{e^{i{\overrightarrow{k}}_{e}.{\overrightarrow{r}}_{e}}}{\sqrt{V}},\ \\ \varphi_{{\overrightarrow{k}}_{h}}\left( {\overrightarrow{r}}_{h} \right) &\approx& u_{{\overrightarrow{k}}_{h} = 0}\left( {\overrightarrow{r}}_{h} \right)\frac{e^{i{\overrightarrow{k}}_{h}.{\overrightarrow{r}}_{h}}}{\sqrt{V}}.
\end{eqnarray} 
\noindent
Here, ${\overrightarrow{k}}_{e}$ (${\overrightarrow{k}}_{h}$ )
denotes the electron (hole) wavevector and $V$ is the unit cell
volume. The details of the lattice periodicity and atomic structure are
accounted by the Bloch function
$u_{{\overrightarrow{k}}_{e} = 0}\left( {\overrightarrow{r}}_{e} \right)$
($u_{{\overrightarrow{k}}_{h} = 0}\left( {\overrightarrow{r}}_{h} \right)$), 
while the effective mass $m_{e}$ ($m_{h}$) is
derived from assumed parabolic electronic dispersion. Within this
approximation, the wavevector
${\overrightarrow{k}}_{e}\ ({\overrightarrow{k}}_{h}$) is expressed in
spherical coordinates with a modulus extended to $+ \infty$. The model can 
be further simplified by adopting center of
mass coordinates, since the total momentum $\widehat{P}$ commutes with
the Hamiltonian,

\begin{eqnarray} \label{eq5}
{\widehat{H}}_{X} = E_{g} + \frac{{\widehat{P}}^{2}}{2M} + \frac{{\widehat{p}}^{2}}{2\mu} - \frac{e^{2}}{4\pi\varepsilon_{0}\varepsilon_{\text{eff}}r},
\end{eqnarray} 

\noindent
where $M = m_{e} + m_{h}$ and
$\mu = \frac{1}{m_{e}} + \frac{1}{m_{h}}$ are the total and reduced
masses, respectively. Solutions are thus products of wavefunctions for
the center of mass coordinate $\overrightarrow{R}$ and for the
relative e-h motion given by $\overrightarrow{r}$
($\overrightarrow{r} = {\overrightarrow{r}}_{e} - {\overrightarrow{r}}_{h}$),
with a well-known expression for the 1S exciton ground
state~\cite{Knox_1983}:

\begin{eqnarray} \label{eq6}
\varphi_{1S,K}\left( \overrightarrow{r} \right) = \frac{e^{- r / a_{B}}}{{a_{B}}^{3/2}\pi^{1/2}} \times \frac{e^{i\overrightarrow{K}.\overrightarrow{R}}}{\sqrt{V}},
\end{eqnarray}
\noindent
where $a_{B}$ is the 1S exciton Bohr radius, which scales as
\begin{eqnarray} \label{eq7}
a_{B}\left( \varepsilon_{\text{eff}},\mu \right) = a_{B}^{\text{vac}} \times \left( \frac{\varepsilon_{\text{eff}}}{\mu} \right)
\end{eqnarray}
and $\overrightarrow{K}$ is the sum of wavevectors. In the quest for a
simple picture for the Wannier exciton, it is tempting to assume that
the electron and hole are point charges. 
It should be noted here that, in this particular case, the same 
validity limits as those of Bohr's model of the hydrogen
atom apply.~\cite{Wannier_1937,Dresselhaus_1956,Knox_1983} 
Therefore, constructing e-h potential
including effects of the Fr\"ohlich interaction by assuming point
charges~\cite{Emin_2018} must be handled very carefully, especially
for 1S exciton in the intermediate coupling regime.

\begin{figure*}[]
\includegraphics[width=0.99\textwidth]{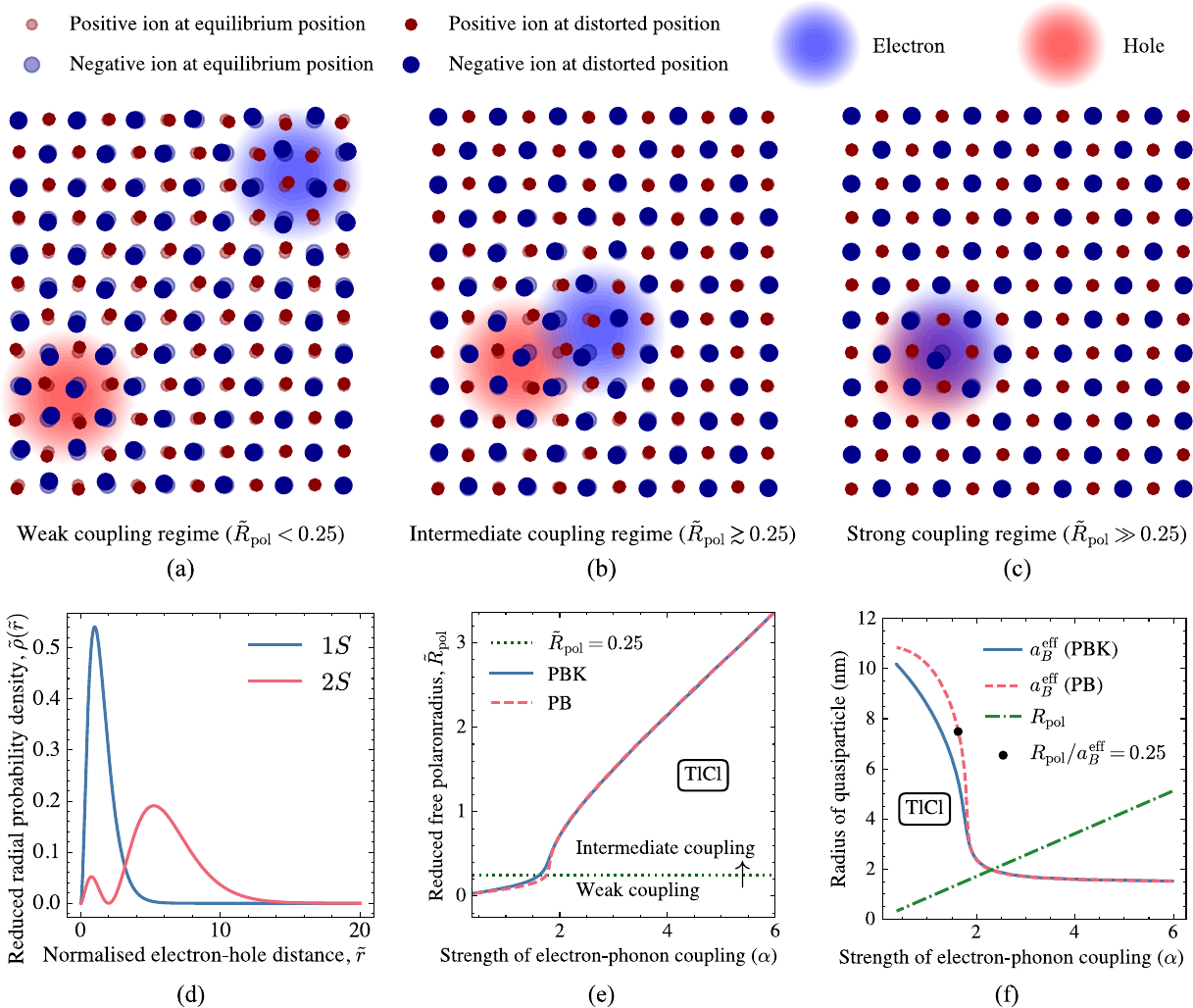}
\caption{Illustration of the various regimes where the e-h
correlations influence free polaronic distortions. The regimes can be qualitatively 
separated based on the ratio between the free polaron radii (here we use the same radius 
$R_{\text{pol}} = R_{e} = R_{h}$ for the electron and the hole) and the 1S exciton Bohr 
radius $a_{B}$. $\ {\widetilde{R}}_{\text{pol}}$ denotes the reduced free polaron radius
${\widetilde{R}}_{\text{pol}} = \frac{R_{\text{pol}}}{a_{B}}\ $.
(a-c)~Artist view of various interaction regimes leading to excitonic polarons. 
(a) Weak coupling regime where the e-h correlations weakly influence the free polaronic
distorsions. In these regime, empirical models detailed in Sec.~\ref{SecV} apply. 
(b) Intermediate coupling regime where the polaronic distortion differs from the simple 
superposition of the distortions related to free polarons. Models developed in  
Sec.~\ref{SecIV} are needed.
(c) Strong coupling regime where none of the empirical models described in the review 
apply. Alternative approaches based on path-integrals or first principles calculations 
may be implemented, see Sec.~\ref{SecVI}.
(d)~Reduced radial probability densities $\widetilde{\rho}\left( \widetilde{r} \right) = 
\rho\left( r \right)*a_{B}\ $ of hydrogenic 1S and 2S Wannier exciton wavefunctions as a 
function of the reduced e-h distance $\widetilde{r} = \frac{r}{a_{B}}\ $ (the same 
value of $a_{B}$ is considered here for both 1S and 2S excitons). 
(e) Reduced free polaron radius as obtained within the PBK model 
(blue straight line) and the PB model (red dashed line) as a function of $\alpha$ taking
materials parameters for TlCl (Tab.~\ref{tab:Table1}). The dashed horizontal line indicates
that the crossover from the weak to intermediate exciton-polaron coupling regime
occurs for ${\widetilde{R}}_{\text{pol}} = 1/4$, i.e.
${4R}_{\text{pol}} = a_{B}$. The differences between the PB and
PBK models stem from a small difference between the computed Bohr radii $a_{B}$.
(f) Free polaron radius (green dashed line), Bohr radius within the PBK model (blue line) and the PB model (red line) as a function of $\alpha$ taking
materials parameters for TlCl (Tab.~\ref{tab:Table1}). The variation of $\alpha$ is
obtained by varying the effective phonon frequency $\hbar \omega_{\text{LO}}$. 
\label{fig4}}
\end{figure*}

The 1S ground state (GS) exciton energy is given by:
\begin{eqnarray} \label{eq8}
E_{GS,\ K} = E_{\text{uncorr}} - Ry
\end{eqnarray}
where:
\begin{eqnarray} \label{eq9}
E_{\text{uncorr}} = E_{g} + \frac{\hslash^{2}K^{2}}{2M}
\end{eqnarray}
denotes the energy of uncorrelated carriers, and $\text{Ry}$ is the
Rydberg energy of the Wannier exciton, which scales as
\begin{eqnarray} \label{eq10}
\text{Ry}\left( \varepsilon_{\text{eff}},\mu \right) = \text{Ry}^{\text{vac}} \times \left( \frac{\mu}{{\varepsilon_{\text{eff}}}^{2}} \right).
\end{eqnarray}
The 1S exciton binding energy is $E_{b,1S} = Ry$ in this Wannier
model. 

Independent measurements of Landau levels and exciton binding
energy allow the determination of the effective reduced mass
$\mu^{\text{eff}}$ and dielectric constant
$\varepsilon_{\text{eff}}$ (see Appendix~\ref{AppA}).~\cite{Miyata_2015} However,
within the framework of the Wannier model and from the perspective of applying the model, the choice of the value of
$\varepsilon_{\text{eff}}$ is not obvious. 
Furthermore, it has been observed experimentally
that $\mu^{\text{eff}}$ can deviate from the bare reduced mass 
$\mu$.~\cite{Baranowski_2024} To get more insight into the physics of
excitons and the various consequences of the coupling of charge carriers
to the lattice, it is necessary to switch to more advanced theoretical
descriptions. Indeed, these quantities can be derived using an approach
that explicitly includes electron-hole (e-h) correlations as well as the
interactions between charge carriers and phonons (Fig.~\ref{fig3}, Tab.~\ref{tab:table2} and vide infra).

\subsection{First approach to polar exciton-lattice coupling and the various interaction regimes}

While for the classification of polarons the coupling regime 
essentially refers to the value of the constant $\alpha_{e}$ 
Eq.~\eqref{eq1}),~\cite{Devreese2010_arxiv} for excitonic polarons the separation between
the various coupling regimes is essentially qualitative (Fig.~\ref{fig4}). 
It may rely on the physics of excitonic polarons, which is related to the interplay 
between the effect of electron-hole correlations due to the electrostatic interaction 
and the amount of overlap between the lattice distortion fields produced and the electron
and the hole. One way to gain qualitative understanding of the difference between
weak and intermediate coupling regimes, is to inspect the probability density of the 1S 
exciton wavefunction for variable e-h distances (Fig.~\ref{fig4}d) with respect to the free 
polaron radius ($R_{\text{pol}}$). This cross-over will be explored mathematically in  
Sec.~\ref{SecIV}. The dashed line in Fig.~\ref{fig4}e indicates that the crossover from the 
weak to intermediate excitonic polaron coupling regime occurs for 
${4R}_{\text{pol}} = a_{B}$, 
when the sum of the two polaron diameters matches the Bohr radius.

In the weak coupling limit an exciton can be qualitatively understood as 
two weakly interacting almost free polarons (Fig.~\ref{fig4}a). The two charges surrounded 
by their distortion fields (almost free polarons) are loosely bound by a weak electrostatic
interaction.  This is the reason why the polarisation effect of the lattice is maximum and 
the dielectric constant relevant to this regime is $\varepsilon_s$, instead of 
$\varepsilon_{\infty}$. GaAs is a good example for this regime.

In the strong coupling limit (Fig.~\ref{fig4}c), the distortion fields 
produced by the two opposite charges exactly compensate. In other words, the polaron 
effect vanishes for an excitonic polaron, while it is maximum for free polarons. 
LiF can be considered as a prototype for this limit, although other interactions may play 
a role and atomistic techniques may be required to accurately treat localisation effects 
(see Sec.~\ref{SecVI}). 

The intermediate excitonic polaron coupling regime (Fig.~\ref{fig4}b) can 
be schematically envisioned as a regime where the two dressed quasi-particles approximated 
by to spheres of radius $R_{\text{pol}}$ cannot be encompassed, without overlapping, by the 
sphere that characterizes the exciton, whose characteristic diameter is $a_{B}$. From the 
perspective of physical properties, the intermediate excitonic polaron coupling regime has
the most complex interplay between e-h correlations and couplings to the lattice. TlCl and
metal halide perovskites are good case studies for this regime. Technically speaking, the 
issue with the numerical comparison of the Bohr radius with the sum of the polaron radii, 
is that the actual Bohr radius is unknown a priori, but is actually an output of the 
excitonic polaron calculation if not extracted from experimental observables based on quite 
a few assumptions. Still, the actual Bohr radius is bound by the values related to the weak
($\varepsilon_s$) and strong ($\varepsilon_{\infty}$) limit cases. 

The cross-over (vide infra) from the weak coupling regime 
($\alpha \rightarrow 0,\ a_{B}^{\text{eff}} \gg R_{\text{pol}}$) to
the intermediate coupling regime
($a_{B}^{\text{eff}} < 4R_{\text{pol}}$) is obtained in the present
work by tuning the effective phonon frequency in the expression of
$\alpha$ [Eq.~\eqref{eq1}], as illustrated in (Fig.~\ref{fig4}e,f).

\begin{table*}[]
\def\arraystretch{1.5}
\centering
\begin{tabular}{c||c|c|c|c|c|c|c|c|c|c|c|c}
\hline \hline 
\textbf{Semiconductor} & $\bm{\mu}$ & $\bm{\mu^{\text{\textbf{PBK}}}}$ & $\bm{\mu^{\text{\textbf{LLP}}}}$ & $\bm{E^{\infty}_b}$ & $\bm{E^{\text{\textbf{Hak}}}_b}$  & $\bm{E^{\text{PBK}_\text{app}}_b}$  & $\bm{E^{\text{\textbf{PBK}}}_b}$  & $\bm{E^{\text{PB}}_b}$  & $\bm{E^{\text{PB}_\text{app}}_b}$  & $\bm{E^{s}_b}$ & $\bm{\varepsilon^{\text{\textbf{PBK}}}}$ & $\bm{a_{B^{\text{\textbf{PBK}}}}}$ \\

 &  &  &  & (meV) &  (meV) & (meV) &(meV) &  (meV) &  (meV) &  (meV) & &  \\ \hline \hline 

\textbf{GaAs}          & 0.0614     & 0.0619                  & 0.0623                  & 6.8                  & 5.1                      & 5.0                                 & 5.0                      & 4.9                     & 4.9                                & 4.9             & 13.0                       & 11.0                 \\
\textbf{CdS}           & 0.158      & 0.161                   & 0.180                   & 78                   & 50                       & 40                                  & 38                       & 38                      & 35                                 & 29              & 7.6                        & 2.1                  \\
\textbf{TlCl}          & 0.183      & 0.191                   & 0.290                    & 96                   & 52                       & 28                                  & 16                       & 16                      & 7.3                                & 1.8             & 12.6                       & 1.8                  \\
\textbf{MAPbI$_3$}     & 0.104      & 0.106                   & 0.184                   & 57                   & 40                       & 27                                  & 16                       & 16                      & 9.4                                & 1.1             & 9.4                        & 2.9                  \\
\textbf{MAPbBr$_3$}    & 0.117      & 0.119                   & 0.200                   & 86                   & 60                       & 41                                  & 25                       & 25                      & 15                                 & 2.4             & 8.0                        & 2.2                  \\
\textbf{CsPbCl$_3$}    & 0.202      & 0.205                   & 0.372                   & 201                  & 149                      & 105                                 & 64                       & 63                      & 38                                 & 4.1             & 6.6                        & 1.1          \\ \hline \hline         
\end{tabular}
\caption{ Excitonic polaron properties.  Bare, PBK and LLP
reduced masses. Exciton binding energies $E_{b}^{\infty}$ and
$E_{b}^{s}$ are computed using the bare reduced mass $\mu$ with
$\varepsilon_{\infty}$ or $\varepsilon_{s}$, respectively.
$E_{b}^{\text{Hak}}$ , $E_{b}^{\text{PBK}_{\text{app}}}$ are
computed using the LLP reduced mass $\mu_{\text{LLP}}$ and
$V_{\text{latt}}^{\text{Hak}}\left( \overrightarrow{r}\  \right)$ (Eqs.~\eqref{eq64}/\eqref{eq65}),
$V_{\text{latt}}^{\text{PB}_{\text{app}}}\left( \overrightarrow{r}\  \right)$
(Eqs.~\eqref{eq63}/\eqref{eq66}), respectively. $E_{b}^{\text{PB}_{\text{app}}}$ are
computed using the bare reduced mass $\mu$ with
$V_{\text{latt}}^{\text{PB}_{\text{app}}}\left( \overrightarrow{r}\  \right)$(Eq.
63/66). The reduced mass $\mu^{\text{PBK}}\ $is obtained together with
the binding energy $E_{b}^{\text{PBK}}$ , the effective dielectric
constant $\varepsilon^{\text{PBK}}$ and the Bohr radius
$a_{B}^{\text{PBK}}$ in the PBK model.~\cite{Kane_1978}}
\label{tab:Table3}
\end{table*}

\section{Free polaron in a polar semiconductor}  \label{SecIII}

\subsection{Fr\"ohlich electron-phonon coupling (Fr\"o)}

A compact form of the Fr\"ohlich (abbreviated by Fr\"o) e-ph coupling
Hamiltonian reads:
\begin{eqnarray} \label{eq11}
{\widehat{H}}^{\text{ Fr\"o}} = \sum_{\overrightarrow{k}}^{}\left( {e^{i\overrightarrow{k}.{\overrightarrow{r}}_{e}}g}_{\overrightarrow{k}}{\widehat{a}}_{\overrightarrow{k}} + {e^{- i\overrightarrow{k}.{\overrightarrow{r}}_{e}}g}_{\overrightarrow{k}}^{*}{\widehat{a}}_{\overrightarrow{k}}^{+} \right),
\end{eqnarray} 
where $g_{\overrightarrow{k}}$ is a scalar entering the Fr\"{o}hlich e-ph
coupling matrix element:
\begin{eqnarray} \label{eq12}
g_{\overrightarrow{k}} = - \frac{i}{k}\left( \frac{{e^{2}\hbar \omega}_{\text{LO}}}{2V\varepsilon_{0}\varepsilon^{*}} \right)^{1/2},
\end{eqnarray}
${\widehat{a}}_{\overrightarrow{k}}\ $(${\widehat{a}}_{\overrightarrow{k}}^{+}$)
is the creation (annihilation) operator for a longitudinal optical
phonon of wave vector $\overrightarrow{k}$ and energy
$\hbar \omega_{\text{LO}}$.

A continuum approach such as the one chosen by Fr\"ohlich reasonably
grasps the physics because interaction has a long-range character
through the $\frac{1}{k}$ term in Eq.~(\ref{eq12}).~\cite{Verdi_2015,Sio_2022} More, it
allows performing analytic or semi-analytic summations over $\overrightarrow{k}$ in 
spherical coordinates by extending the modulus to $+ \infty$, 
instead of limiting the summation to the first BZ, 
since the greatest contributions are related to
$k \rightarrow 0$.~\cite{Lee_1953,Sio_2022} For more complex situations
where multiple optical phonon branches contribute, the longitudinal optical phonon 
$\hbar \omega_{\text{LO}}$ in Eq.~(\ref{eq12}) is, as a first
approximation, considered to be an effective phonon. \

The macroscopic and microscopic physical quantities entering the 
Fröhlich Hamiltonian are related by
the Lyddane-Sachs-Teller (LST) equation:~\cite{Lyddane_1941}
\begin{eqnarray} \label{eq13}
\frac{\varepsilon_{s}}{\varepsilon_{\infty}} = \left( \frac{\omega_{\text{LO}}}{\omega_{\text{TO}}} \right)^{2},
\end{eqnarray}
transverse optical (TO) modes being associated to the longitudinal one. Using
the Fr\"ohlich Hamiltonian [Eq.~\eqref{eq11}] requires the knowledge of only a few
basic quantities namely the effective masses, the frequency of an
effective polar optical mode and the static and infinite limits of the
dielectric constant.~\cite{Poglitsch_1987,Onoda_Yamamuro_1992} However, the precise
experimental determination of these quantities is not an easy task in
many materials, and for complex materials such as 3D halide perovskites
the definition of the effective polar optical mode is not
straightforward. For these semiconductors, existing data are often
limited to bare carrier effective masses and effective dielectric
constants for the exciton. Available data on optical phonons are more
numerous, including results from Raman and neutron scattering, 
infrared and THz spectroscopies, as well as photoluminescence phonon side bands.  
Meanwhile, reported values show many inconsistencies. 
The analysis deserves to be placed in a wider 
context that includes at least the multimode character of halide perovskites, if 
not also its polymorphous nature with strong anharmonicity and local 
disorder.~\cite{Zacharias_2023b,Zacharias_2023}

The empirical e-ph Hamiltonian reads:
\begin{eqnarray} \label{eq14}
{\widehat{H}}_{\rm e - ph} &=& \frac{{{\widehat{p}}_{e}}^{2}}{2m_{e}} + \sum_{\overrightarrow{k}}^{}\left( {e^{i\overrightarrow{k}.{\overrightarrow{r}}_{e}}g}_{\overrightarrow{k}}{\widehat{a}}_{\overrightarrow{k}} + {e^{- i\overrightarrow{k}.{\overrightarrow{r}}_{e}}g}_{\overrightarrow{k}}^{*}{\widehat{a}}_{\overrightarrow{k}}^{+} \right) \nonumber \\ &+& \hbar \omega_{\text{LO}}\sum_{\overrightarrow{k}}^{}{{\widehat{a}}_{\overrightarrow{k}}^{+}{\widehat{a}}_{\overrightarrow{k}}}.
\end{eqnarray} 
Here, the term $\frac{\hbar \omega_{\text{LO}}}{2}$ has been left out, as
in the rest of the paper. From a perturbative calculation suitable for
the regime of weak e-ph interaction, Fr\"ohlich proposed to renormalize
the expression of the energy of the electron as follows:
\begin{eqnarray} \label{eq15}
E_{e}\left( k_{e} \right) \approx \sigma_{e} + \frac{\hslash^{2}{k_{e}}^{2}}{2m_{e}}\left( 1 - \frac{\alpha_{e}}{6} \right) + \ldots
\end{eqnarray}
The electron ground state is stabilized by a self-energy
term:~\cite{Frohlich_1950}
\begin{eqnarray} \label{eq16}
\sigma_{e} \approx - \alpha_{e}\hbar \omega_{\text{LO}}.
\end{eqnarray}
A similar expression can be considered for a hole, leading together to
corrections for the electronic band gap of the semiconductor.
Noteworthy, these corrections are missing in standard DFT calculations.
In addition, the quasiparticle acquires a heavier effective mass:
\begin{eqnarray} \label{eq17}
m_{e}^{*} \approx \frac{m_{e}}{\left( 1 - \frac{\alpha_{e}}{6} \right)},
\end{eqnarray} 
\noindent
where $m_{e}$ is referred to as the `bare' effective mass, that can be
computed within DFT. $m_{e}^{*}$ is the effective mass of the dressed
quasiparticle.

The perturbation approach of Fr\"ohlich turned out to provide reasonable
approximations for the carrier self-energies and renormalized masses,
not only in the free polaron weak coupling regime but also in the intermediate
one.~\cite{Devreese2010_arxiv} From the materials point of view, the intermediate
regime is usually present in polar ionic materials due to large
differences between $\varepsilon_{\infty}$ and $\varepsilon_{s}$.
Noteworthy, soft materials such as 3D perovskites with low energy
optical modes ($\hbar \omega_{\text{LO}}\ $\textless{} 10meV) are more
prone to exhibit a free polaron intermediate coupling regime.

\subsection{Empirical Hamiltonian for free polarons (LLP)}
Despite its apparent simplicity, Eq.~\eqref{eq14} has no analytical solutions,
and was the subject of intense research over decades due to its
experimental and theoretical implications. Interested readers may
refer to the series of review papers by J. T.
Devreese.~\cite{Devreese2010_arxiv} 

In their initial analysis, assuming that
the electrons adiabatically follow the ionic motion, Landau and Pekar
obtained expressions for the polaron self-energy and effective
masses.~\cite{LandauPekar1948} This approach valid for the strong e-ph
coupling regime can be tackled by unitary transformations of the
Hamiltonian represented in the basis of Gaussian trial functions for the
electron wavefunction.~\cite{LandauPekar1948} In the fifties, various
theoretical approaches were developed to reshape the problem into forms
suitable for efficient and accurate variational approaches. In 1953, LLP
introduced a series of unitary transformations connected to shift
operators that are determined from variational considerations. It also
questioned the validity of the assumption of adiabaticity for low energy
electrons.~\cite{Lee_1953} As stated by Devreese and coworkers, the
polaron theory of LLP {\it ``essentially treats the opposite case, where the
phonon field tends to follow the displacements of the
electron''}.~\cite{Devreese_1963} 

In 1955,~\cite{Aldrich_1977} Feynman
formulated the polaron problem in a different way using a Lagrangian
form of quantum mechanics, referred to as the path-integral method, also
coupled with a variational approach. The later has the advantage of
recovering also the limits known for the strong coupling regime
($\alpha_{e} \rightarrow + \infty$) and can be extended to tackle
exciton-phonon coupling.~\cite{Adamowski_1981} Unitary transformations
were also used to bridge the gap between weak and strong coupling
regimes for free polarons.~\cite{Huybrechts_1977,Tokuda_1980,Filippis_2003} The field remains very active with various extensions, e.g. with the exploration of topological~\cite{Lafuente2024,Luo2026} or chiral~\cite{Li2024,Wang2024} polarons. 

Here, we detail the method developed by LLP for free polarons, 
since its variational solution (Sec.~\ref{SecIIIC}) will serve as our reference
when we examine the empirical models developed for excitonic polarons 
(Sec.~\ref{SecIV}-\ref{SecVII}).  
Before commenting on the LLP approach for free polarons, let us consider
the infinite mass limit of ${\widehat{H}}_{\rm e - ph}$
\begin{eqnarray} \label{eq18}
{\widehat{H}}_{e - ph}^{\infty} &=& \sum_{\overrightarrow{k}}^{}\left( {e^{i\overrightarrow{k}.{\overrightarrow{r}}_{e}}g}_{\overrightarrow{k}}{\widehat{a}}_{\overrightarrow{k}} + {e^{- i\overrightarrow{k}.{\overrightarrow{r}}_{e}}g}_{\overrightarrow{k}}^{*}{\widehat{a}}_{\overrightarrow{k}}^{+} \right) \nonumber \\ &+& {\hslash\omega}_{\text{LO}}\sum_{\overrightarrow{k}}^{}{{\widehat{a}}_{\overrightarrow{k}}^{+}{\widehat{a}}_{\overrightarrow{k}}}.
\end{eqnarray}
As shown by Devreese,~\cite{Devreese2010_arxiv} using the unitary
transformation (See Appendix~\ref{AppB})
\begin{eqnarray} \label{eq19}
\widehat{W} = e^{\sum_{\overrightarrow{k}}^{}\left( {F_{\overrightarrow{k}}^{*}\widehat{a}}_{\overrightarrow{k}} - F_{\overrightarrow{k}}{\widehat{a}}_{\overrightarrow{k}}^{+} \right)},
\end{eqnarray}
with a shift operator $F_{\overrightarrow{k}}$~\cite{Devreese2010_arxiv} that can be understood as the amplitude of the lattice distortion
around the charge carrier (vide infra):
\begin{eqnarray} \label{eq20}
F_{\overrightarrow{k}} = \frac{g_{\overrightarrow{k}}^{*}}{{\hslash\omega}_{\text{LO}}},
\end{eqnarray}
yields a transformed Hamiltonian:
\begin{eqnarray} \label{eq21}
{\widehat{H'}}_{e - ph}^{\infty} &=& {\widehat{W}}^{+}{\widehat{H}}_{e - ph}  ^{\infty}\widehat{W} \nonumber \\ &=& - \sum_{\overrightarrow{k}}^{}\frac{\left| g_{\overrightarrow{k}} \right|^{2}}{{\hslash\omega}_{\text{LO}}} + {\hslash\omega}_{\text{LO}}\sum_{\overrightarrow{k}}^{}{{\widehat{a}}_{\overrightarrow{k}}^{+}{\widehat{a}}_{\overrightarrow{k}}},
\end{eqnarray}
where the first term represents self-energy. The sign convention for the
shift operator and the related unitary transformation follows the work
by PB. Eigenvalues can be derived and read:
\begin{eqnarray} \label{eq22}
E_{\left\{ n_{\overrightarrow{k}} \right\}} = - \sum_{\overrightarrow{k}}^{}\frac{\left| g_{\overrightarrow{k}} \right|^{2}}{{\hslash\omega}_{\text{LO}}} + {\hslash\omega}_{\text{LO}}\sum_{\overrightarrow{k}}^{}n_{\overrightarrow{k}},
\end{eqnarray}
where $n_{\overrightarrow{k}}$ is the phonon population. Phonon operators are shifted by 
$F_{\overrightarrow{k}}$ (see Appendix~\ref{AppB}):
\begin{eqnarray} \label{eq23a}
{\widehat{W}}^{+}{\widehat{a}}_{\overrightarrow{k}}\widehat{W} = {\widehat{a}}_{\overrightarrow{k}} - F_{\overrightarrow{k}}
\end{eqnarray}
and
\begin{eqnarray} \label{eq23b}
{\widehat{W}}^{+}{\widehat{a}}_{\overrightarrow{k}}^{+}\widehat{W} = {\widehat{a}}_{\overrightarrow{k}}^{+} - F_{\overrightarrow{k}}^{*}.
\end{eqnarray}
When considering the complete problem, LLP first noted that, when the
e-ph interaction is included, the momentum ${\widehat{p}}_{e}$ does
not commute with ${\widehat{H}}_{e - ph}$. For this reason, the total momentum
operator $\widehat{\wp}$ is introduced and reads:~\cite{Devreese2016}
\begin{eqnarray} \label{eq24}
\widehat{\wp} = {\widehat{p}}_{e} + \sum_{\overrightarrow{k}}^{}{\hslash\overrightarrow{k}{\widehat{a}}_{\overrightarrow{k}}^{+}{\widehat{a}}_{\overrightarrow{k}}},
\end{eqnarray}
with
$\left\lbrack \widehat{\wp},{\widehat{H}}_{e - ph} \right\rbrack = 0$,
${\widehat{H}}_{e - ph}\left|\left. \ \varphi\right\rangle\right.\  = E\left| \left. \ \varphi\right\rangle\right.\ $
and
$\widehat{\wp}\left|\left. \ \varphi\right\rangle\right.\  = \hslash\overrightarrow{Q}\left|\left. \ \varphi\right\rangle\right.$.
Therefore, $\hslash\overrightarrow{Q}$, the eigenvalue of
$\widehat{\wp}$, is introduced into the Hamiltonian thanks to the
unitary transformation (see Appendix~\ref{AppC}):
\begin{eqnarray} \label{eq25}
\widehat{U} = e^{i\left( \overrightarrow{Q} - \sum_{\overrightarrow{k}}^{}{\overrightarrow{k}{\widehat{a}}_{\overrightarrow{k}}^{+}{\widehat{a}}_{\overrightarrow{k}}} \right).{\overrightarrow{r}}_{e}}
\end{eqnarray}
\begin{widetext}
\begin{eqnarray} \label{eq26}
{\widehat{H}}_{\text{CM}}^{\text{LLP}} = {\widehat{U}}^{+}{\widehat{H}}_{e - ph}\widehat{U} = \frac{\left( \hslash\overrightarrow{Q} - \sum_{\overrightarrow{k}}^{}{\hslash\overrightarrow{k}{\widehat{a}}_{\overrightarrow{k}}^{+}{\widehat{a}}_{\overrightarrow{k}}} \right)^{2}}{2m_{e}} + \sum_{\overrightarrow{k}}^{}\left( g_{\overrightarrow{k}}{\widehat{a}}_{\overrightarrow{k}} + g_{\overrightarrow{k}}^{*}{\widehat{a}}_{\overrightarrow{k}}^{+} \right) + \hbar \omega_{\text{LO}}\sum_{\overrightarrow{k}}^{}{{\widehat{a}}_{\overrightarrow{k}}^{+}{\widehat{a}}_{\overrightarrow{k}}}.
\end{eqnarray}
\end{widetext}
The electron position and momentum have been removed from the
Hamiltonian and replaced by the polaron momentum, provided that the
unitary transformation is also applied to the eigenstates
$\left| \left. \ \varphi_{\text{CM}} \right\rangle \right.\  = {\widehat{U}}^{+}\left| \left. \ \varphi \right\rangle \right.\ $.
Then a second unitary transformation (Eq.~\eqref{eq19}), which includes the shift
operators, is applied. The LLP Hamiltonian reads:
\begin{widetext}
\begin{eqnarray} \label{eq27}
{\widehat{H}}^{\text{LLP}}\left( F_{\overrightarrow{k}}\left( \overrightarrow{Q} \right) \right) = {\widehat{W}}^{+}{\widehat{U}}^{+}{\widehat{H}}_{e - ph}\widehat{U}\widehat{W} = {\widehat{H}}_{0_{\text{ph}}}^{\text{LLP}} + {\widehat{H}}_{1_{\text{ph}}}^{\text{LLP}} + {\widehat{H}}_{2_{\text{ph}}}^{\text{LLP}} + {\widehat{H}}_{3_{\text{ph}}}^{\text{LLP}} + {\widehat{H}}_{4_{\text{ph}}}^{\text{LLP}}.
\end{eqnarray}
\end{widetext}
where
${\widehat{H}}_{n_{\text{ph}}}^{\text{LLP}}\left( F_{\overrightarrow{k}}\left( \overrightarrow{Q} \right) \right)$
are the n-phonon~($n_{\text{ph}}$) contributions to the Hamiltonian.
The eigenstates are now deduced from
$\left| \left. \ \psi \right\rangle \right.\  = {{\widehat{W}}^{+}\widehat{U}}^{+}\left| \left. \ \varphi \right\rangle \right.\ $.
$F_{\overrightarrow{k}}\left( \overrightarrow{Q} \right)$ is expected
to be determined by minimization of energy. The resulting one-electron
Hamiltonian is independent of the electronic coordinate
$\overrightarrow{r}$. This preserves the polaron total momentum as a
constant of motion and enables exact diagonalization of Hamiltonians of
Fr\"ohlich type with regard to the electronic
subspace.~\cite{Rapp_2000} Noteworthy, exact diagonalization no longer applies
for an equivalent treatment of the excitonic polarons by PB (vide
infra). We also note that the exact diagonalization of the one-electron
Hamiltonian through the LLP transformation is connected to the implicit
absence of an electronic periodic potential in this Hamiltonian, and
thus to the effective mass approximation [Eq.~\eqref{eq3}]. Alternative complex
approaches~\cite{Burovski_2008}, or unitary transformation such as the
Fulton--Gouterman (FG) transformation capable of handling periodicity of
potential, Umklapp terms and phonon anharmonicity do
exist.~\cite{Rapp_2000,Fulton_1961} Introducing lattice periodicity may
further reconcile semi-empirical approaches with on-going developments
connected to DFT calculations in the field of free 
polarons.~\cite{Verdi_2015,Vasilchenko_2024,Lafuente_Bartolome_2022,Lafuente_Bartolome_2024}

\subsection{Variational solution to the free polaron problem (LLP)} \label{SecIIIC}

In order to derive the variational solution within LLP, 
the Hamiltonian
${\widehat{H}}^{\text{LLP}}\left( F_{\overrightarrow{k}}\left( \overrightarrow{Q} \right) \right)\ $
is applied to a `free vacuum' e-ph ground state with zero (real) phonon
($0_{\text{ph}}$), retaining only
${\widehat{H}}_{0_{\text{ph}}}^{\text{LLP}}\left( F_{\overrightarrow{k}}\left( \overrightarrow{Q} \right) \right)$. This leads to the energy dispersion of the polaron:
\begin{widetext}
\begin{eqnarray} \label{eq28}
E_{\text{GS}}\left( \overrightarrow{Q} \right) &=& \frac{\hslash^{2}Q^{2} - 2\overrightarrow{Q}.\left( \sum_{\overrightarrow{k}}^{}{\hslash\overrightarrow{k}\left| F_{\overrightarrow{k}}\left( \overrightarrow{Q} \right) \right|^{2}} \right) + \left( \sum_{\overrightarrow{k}}^{}{\hslash\overrightarrow{k}\left| F_{\overrightarrow{k}}\left( \overrightarrow{Q} \right) \right|^{2}} \right)^{2}}{2m_{e}} \nonumber \\ &-& \sum_{\overrightarrow{k}}^{}\left( g_{\overrightarrow{k}}F_{\overrightarrow{k}}\left( \overrightarrow{Q} \right) + c.c \right) + \sum_{\overrightarrow{k}}^{}{\left| F_{\overrightarrow{k}}\left( \overrightarrow{Q} \right) \right|^{2}\left( \hbar \omega_{\text{LO}} + \frac{\hslash^{2}k^{2}}{2m_{e}} \right)}.
\end{eqnarray}
For the validity of this approximate representation of the free polaron wavefunction beyond the weak coupling regime, readers may refer to  ref.~\cite{Devreese_1963,Filippis_2003}). 
An additional relation between $g_{\overrightarrow{k}}$ and
$F_{\overrightarrow{k}}^{*}$ can be derived by energy minimization
$\frac{\partial E_{\text{GS}}\left( \overrightarrow{Q} \right)}{\partial F_{\overrightarrow{k}}\left( \overrightarrow{Q} \right)} = \ \frac{\partial E_{\text{GS}}\left( \overrightarrow{Q} \right)}{\partial F_{\overrightarrow{k}}^{*}\left( \overrightarrow{Q} \right)} = 0$
(see Appendix~\ref{AppD}):

\begin{eqnarray} \label{eq29}
F_{\overrightarrow{k}}^{\min}\left( \overrightarrow{Q} \right) =\frac{g_{\overrightarrow{k}}^{*}}{\hbar \omega_{\text{LO}} - \hslash^{2}\frac{\overrightarrow{Q}.\overrightarrow{k}}{m_{e}} + \frac{\hslash^{2}k^{2}}{2m_{e}} + \frac{\hslash^{2}}{m_{e}}\overrightarrow{k}.\sum_{\overrightarrow{k'}}^{}{\overrightarrow{k'}\left| F_{\overrightarrow{k'}}^{\min}\left( \overrightarrow{Q} \right) \right|^{2}}}  = \frac{g_{\overrightarrow{k}}^{*}}{{\hslash\omega}_{\text{LO}}\left( 1 + K^{2} - 2\overrightarrow{q},\overrightarrow{K} \right)}.
\end{eqnarray}
\end{widetext}
where $\overrightarrow{K} = R_{e}\overrightarrow{k}$ and
$\overrightarrow{q} = R_{e}\overrightarrow{Q}\left( 1 - \eta \right)$.

The total momentum is the only physical quantity associated with a symmetry-breaking.
$\sum_{\overrightarrow{k}}^{}{\overrightarrow{k}\left| F_{\overrightarrow{k}}^{\min}\left( \overrightarrow{Q} \right) \right|^{2}}$
is thus set equal to $\eta\overrightarrow{Q}$, with a constant
$\eta$ that is determined self-consistently. For zero (real)
phonon, the mean number of phonons in the cloud around the
electron~\cite{Lee_1953} (also called virtual phonon 
population~\cite{Lee_1953,Pollmann_1977}) can be related to the distortion field
$F_{\overrightarrow{k}}^{\min}$ (see Appendix~\ref{AppD}):

\begin{eqnarray} \label{eq30}
N_{e}^{\text{LLP}}\left( \overrightarrow{Q} \right) =& \left\langle \varphi_{\text{GS}} \middle| \sum_{\overrightarrow{k}}^{}{{\widehat{a}}_{\overrightarrow{k}}^{+} {\widehat{a}}_{\overrightarrow{k}}} \middle| \varphi_{\text{GS}} \right\rangle \nonumber \\
=& \sum_{\overrightarrow{k}}^{}\left| F_{\overrightarrow{k}}^{\min}\left( \overrightarrow{Q} \right) \right|^{2}. 
\end{eqnarray}
At this stage the problem remains highly nonlinear and further
developments are needed to find analytical expressions for the energy
dispersion $E_{\text{GS}}\left( \overrightarrow{Q} \right)$ of the
polaron.~\cite{Lee_1953} Finding the energy minimum
$E_{\text{GS}}\left( \overrightarrow{0} \right)$ at the bottom of the
polaron dispersion is easier. In fact, it is possible to drastically
simplify the problem by setting
$\sum_{\overrightarrow{k}}^{}{\overrightarrow{k}\left| F_{\overrightarrow{k}}^{\min}\left( \overrightarrow{0} \right) \right|^{2}} = \overrightarrow{0}$,
since for zero center of mass momentum the free polaron wavefunction is
symmetric. In that case,
$F_{\overrightarrow{k}}^{\min}\left( \overrightarrow{0} \right) = \frac{g_{\overrightarrow{k}}^{*}}{{\hslash\omega}_{\text{LO}}\left( 1 + R_{e}^{2}k^{2} \right)}$
and the energy after integration is given by
$E_{\text{GS}}\left( \overrightarrow{0} \right) = \sigma_{e}^{\text{LLP}} = - \alpha_{e}\hbar \omega_{\text{LO}}$,
which is the self-energy correction considered by Fr\"ohlich. This
population of virtual phonons yields a simple metric for the amplitude
of the lattice distortion around the charge carrier, leading to the
elegant formula proposed by LLP:
$N_{e}^{\text{LLP}}\left( \overrightarrow{0} \right)={\alpha}_{e}/{2}.$
This formula highlights the connection between the strength of coupling
and the induced lattice distortion.

\begin{figure*}[]
\includegraphics[width=0.8\textwidth]{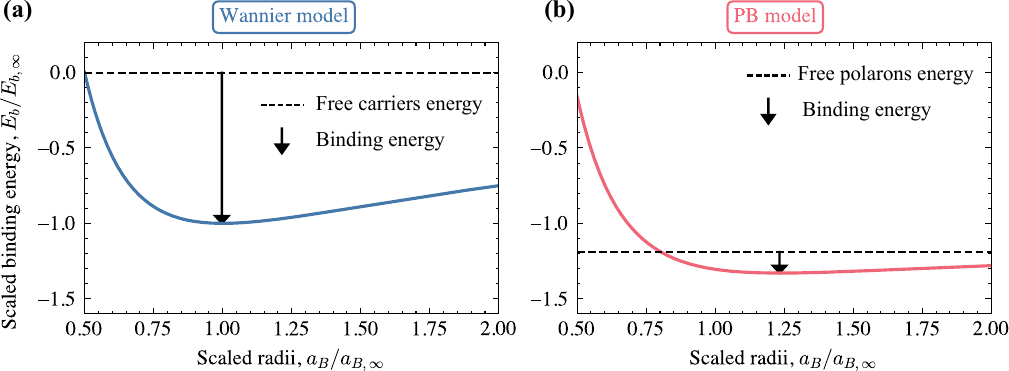}
\caption{Illustration of the calculation of 1S exciton binding
energies for TlCl. The variations of exciton energies are represented by
straight lines as a function of the Bohr radius, for (a) a Wannier exciton
model with $\varepsilon_{\text{eff}} = \varepsilon_{\infty}$ and (b) a
PB excitonic polaron model. The Bohr radii are scaled by the value
$a_{B,\infty}$ at the minimum energy. Energies are scaled by
the minimum energy $E_{b}^{\infty}$. The binding energies
represented by arrows are computed at the energy minima by subtracting
the energies for (a) free carriers or (b) free polarons. The minimum
energy for free carriers (band gap) is set as a reference to 0 to define
the vertical axis. The minimum energy used for free polarons is lowered by
self-energy terms computed within LLP theory.\label{fig5}}
\end{figure*}

Expressions for the polaron energy and virtual phonon population close to the 
minimum of $E_{\text{GS}}\left( \overrightarrow{Q} \right)$, including the polaron effective mass, are obtained when going beyond the simple approximation 
$\sum_{\overrightarrow{k}}^{}{\overrightarrow{k}\left| F_{\overrightarrow{k}}^{\min}\left( \overrightarrow{Q} \right) \right|^{2}} = \overrightarrow{0}$.
To do so, one introduces the above mentioned variable $\eta$ in the formula
$\sum_{\overrightarrow{k}}^{}{\overrightarrow{k}\left| F_{\overrightarrow{k}}^{\min}\left( \overrightarrow{Q} \right) \right|^{2}} = \eta\overrightarrow{Q}$. Full analytical integrations leads to (see
Appendix~\ref{AppD}):
$\eta = \frac{\alpha_{e}\left( 1 - \eta \right)}{2q^{3}}\left( \frac{q}{\sqrt{1 - q^{2}}} - {\rm asin}(q) \right)$,
$N_{e}^{\text{LLP}}\left( \overrightarrow{Q} \right) = \frac{\alpha_{e}}{2\sqrt{1 - q^{2}}}$
and
$E_{\text{GS}}\left( \overrightarrow{Q} \right) = \frac{\hslash^{2}Q^{2}\left( 1 - \eta^{2} \right)}{2m_{e}} - \alpha_{e}\hbar \omega_{\text{LO}}\frac{\text{asin}(q)}{q}$,
with $q = R_{e}Q\left( 1 - \eta \right)$. Close to the bottom of the
energy dispersion, a parabolic expression can be derived, leading to a
correction of the charge carrier mass:

\begin{eqnarray} \label{eq31}
E_{\text{GS}}\left( \overrightarrow{Q} \right) \approx - \alpha_{e}\hbar \omega_{\text{LO}} + \frac{\hslash^{2}Q^{2}}{2m_{e}\left( 1 + \frac{\alpha_{e}}{6} \right)} + \ldots
\end{eqnarray} 
Thus, the effective mass expression is slightly different from the
one derived within the Fr\"ohlich model:

\begin{eqnarray} \label{eq32}
m_{e}^{\text{LLP}} \approx m_{e}\left( 1 + \frac{\alpha_{e}}{6} \right).
\end{eqnarray} 
When treating the excitonic polarons within PB and PBK models latter in this review
(Sec.~\ref{SecIV}), Eqs.~\eqref{eq28},\eqref{eq29} of the LLP model will be our reference 
for uncorrelated e-h motion (Fig.~\ref{fig5}).

It is interesting to note that the LLP expressions for self-energy
correction and virtual phonon population at the bottom of the polaron
dispersion can be derived by considering:
\begin{eqnarray} \label{eq33}
F_{\overrightarrow{k}}^{\min}\left( \overrightarrow{0} \right) = \frac{g_{\overrightarrow{k}}^{*}}{{\hslash\omega}_{\text{LO}}\left( 1 + R_{e}^{2}k^{2} \right)}.
\end{eqnarray}
Notice that the denominator of Eq.~\eqref{eq33} captures the tendency of the
lattice (spatial extension) to undergo a distortion along a polar
optical eigenvector. However, $R_{e}\ $does not account for the
amplitude of the distortion, which is related to
$g_{\overrightarrow{k}}$ included in the numerator of Eq.~\eqref{eq33}. This
is also true for the population of the virtual phonon cloud which is
directly related to$\ \alpha_{e}$ for a free polaron.

\section{Pollman-B\"uttner model for excitonic polarons in the weak and intermediate coupling regimes}  \label{SecIV}

\subsection{Pollman-B\"uttner empirical Hamiltonian for excitonic polarons in a polar semiconductor (PB)}

Including explicitly the coupling between the e-h pairs and the lattice
at the same level of theory is possible with an extension of the Wannier
Hamiltonian in the center of mass coordinates:
\begin{eqnarray} \label{eq34}
{\widehat{H}}_{\rm X - ph} &=& E_{g} + \frac{{\widehat{P}}^{2}}{2M} + \frac{{\widehat{p}}^{2}}{2\mu} - \frac{e^{2}}{4\pi\varepsilon_{0}\varepsilon_{\infty}r} + \hbar \omega_{\text{LO}}\sum_{\overrightarrow{k}}^{}{{\widehat{a}}_{\overrightarrow{k}}^{+}{\widehat{a}}_{\overrightarrow{k}}} \nonumber \\ &+& \sum_{\overrightarrow{k}}^{}\left\lbrack \rho_{\overrightarrow{k}}{e^{i\overrightarrow{k}.\overrightarrow{R}}g}_{\overrightarrow{k}}{\widehat{a}}_{\overrightarrow{k}} + \text{c.c} \right\rbrack,
\end{eqnarray}
where the interplay between the correlated e-h motion and the coupling
to the lattice is related to the exciton charge density
operator~\cite{Hermanson_1970}:
\begin{eqnarray} \label{eq35}
\rho_{\overrightarrow{k}} = e^{is_{h}\overrightarrow{k}.\overrightarrow{r}} - e^{- is_{e}\overrightarrow{k}.\overrightarrow{r}},
\end{eqnarray}
and $s_{h} = \frac{m_{h}}{M}$ and $s_{e} = \frac{m_{e}}{M}$. As such, another
characteristic length, the exciton Bohr radius $a_{B}$ enters the
physics of polarons. The physics of the associated neutral
quasiparticle, called excitonic polaron, strongly depends on the ratio
of the free polaron radius $R_{\text{pol}}$ to the exciton Bohr
radius, motivating the definition of the reduced free polaron radius (Fig.~\ref{fig4}a-c):
\begin{eqnarray} \label{eq36}
{\widetilde{R}}_{\text{pol}} = \frac{R_{\text{pol}}}{a_{B}}
\end{eqnarray} 
This can be understood qualitatively by considering the free polaron
radius with respect to the probability density curves of 1S and 2S
excitons (Fig.~\ref{fig4}d). In the case where the lattice distortions produced
by a 1S excitonic polaron are strongly reduced compared to the sum
of the distortions produced by the related free-like polarons, then we
have ${\widetilde{R}}_{\text{pol}} \geq 0.25$ (vide infra). This
reduction is weaker for the 2S excitonic polaron. The variable extent of
distortions underpins deviations from the Rydberg like series for nS
excitons. This will be quantified later based on the virtual phonon
populations and self-energies of the quasiparticles. 

Early models developed to assess the physics of excitonic 
polarons were proposed by Meyer and Haken,~\cite{Meyer_1956,Haken_1956,Haken1978}.
Later, Toyozawa and Hermanson (TH) proposed models for excitonic
polarons~\cite{Toyozawa_1968,Hermanson_1970} that established a clear connection to
the LLP theory~\cite{Lee_1953} for free polarons, thanks to two
unitary transformations of the initially proposed exciton-phonon
Hamiltonian. PB~\cite{Pollmann_1975,Pollmann_1977} followed by
Kane~\cite{Kane_1978} and few years later by Iadonisi~\cite{Iadonisi_1983} further
refined this approach to quantitatively predict the properties of
excitonic polarons in simple semiconductors. TH introduced a unitary
transformation, analogous to Eq.~\eqref{eq25}, to account for the excitonic polaron total momentum:
\begin{eqnarray} \label{eq37}
\widehat{\wp} = \widehat{P} + \sum_{\overrightarrow{k}}^{}{\hslash\overrightarrow{k}{\widehat{a}}_{\overrightarrow{k}}^{+}{\widehat{a}}_{\overrightarrow{k}}}
\end{eqnarray} 
by choosing:
\begin{eqnarray}\label{eq38}
\widehat{U} = e^{i\left( \overrightarrow{Q} - \sum_{\overrightarrow{k}}^{}{\overrightarrow{k}
{\widehat{a}}_{\overrightarrow{k}}^{+}{\widehat{a}}_{\overrightarrow{k}}}
\right)\cdot\overrightarrow{R}},
\end{eqnarray} 
where $\hslash\overrightarrow{Q}$ is the eigenvalue of
$\widehat{\wp}$. The resulting Hamiltonian is comparable to
${\widehat{H}}_{\text{CM}}^{\text{LLP}}$ with additional terms (see
Appendix~\ref{AppD}):
\begin{widetext}
\begin{eqnarray}\label{eq39}
{\widehat{H}}_{\text{CM}}^{\text{TH}} &=& {\widehat{U}}^{+}{\widehat{H}}_{X - ph}\widehat{U} = E_{g} + \frac{\hslash^{2}Q^{2}}{2M} + \frac{{\widehat{p}}^{2}}{2\mu} - \frac{e^{2}}{4\pi\varepsilon_{0}\varepsilon_{\infty}r} + \hslash\Omega_{\text{CM}}^{\text{TH}}\sum_{\overrightarrow{k}}^{}{{\widehat{a}}_{\overrightarrow{k}}^{+}{\widehat{a}}_{\overrightarrow{k}}} \nonumber \\ &+& \sum_{\overrightarrow{k}}^{}\left\lbrack \rho_{\overrightarrow{k}}{e^{i\overrightarrow{k}.\overrightarrow{R}}g}_{\overrightarrow{k}}{\widehat{a}}_{\overrightarrow{k}} + \text{c.c} \right\rbrack + \sum_{\overrightarrow{k}}^{}{\sum_{\overrightarrow{k'}}^{}{\frac{\hslash^{2}\overrightarrow{k}.\overrightarrow{k}}{2M}{\widehat{a}}_{\overrightarrow{k}}^{+}{\widehat{a}}_{\overrightarrow{k'}}^{+}{\widehat{a}}_{\overrightarrow{k}}}}{\widehat{a}}_{\overrightarrow{k'}}
\end{eqnarray} 
\end{widetext}
with
$\hslash\Omega_{\text{CM}}^{\text{TH}} = \hbar \omega_{\text{LO}} - \frac{\hslash^{2}}{M}\overrightarrow{k}.\overrightarrow{Q} + \frac{\hslash^{2}k^{2}}{2M}$
.

This transformed Hamiltonian was used before PB's paper as a
starting point for perturbative calculations.~\cite{Wang_1974} The
expression of $\hslash\Omega_{\text{CM}}^{\text{TH}}$ indicates that
the phonon spectrum may deviate from the one of the crystal ground
state.~\cite{Iadonisi_1989} The last term also shows that for excited
states beyond the excitonic polaron ground state, additional small
energy contributions might be included~\cite{Pollmann_1977,Iadonisi_1989}. The
second unitary transformation was subsequently 
proposed by PB
allowing a variational approach:
\begin{eqnarray}\label{eq40}
{\widehat{W}\left( \overrightarrow{r} \right)} = e^{\sum_{\overrightarrow{k}}^{}\left( {F_{\overrightarrow{k}}^{*}\left( \overrightarrow{Q},\overrightarrow{r} \right)\widehat{a}}_{\overrightarrow{k}} - F_{\overrightarrow{k}}\left( \overrightarrow{Q},\overrightarrow{r} \right){\widehat{a}}_{\overrightarrow{k}}^{+} \right)}.
\end{eqnarray} 
It introduces a lattice distortion amplitude
$F_{\overrightarrow{k}}\left( \overrightarrow{Q},\overrightarrow{r} \right)$
function of the e-h relative positions that allows accounting for the
influence of their correlated motions. This function replaces the
distortion parameter
$F_{\overrightarrow{k}}\left( \overrightarrow{Q} \right)$ of LLP, and
must thus be determined by functional minimization, instead of the
standard minimization against a parameter. 
The PB Hamiltonian reads:
\begin{widetext}
\begin{eqnarray}\label{eq41}
{\widehat{H}}^{\text{PB}}\left( F_{\overrightarrow{k}}\left( \overrightarrow{Q},\overrightarrow{r} \right) \right) &=& {\widehat{W}}^{+}{\widehat{U}}^{+}{\widehat{H}}_{X - ph}\widehat{U}\widehat{W} \\ &=& {\widehat{H}}_{0_{\text{ph}}}^{\text{PB}} + {\widehat{H}}_{1_{\text{ph}}}^{\text{PB}} + {\widehat{H}}_{2_{\text{ph}}}^{\text{PB}} + {\widehat{H}}_{3_{\text{ph}}}^{\text{PB}} + {\widehat{H}}_{4_{\text{ph}}}^{\text{PB}},\nonumber
\end{eqnarray} 
\end{widetext}
where
${\widehat{H}}_{n_{\text{ph}}}^{\text{PB}}\left( F_{\overrightarrow{k}}\left( \overrightarrow{Q},\overrightarrow{r} \right) \right)$
are the n-phonon~contributions to the Hamiltonian. The explicit
expression of PB Hamiltonian for a GS with zero-phonon 
${\widehat{H}}_{0_{\text{ph}}}^{\text{PB}}\left( F_{\overrightarrow{k}}\left( \overrightarrow{Q},\overrightarrow{r} \right) \right)$
is (Appendix~\ref{AppF}): 

\begin{widetext}
\begin{eqnarray}\label{eq_ajout1b}
{\widehat{H}}_{0_{\text{ph}}}^{\text{PB}}\left( F_{\overrightarrow{k}}\left( \overrightarrow{Q},\overrightarrow{r} \right) \right) &=& E_{g} + \frac{\hslash^{2}Q^{2} - 2\hslash^{2}\overrightarrow{Q}.\sum_{\overrightarrow{k}}^{}{\overrightarrow{k}\left| F_{\overrightarrow{k}}\left( \overrightarrow{Q},\overrightarrow{r} \right) \right|^{2}} + \hslash^{2}\left| \sum_{\overrightarrow{k}}^{}{\overrightarrow{k}\left| F_{\overrightarrow{k}}\left( \overrightarrow{Q},\overrightarrow{r} \right) \right|^{2}} \right|^{2}}{2M} \\ \nonumber +
\frac{\left( \widehat{p} + \mu\overrightarrow{j} \right)^{2}}{2\mu}&+&
\sum_{\overrightarrow{k}}^{}{\frac{\hslash^{2}\left| \nabla F_{\overrightarrow{k}}\left( \overrightarrow{Q},\overrightarrow{r} \right) \right|^{2}}{2\mu}}
- \sum_{\overrightarrow{k}}^{}\left( \rho_{\overrightarrow{k}}g_{\overrightarrow{k}}F_{\overrightarrow{k}}\left( \overrightarrow{Q},\overrightarrow{r} \right) + c.c \right) \\ \nonumber &+& \sum_{\overrightarrow{k}}^{}{\left| F_{\overrightarrow{k}}\left( \overrightarrow{Q},\overrightarrow{r} \right) \right|^{2}\left( \hbar \omega_{\text{LO}} + \frac{\hslash^{2}k^{2}}{2M} \right)},
\end{eqnarray}
\end{widetext}
with
$\overrightarrow{j} =\sum_{\overrightarrow{k}}^{}{\frac{F_{\overrightarrow{k}}^{*}\left( \overrightarrow{Q},\overrightarrow{r} \right)\left( - i\hslash\nabla F_{\overrightarrow{k}}\left( \overrightarrow{Q},\overrightarrow{r} \right) \right) + c.c}{2\mu}}$.

It should be noted that the
PB Hamiltonian has a more complex form than the LLP one. In the
intermediate coupling regime, it does not allow full analytical
derivation of approximate expressions for the excitonic polaron energy,
effective mass or virtual phonon population.

\subsection{Pollman-B\"uttner variational solution (PB)}

PB's initial work is focused on the 1S exciton-phonon ground state
($0_{\text{ph}}$) with
$\left| \left. \ \psi_{\text{GS}} \right\rangle \right.\  = \phi_{\ 1S}\left( \overrightarrow{r} \right)\left| \left. \ 0 \right\rangle \right.\ $
related to the energy minimum at the bottom of the excitonic polaron
dispersion ($\overrightarrow{Q} = \overrightarrow{0}$).
In this case, a small first order contribution
(${\widehat{H}}_{1_{\text{ph}}}^{\text{PB}}$) and very small higher
order terms can be derived after functional minimization:
\begin{eqnarray}\label{eq42}
\frac{\delta}{\delta F_{\overrightarrow{k}}\left( \overrightarrow{r} \right)}\left\langle \phi_{1S}\left( \overrightarrow{r} \right) \middle| {\widehat{H}}_{0_{\text{ph}}}^{\text{PB}}\left( F_{\overrightarrow{k}}\left( \overrightarrow{0},\overrightarrow{r} \right) \right) \middle| \phi_{\ 1S}\left( \overrightarrow{r} \right) \right\rangle = 0.\nonumber
\\
\end{eqnarray} 

The exact expression for the function
$F_{\overrightarrow{k}}^{\min}\left( \overrightarrow{0},\overrightarrow{r} \right)$
 was obtained in 1983 by Iadonisi and coworkers for the
ground state (quoted as Iad),~\cite{Iadonisi_1983}
thanks to the infinite series of Legendre polynomials and integrals of
confluent hypergeometric functions. Exact solutions for excited states
were provided a few years later by the same
authors.~\cite{Iadonisi_1989,Strinati_1987} Iad's full analytical solution yields
numerical results close to the initial PB and PBK semi-analytical
implementation, but at much higher computational cost (vide infra). In 1975, to
perform a semi-analytical treatment of the problem for
$F_{\overrightarrow{k}}\left( \overrightarrow{0},\overrightarrow{r} \right)$,
PB indeed assumed a simplified form for the function, namely:
\begin{widetext}
\begin{eqnarray}\label{eq44}
F_{\overrightarrow{k}}\left( \overrightarrow{0},\overrightarrow{r} \right) \approx \frac{g_{\overrightarrow{k}}^{*}}{\hbar \omega_{\text{LO}}}\left( {f_{e,\overrightarrow{k}}\left( \overrightarrow{0} \right)e}^{- is_{h}\overrightarrow{k}.\overrightarrow{r}}- f_{h,\overrightarrow{k}}\left( \overrightarrow{0} \right)e^{is_{e}\overrightarrow{k}.\overrightarrow{r}} \right),
\end{eqnarray} 
replacing the complex functional minimization against the amplitude
function
$F_{\overrightarrow{k}}\left( \overrightarrow{0},\overrightarrow{r} \right)$
by minimization with respect to the parameters
$f_{e(h),\overrightarrow{k}}$:
\begin{eqnarray}\label{eq45}
\frac{\delta}{\delta f_{e(h),\overrightarrow{k}}\left( \overrightarrow{0} \right)}\left\langle \phi_{\ 1S}\left( \overrightarrow{r} \right) \middle| {\widehat{H}}_{0_{\text{ph}}}^{\text{PB}}\left( f_{e,\overrightarrow{k}}\left( \overrightarrow{0} \right),f_{h,\overrightarrow{k}}\left( \overrightarrow{0} \right) \right) \middle| \phi_{,\ 1S}\left( \overrightarrow{r} \right) \right\rangle = 0.
\end{eqnarray} 
It allows retrieving analytical expressions for the k-dependent
parameters:
\begin{eqnarray}\label{eq46}
f_{e\left( h \right),\overrightarrow{k}}^{\min}\left( \overrightarrow{0} \right) = \frac{\left( 1 - G_{1S}\left( \overrightarrow{k} \right) \right)\left( 1 + R_{h(e)}^{2}k^{2} + G_{1S}\left( \overrightarrow{k} \right) \right)}{\left( 1 + R_{h(e)}^{2}k^{2} \right)\left( 1 + R_{e(h)}^{2}k^{2} \right) - {G_{1S}\left( \overrightarrow{k} \right)}^{2}}
\end{eqnarray} 
with
\begin{eqnarray}\label{eq47}
G_{1S}\left( \overrightarrow{k} \right) = \iiint_{}^{}{e^{i\overrightarrow{k}.\overrightarrow{r}}\left| \phi_{\ 1S}\left( \overrightarrow{r} \right) \right|^{2}d^{3}\overrightarrow{r}}. 
\end{eqnarray} 
The resulting 1S excitonic polaron ground state energy for
$\overrightarrow{Q} = \overrightarrow{0}$ reads:
\begin{eqnarray}\label{eq48}
E_{GS}\left(\overrightarrow{0} \right) &=& \left\langle \phi_{\ 1S}\left( \overrightarrow{r} \right) \middle| E_{g} + \frac{p^{2}}{2\mu} - \frac{e^{2}}{4\pi\varepsilon_{0}\varepsilon_{\infty}r}\text{+}V_{\text{latt}}^{\text{PB}}\left( \overrightarrow{r},\ f_{e,\overrightarrow{k}}^{\min}\left( \overrightarrow{0} \right),f_{h,\overrightarrow{k}}^{\min}\ \left( \overrightarrow{0} \right) \right)\text{~} \middle| \phi_{\ 1S}\left( \overrightarrow{r} \right) \right\rangle \nonumber \\ &+& \sigma_{h,1S}\left( f_{e,\overrightarrow{k}}^{\min}\left( \overrightarrow{0} \right),f_{h,\overrightarrow{k}}^{\min}\ \left( \overrightarrow{0} \right) \right) + \sigma_{e,1S}\left( f_{e,\overrightarrow{k}}^{\min}\left( \overrightarrow{0} \right),f_{h,\overrightarrow{k}}^{\min}\ \left( \overrightarrow{0} \right) \right),
\end{eqnarray} 
including an effective lattice-mediated repulsive interaction in
addition to the e-h Coulomb attraction:
\begin{eqnarray}\label{eq49}
V_{\text{latt}}^{\text{PB}}\left( \overrightarrow{r},\ f_{e,\overrightarrow{k}}^{\min}\left( \overrightarrow{0} \right),f_{h,\overrightarrow{k}}^{\min}\left( \overrightarrow{0} \right) \right) = \text{2}\sum_{\overrightarrow{k}}^{}{\frac{\left| g_{\overrightarrow{k}} \right|^{2}}{\hbar \omega_{\text{LO}}}\left( f_{e,\overrightarrow{k}}^{\min}\left( \overrightarrow{0} \right) + f_{h,\overrightarrow{k}}^{\min}\left( \overrightarrow{0} \right) - f_{e,\overrightarrow{k}}^{\min}\left( \overrightarrow{0} \right)f_{h,\overrightarrow{k}}^{\min}\left( \overrightarrow{0} \right) \right)\cos\left( \overrightarrow{k}.\overrightarrow{r} \right)}\text{~}
\end{eqnarray} 
and a sum of self-energy corrections
$\sigma_{h,1S}\left( \overrightarrow{0} \right) + \sigma_{e,1S}\left( \overrightarrow{0} \right)$
in the presence of the exciton with:
\begin{eqnarray}\label{eq50}
\sigma_{e(h),1S}\left( \overrightarrow{0} \right) = \text{-}\sum_{\overrightarrow{k}}^{}{\frac{\left| g_{\overrightarrow{k}} \right|^{2}}{\hbar \omega_{\text{LO}}}\left( 2f_{e\left( h \right),\overrightarrow{k}}^{\min}\left( \overrightarrow{0} \right) - \left( 1 + R_{e(h)}^{2}k^{2} \right){f_{e\left( h \right),\overrightarrow{k}}^{\min}\left( \overrightarrow{0} \right)}^{2} \right)}.
\end{eqnarray}
\end{widetext}

In a second step, a 1S excitonic polaron wavefunction form 
\begin{eqnarray}\label{eq43}
\phi_{\ 1S}\left( \overrightarrow{r} \right) = \frac{e^{- r /a_{B}^{\text{eff}}}}{{a_{B}^{\text{eff}}}^{3/2}\pi^{1/2}}
\end{eqnarray} yields $G_{1S}\left( \overrightarrow{k},a_{B}^{\text{eff}} \right) = \frac{1}{\left( 1 + \left( \frac{ka_{B}^{\text{eff}}}{2} \right)^{2} \right)^{2}}$ and allows introducing  the expected effective Bohr radius $a_{B}^{\text{eff}}$ as a parameter for energy minimization $E_{GS}\left( a_{B}^{\text{eff}},\overrightarrow{0} \right)$. Noteworthy, the expression of
$E_{GS}\left( a_{B}^{\text{eff}},\overrightarrow{0} \right)$ depends
self-consistently on $a_{B}^{\text{eff}}$ through
$\phi_{1S}\left( \overrightarrow{r} \right)$ and
$f_{e\left( h \right),\overrightarrow{k}}^{\min}\left( \overrightarrow{0} \right)$. 

The case of a bipolaron (either two electrons or two holes) is instructive. 
Compared to an excitonic polaron, the Coulomb interaction is repulsive, but the lattice 
mediated interaction is attractive. At this level of theory, the effect of this
additional interaction is not sufficient to create bound states as in
the case of excitonic polarons (Sec.~\ref{AppI}, Fig.~\ref{figApp1}). 
Interestingly, approximate methods derived from Pollman-B\"uttner variational 
solution can be extended to add the effect of quantum confinement in nanostructures
on excitonic polarons in the intermediate coupling regime (vide infra).

\subsection{Analytical expressions for identical electron and hole masses}

For $m_{h} = m_{e}$, Kane derived in 1978 a compact analytical expression of 
Pollmann-B\"uttner 1S excitonic polaron ground state energy.~\cite{Kane_1978} 
This compact form does not separate the lattice mediated interaction from the self energy 
part, as proposed in Eqs.~\eqref{eq49} and ~\eqref{eq50}, respectively. 

Here, we first derive (contribution n° 1) an exact analytical expression of the effective 
interaction $V_{\text{latt}}^{\text{PB}}\left( \overrightarrow{r},\ f_{e,\overrightarrow{k}}^{\min}\left( \overrightarrow{0} \right),f_{h,\overrightarrow{k}}^{\min}\left( \overrightarrow{0} \right) \right)$, assuming
Eq.~\eqref{eq43} for the 1S wavefunction and simplifying the expressions of
$f_{e\left( h \right),\overrightarrow{k}}^{\min}\left( \overrightarrow{0} \right)$ to:
\begin{eqnarray}\label{eq51}
f_{e\left( h \right),\overrightarrow{k}}^{\min}\left( \overrightarrow{0} \right) = \frac{1 - G_{1S}\left( \overrightarrow{k},a_{B}^{\text{eff}} \right)}{1 + {R_{\text{pol}}}^{2}k^{2} - G_{1S}\left( \overrightarrow{k},a_{B}^{\text{eff}} \right)}.
\end{eqnarray} 
In fact, after analytical integration over $\overrightarrow{k}$, we find:
\begin{widetext}
\begin{eqnarray}\label{eq52}
{\widetilde{V}}_{latt,m_{e} = m_{h}}^{\text{PB}}\left( \overrightarrow{r}\ ,\overrightarrow{0} \right) &=& \frac{V_{latt,m_{e} = m_{h}}^{\text{PB}}\left( \overrightarrow{r},\overrightarrow{0}\  \right)}{\frac{e^{2}}{4\pi\varepsilon_{0}\varepsilon^{*}R_{\text{pol}}}} \\ &=& f_{a}^{+}\frac{1 - e^{- 2\sqrt{- \lambda_{+}\ }\ \widetilde{r}{\widetilde{R}}_{\text{pol}}}}{\ \widetilde{r}} - f_{b}^{+}\sqrt{- \lambda_{+}\ }\ {\widetilde{R}}_{\text{pol}}e^{- 2\sqrt{- \lambda_{+}\ }\ \widetilde{r}{\widetilde{R}}_{\text{pol}}} \nonumber \\
&+& f_{a}^{-}\frac{1 - e^{- 2\sqrt{- \lambda_{-}\ }\ \widetilde{r}{\widetilde{R}}_{\text{pol}}}}{\ \widetilde{r}} - f_{b}^{-}\sqrt{- \lambda_{-}\ }\ {\widetilde{R}}_{\text{pol}}e^{- 2\sqrt{- \lambda_{-}\ }\ \widetilde{r}{\widetilde{R}} _{\text{pol}}}, \nonumber
\end{eqnarray}
with $- \lambda_{\pm}f_{b}^{\pm} = \frac{\left( 2 - 1/\eta \right)}{1 - 4\eta}\frac{\lambda_{\pm} + \frac{3\eta - 2}{2\eta - 1}}{\lambda_{\pm}}$
and $- \lambda_{\pm}f_{a}^{\pm} = - \lambda_{\pm}f_{b}^{\pm} - \frac{\left( 2 - 1/\eta \right)}{1 - 4\eta} \pm \frac{2\left( \lambda_{\pm} + 2 - 1/2\eta \right)}{\sqrt{1 - 4\eta}} \pm \frac{2\eta\left( 2 - 1/\eta \right)\left( \lambda_{\pm} + \frac{3\eta - 2}{2\eta - 1} \right)}{\left( 1 - 4\eta \right)^{3/2}}$.

\noindent
The internal excitonic polaron energy first reads:
\begin{eqnarray}\label{eq53}
E_{GS}\left( a_{B}^{\text{eff}},\overrightarrow{0} \right) = E_{g} + \frac{\hslash^{2}}{2\mu{a_{B}^{\text{eff}}}^{2}} - \frac{e^{2}}{4\pi\varepsilon_{0}\varepsilon_{\infty}a_{B}^{\text{eff}}} - \frac{e^{2}}{2\pi^{2}\varepsilon_{0}\varepsilon^{*}R_{\text{pol}}}\int_{0}^{+ \infty}{\text{dK}\frac{\left( 1 - G_{1S} \right)^{2}}{\left( 1 + K^{2} - G_{1S} \right)}},
\end{eqnarray} 
where $\overrightarrow{K} = R_{\text{pol}}\overrightarrow{k}$.
Effective interaction and self-energies have been merged into a single
integral.

\noindent
The solutions can be decomposed in two parts relevant to the weak and intermediate coupling regimes (contribution n° 2):
\begin{eqnarray}
{\widetilde{R}}_{\text{pol}} \leq \frac{1}{4} &&\nonumber\\
&&\frac{E_{GS}\left( a_{B}^{\text{eff}},\overrightarrow{0} \right)}{\text{Ry}^{\text{vac}}} = \frac{{a_{B}^{\text{vac}}}^{2}}{\mu{a_{B}^{\text{eff}}}^{2}} - \frac{2a_{B}^{\text{vac}}}{\varepsilon_{\infty}a_{B}^{\text{eff}}} - \frac{4a_{B}^{\text{vac}}}{\varepsilon^{*}a_{B}^{\text{eff}}}\left( \frac{1}{\sqrt{1 - 4\eta}}\left( \frac{\lambda_{+} + \eta + 2}{\sqrt{- \lambda_{+}}} - \frac{\lambda_{-} + \eta + 2}{\sqrt{- \lambda_{-}}} \right) - \frac{1}{2} \right), 
\label{eq54}\\
{\widetilde{R}}_{\text{pol}} \geq \frac{1}{4} &&\nonumber\\
&&\frac{E_{GS}\left( a_{B}^{\text{eff}},\overrightarrow{0} \right)}{\text{Ry}^{\text{vac}}} = \frac{{a_{B}^{\text{vac}}}^{2}}{\mu{a_{B}^{\text{eff}}}^{2}} - \frac{2a_{B}^{\text{vac}}}{\varepsilon_{\infty}a_{B}^{\text{eff}}} - \frac{4a_{B}^{\text{vac}}}{\varepsilon^{*}a_{B}^{\text{eff}}}\left( \frac{1}{2\eta \, {\rm sin}\left( \vartheta/2 \right)}\left( \frac{1}{Z^{\frac{1}{2}}} - \frac{\eta + 2}{Z^{\frac{3}{2}}} \right) - \frac{1}{2} \right).
\label{eq55}
\end{eqnarray} 
Here, $\eta = 4{{\widetilde{R}}_{\text{pol}}}^{2}$\emph{,}
$\lambda_{\pm} = \big[ - \left( 1 + 2\eta \right) \pm \sqrt{1 - 4\eta}\big]/2\eta$
,
$\sin\left( \vartheta/2 \right) = \sqrt{(1 - {\rm cos}\left( \vartheta \right))/2}$
,
$\cos\left( \vartheta \right) = - \left( 1 + \frac{1}{2\eta} \right)/Z$,
$Z = \sqrt{2 / \eta + 1}$.
\end{widetext}

Analyzing the effects of the excitonic polaron center of mass motion on
the excitonic polaron energy
$E_{\text{GS}}\left( \overrightarrow{Q} \right)$, virtual phonon
population $N_{1S}\left( \overrightarrow{Q} \right)$ and effective
total mass close to the minimum of
$E_{\text{GS}}\left( \overrightarrow{Q} \right)$ is more complex than
in the case of free polarons and has been seldom implemented.~\cite{Kane_1978,Behnke_1978}
To allow derivation of analytical expressions, a few attempts based on PB approach started
by neglecting terms describing e-h correlations in the expression of
${\widehat{H}}_{0_{\text{ph}}}^{\text{PB}}\left( F_{\overrightarrow{k}}\left( \overrightarrow{Q},\overrightarrow{r} \right) \right)$.~\cite{Behnke_1978,Iadonisi_1987} 
Here, we propose a different strategy (contribution n° 3) , 
also based on the PB approach, but assuming $m_{e} = m_{h}$, 
which works for semiconductors like TlCl or 
metal halide perovskites. In this case, the expressions of the PB parameters
$f_{e\left( h \right),\overrightarrow{k}}^{\min}\left( \overrightarrow{Q} \right)$
can be simplified and read (see Appendix~\ref{AppF}):
\begin{eqnarray}\label{eq56}
f_{e\left( h \right),\overrightarrow{k}}^{\min}\left( \overrightarrow{Q} \right) = \frac{1 - G_{1S}\left( \overrightarrow{k},a_{B}^{\text{eff}} \right)}{1 + K^{2} - \overrightarrow{q}.\overrightarrow{K} - G_{1S}\left( \overrightarrow{k},a_{B}^{\text{eff}} \right)},
\end{eqnarray} 
with $\overrightarrow{K} = R_{\text{pol}}\overrightarrow{k}$,
$\overrightarrow{q} = R_{\text{pol}}\overrightarrow{Q}\left( 1 - \eta \right)$
and
$\eta\overrightarrow{Q} = \left\langle \phi_{1S}\left( \overrightarrow{r} \right) \middle| \sum_{\overrightarrow{k}}^{}{\overrightarrow{k}\left| F_{\overrightarrow{k}}^{\min}\left( \overrightarrow{Q},\overrightarrow{r} \right) \right|^{2}}\text{~} \middle| \phi_{\ 1S}\left( \overrightarrow{r} \right) \right\rangle$.
Semi-analytic expressions for
$E_{GS}\left( a_{B}^{\text{eff}},\overrightarrow{Q} \right)$ and
$\eta\overrightarrow{Q}$ are given in Appendix~\ref{AppF}. For
small Q values, these expressions can be simplified to a parabolic
energy dispersion for the center of mass motion:
\begin{eqnarray}\label{eq57}
E_{GS}\left( a_{B}^{\text{eff}},\overrightarrow{Q} \right) &\approx& E_{GS}\left( a_{B}^{\text{eff}},\overrightarrow{0} \right) + \frac{\hslash^{2}Q^{2}\left( 1 - \eta \right)^{2}}{2M} \nonumber \\ &+& \frac{{\alpha\hslash}^{2}Q^{2}}{2M}\left( I_{3}\left( a_{B}^{\text{eff}} \right) - \eta I_{2}\left( a_{B}^{\text{eff}} \right) \right),
\end{eqnarray} 
with
\begin{eqnarray}\label{eq58}
\eta \approx \frac{\alpha I_{4}\left( a_{B}^{\text{eff}} \right)}{1 + \alpha I_{3}\left( a_{B}^{\text{eff}} \right)}
\end{eqnarray} 
and
\begin{eqnarray}\label{eq59}
I_{n}\left( a_{B}^{\text{eff}} \right) &=& \int_{0}^{+ \infty}{\frac{8K^{2}\text{dK}}{3\pi}\left( \frac{\left( 1 - G_{1S} \right)^{n}}{\left( 1 + K^{2} - G_{1S} \right)^{3}} \right)}, 
\end{eqnarray} 
\begin{eqnarray}\label{eq59bis}
G_{1S} &=& \frac{1}{\left( 1 + \left( \frac{K}{2{\widetilde{R}}_{\text{pol}}} \right)^{2} \right)^{2}}.
\end{eqnarray} 
In the weak coupling limit $G_{1S} \rightarrow 0$. Using
$\int_{0}^{+ \infty}\frac{K^{2}\text{dK}}{\left( 1 + K^{2} \right)^{3}} = \frac{\pi}{16}$,
one retrieves an effective mass for the center of mass motion consistent
with the LLP model (free polarons) and renormalized to
$M^{\text{PB}} \approx M\left( 1 + \frac{\alpha}{6} \right) \approx m_{e}^{\text{LLP}} + m_{h}^{\text{LLP}}$.

\begin{figure}[htb]
\includegraphics[width=0.48\textwidth]{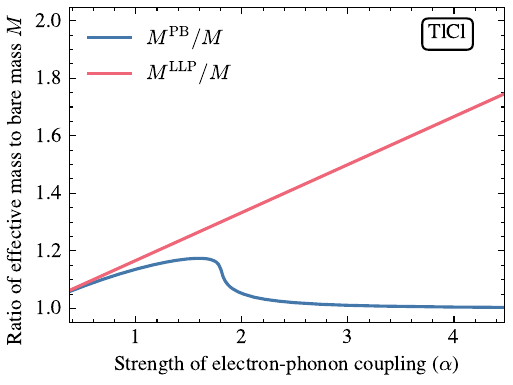}
\caption{ Total excitonic polaron mass (in unit of the bare mass M) computed within PB (blue) model for TlCl. The cross-over from the weak ($\alpha \rightarrow 0,a_{B}^{\text{eff}} \gg R_{\text{pol}}$) to the intermediate coupling regime
($a_{B}^{\text{eff}} < 4R_{\text{pol}}$) is obtained by tuning the effective phonon frequency in the expression of $\alpha$. The red line represents the mass computed using the LLP expression for free polarons
$M^{\text{LLP}} = m_{e}^{\text{LLP}} + m_{h}^{\text{LLP}}$.\label{fig6}}
\end{figure}

\begin{figure*}[htb]
\includegraphics[]{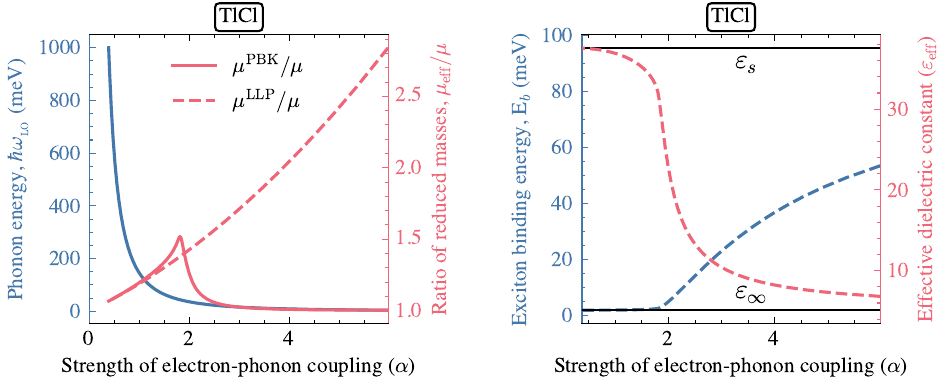}
\caption{Excitonic polaron characteristics computed within the PBK
model considering material parameters for TlCl, from the weak coupling
regime
($\alpha \rightarrow 0,a_{B}^{\text{eff}} \gg R_{\text{pol}}\ $) to
the intermediate coupling regime
($a_{B}^{\text{eff}} < 4R_{\text{pol}}$). (Left, blue line) The
cross-over from the weak to the intermediate coupling regime is obtained
by tuning the effective phonon frequency in the expression of $\alpha$
[Eq.~\eqref{eq1}]. (Left, red line) The exciton reduced mass compared to the
reduced mass of free-like polarons (Left red dashed line). (Right, blue
dashed line) Exciton binding energy and (red dashed line) effective 
dielectric constant as a function of $\alpha$. The lower
($\varepsilon_{\infty}$) and upper ($\varepsilon_{s}$) bounds of the
effective dielectric constant ($\varepsilon_{\text{eff}}$) are
indicated by black horizontal lines.\label{fig7}}
\end{figure*}

Fig.~\ref{fig6} shows the variation of the total excitonic polaron mass
as a function of $\alpha$ illustrated for the specific case of TlCl. 
A cross-over is observed from the weak coupling regime where the LLP limit is valid, 
to the intermediate coupling regime where the mass of bare charges is recovered. 
In the weak coupling limit an exciton can be qualitatively understood as two weakly interacting almost free polarons (Fig.~\ref{fig4}a). For that reason, the total excitonic polaron mass in the PB case is close to the LLP (free polaron) case. 
For similar $\alpha$ values, $\alpha\sim 1 - 2,\ $ a
similar crossover is obtained for the reduced mass and other
physical quantities (Fig.~\ref{fig7} and vide infra). In the strong coupling limit the distorsion fields cancel each other in an excitonic polaron and the polaron effect thus vanishes. For increasing electron-phonon coupling, the PB total excitonic polaron mass strongly diverges from the LLP case, where polaronic effects are not cancelled by e-h correlations.

\subsection{Kane's refinement of Pollman-B\"uttner variational solution (PBK)}
Historically, the excitonic polaron PB model led  to the first
quantitative predictions of the dielectric constant relevant for the
exciton in various semiconductors, ranging from conventional
semiconductors to ionic materials, and from the weak to the
intermediate-coupling regime.~\cite{Pollmann_1975} However, unlike free
polaron self-energies [Eq.~\eqref{eq16}], reduced e-h masses corresponding to the
masses of free polarons [Eq.~\eqref{eq32}] are not recovered in the weak coupling
regime within PB's initial semi-analytical implementation. This is
a direct consequence of the approximate form chosen for
$F_{\overrightarrow{k}}\left( \overrightarrow{r,}\overrightarrow{0} \right)$
with s-like trial functions for the phonon displacements 
(Eq.~\eqref{eq44}).~\cite{Kane_1978} This technical and fundamental issue,
well-identified by PB,~\cite{Pollmann_1977} was solved a few years later
by Kane~\cite{Kane_1978} improving the shift operator variational
expression by going beyond s-like trial functions (PBK):
\begin{eqnarray}\label{eq60}
f_{e(h),\overrightarrow{k}}\left( \overrightarrow{0} \right) \rightarrow f_{e(h),\overrightarrow{k}}\left( \overrightarrow{0} \right)\left( 1 - i\lambda_{e(h),\overrightarrow{k}}\left( \frac{\overrightarrow{k}.\overrightarrow{r}}{r} \right) \right). \nonumber \\
\end{eqnarray} 
Noteworthy, appendix~\ref{Kane} provides a detailed set of 
expressions with the aim of correcting a couple of misprints appearing in the original paper by Kane.~\cite{Kane_1978} 

\begin{figure*}[]
\includegraphics[width=0.9\textwidth]{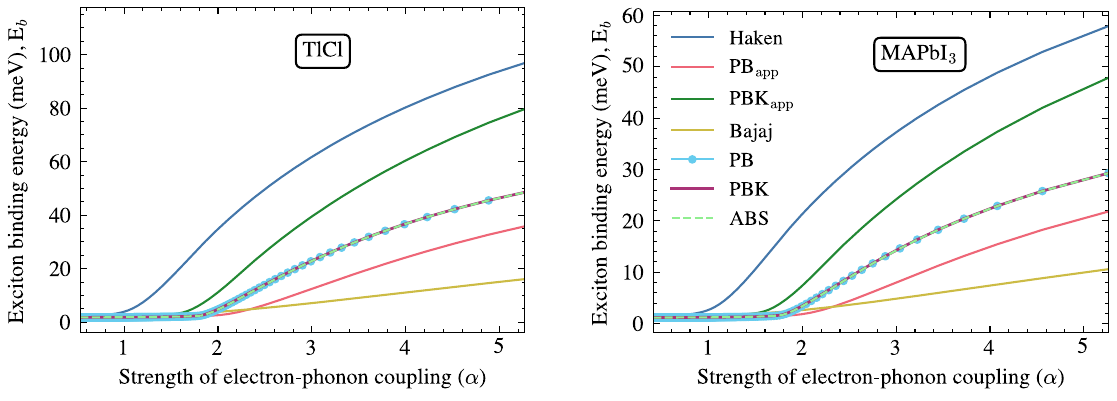}
\caption{Exciton binding energies computed at various levels of theory
for increasing coupling strength $\alpha$. Materials parameters are
taken from Tab.~\ref{tab:Table1} for TlCl (Left) and MAPbI\textsubscript{3} (Right), except 
the effective phonon frequency $\hbar \omega_{\text{LO}}$ of each
material that allows considering different coupling strength
$\alpha$ (Fig.~\ref{fig7} left, blue line). The results of the full PB~\cite{Pollmann_1977} and PBK~\cite{Kane_1978} models (cyan dotted and plum straight lines, Eq.~\eqref{eq48}-\eqref{eq53}, Eq.~\eqref{eq60}) are almost identical at this scale. This is true also for 
the results of the ABS method~\cite{Bednarek_1977} (Eq.~\eqref{eq70})
and a simplified version of the ABS method introduced in this work
(Eq.~\eqref{eq71}) represented by the same light green
dashed line. Green and red lines correspond to the approximation of PB
potential proposed in the present work (Eq.~\eqref{eq66}), respectively with and
without LLP reduced masses. The blue line corresponds to Haken's historical 
derivation, (Eq.~\eqref{eq67}).~\cite{Haken_1956} 
The dark yellow line corresponds to Bajaj's
approximate potential (Eq.~\eqref{eq68}). \label{fig8}}
\end{figure*}

The additional refinement parameters $\lambda_{e(h),\overrightarrow{k}}$ 
in Eq. ~\eqref{eq60} allows to recover a reduced e-h
mass consistent with free polarons (LLP model) in the weak coupling
regime (Fig.~\ref{fig7}, left). When the coupling is increased the e-h reduced
mass recovers its bare value, due to the compensation of the distortion
fields produced by the two charges in the excitonic polaron. The flexibility of the full PBK model is well illustrated with the case of TlCl in Fig.~\ref{fig7}.
The additional corrections defining the PBK model are yielding
sizeable relative deviations from the PB model for the exciton binding
energy only in the weak coupling regime (Fig.~\ref{fig8}).

The PB variational approach and the PBK refined version yield reasonable agreement
with experiments for binding energies of excitonic polarons, in their range of 
applicability (single phonon mode, weak and intermediate coupling 
regimes).~\cite{Pollmann_1975,Pollmann_1977, Kane_1978} 
This is further illustrated for GaAs, CdS and TlCl in 
Tab.~\ref{tab:Table3}, using parameters introduced in Tab.~\ref{tab:Table1}. 
Tab.~\ref{tab:Table3} highlights the scope of the various models: 1) for TlCl 
(intermediate coupling regime), the discrepancy between PB, PBK approaches and 
other approaches only valid for the weak coupling regimes (Hak, PBapp and PBKapp, 
see Sec.~\ref{SecV}) is clear. Hak solution is 
deviating most; 2) a similar discrepancy is obtained for the three halide 
perovskites, where an effective phonon mode energy has been introduced instead 
of an experimental one.

\subsection{Iadonisi and coworkers exact variational solution (Iad)}

At this point, it is interesting to compare the PB or PBK approaches to
alternative solutions for Eq.~\eqref{eq34}. The exact Iad solution for
$F_{\overrightarrow{k}}\left( \overrightarrow{r},\overrightarrow{0} \right)$~\cite{Iadonisi_1983}
can be written in a compact form using Whittaker functions instead of
confluent hypergeometric functions:
\begin{widetext}
\begin{eqnarray}\label{eq61}
F_{\overrightarrow{k}}^{\min}\left( \overrightarrow{r},\overrightarrow{0} \right) &=& \frac{g_{\overrightarrow{k}}^{*}}{\hbar \omega_{\text{LO}}}\sum_{l = 0}^{+ \infty}\begin{matrix}
 \\
\begin{pmatrix}
\left( - 1 \right)^{l + 1}\int_{0}^{+ \infty}{e^{\widetilde{r} - \widetilde{u}}\left( \frac{\widetilde{u}}{s_{e}\widetilde{k}{\widetilde{r}}^{2}} \right)^{\frac{1}{2}}M_{\frac{1}{\widetilde{\zeta}},l + \frac{1}{2}}^{1}\left( 2\widetilde{\zeta}\widetilde{r} \right)M_{\frac{1}{\widetilde{\zeta}},l + \frac{1}{2}}^{2}\left( 2\widetilde{\zeta}\widetilde{u} \right)J_{l + \frac{1}{2}}\left( s_{e}\widetilde{k}\widetilde{u} \right)d\widetilde{u}} \\
 + \int_{0}^{+ \infty}{e^{\widetilde{r} - \widetilde{u}}\left( \frac{\widetilde{u}}{s_{h}\widetilde{k}{\widetilde{r}}^{2}} \right)^{\frac{1}{2}}M_{\frac{1}{\widetilde{\zeta}},l + \frac{1}{2}}^{1}\left( 2\widetilde{\zeta}\widetilde{r} \right)M_{\frac{1}{\widetilde{\zeta}},l + \frac{1}{2}}^{2}\left( 2\widetilde{\zeta}\widetilde{u} \right)J_{l + \frac{1}{2}}\left( s_{h}\widetilde{k}\widetilde{u} \right)d\widetilde{u}} \\
\end{pmatrix}  \\
\end{matrix} \nonumber \\ &\times& \left( \frac{\left( - i \right)^{l}\sqrt{\frac{\pi}{8}}\left( 2l + 1 \right)\Gamma\left( l + 1 - \frac{1}{\widetilde{\zeta}} \right)P_{l}\left( \cos\left( \theta \right) \right)}{\left( {\widetilde{R}}_{e}^{2} + {\widetilde{R}}_{h}^{2} \right)\widetilde{\zeta}\Gamma\left( 2\left( l + 1 \right) \right)} \right)
\end{eqnarray} 
\end{widetext}
\noindent
where $\left( M^{1},M^{2} \right)\ $ stand for the Whittaker functions
$\left( M,W \right)\ /\ \left( W,M \right)\ $ when
$\left( \widetilde{u} > \widetilde{r} \right)$ /
$\left( \widetilde{r} > \widetilde{u} \right)$, $\Gamma$ for the
gamma function, $P_{l}\ $for a Legendre polynomial and
$\widetilde{\zeta} = \sqrt{1 + \frac{1}{{\widetilde{R}}_{e}^{2} + {\widetilde{R}}_{h}^{2}} + \frac{\mu}{M}{\widetilde{k}}^{2}}$.
All the variables are expressed in reduced units with respect to the
effective Bohr radius $a_{B}^{\text{eff}}$. As pointed out by
Iadonisi, this series has unfortunately a very slow convergence as a
function of $l$. Iadonisi performed a partial exact summation over an
infinite number of terms to recover the LLP limit for free
polarons.~\cite{Iadonisi_1983} A similar partial summation using Lommel
integrals  for the Whittaker functions in Eq.~\eqref{eq61} also leads to the LLP
limit (Appendix~\ref{AppE}, contribution n° 4). For $m_{h} = m_{e}$, relevant to
halide perovskites, all the terms with even $l$ values vanish, and
this reduces the computational burden. Noteworthy, these exact solutions
cannot be used for nanostructures.

\subsection{Adamowski Bednarek and Suffczynski approximation 
to Pollman-B\"uttner variational solution (ABS, ABS$_{\text{app}}$)}

A flexible approach was introduced by Adamowski, Bednarek and
Suffczynski (ABS).~\cite{Bednarek_1977,Adamowski_1976,Adamowski_1978} 
It aims at reproducing PB's results with a variational approach for the
$f_{e\left( h \right),\overrightarrow{k}}^{\min}\left( \overrightarrow{0} \right)$
coefficients, replacing the full expression in Eq.~\eqref{eq46}. This method
modifies not only the effective interaction but also the self-energies
and, as such, its validity extends to the intermediate coupling regime. 
It may appear at first sight surprising to further approximate Pollman-B\"uttner 
variational solution, while in the meantime Kane and Iadonisi and coworkers have derived 
improved solutions. However, ABS have shown that beyond the exciton ground state, 
their approach affords a convenient
way to simulate the excited states of the exciton as well as exciton complexes. To
illustrate the ABS method, we consider here excitonic polarons and use
notations consistent with the rest of the paper. The derivation of the effective 
interaction and self-energies can be found in Appendix~\ref{AppJ} for general cases. 
In the main text, we illustrate the specific case where $m_{e} = m_{h}$, for which the 
full expression of PB,
$f_{e\left( h \right),\overrightarrow{k}}^{\min}\left( \overrightarrow{0} \right) = \frac{1 - G_{1S}\left( \overrightarrow{k},a_{B}^{\text{eff}} \right)}{1 + {R_{\text{pol}}}^{2}k^{2} - G_{1S}\left( \overrightarrow{k},a_{B}^{\text{eff}} \right)}$,
is replaced by:
\begin{eqnarray}\label{eq69}
f_{e\left( h \right),\overrightarrow{k}}^{\min}\left( \overrightarrow{0} \right) = \frac{\lambda \rho}{1 + \left( \rho R_{\text{pol}} \right)^{2}k^{2}}.
\end{eqnarray} 
The two parameters $\left( \lambda,\rho \right)$ must be determined by
minimization of the total energy. For $m_{e} = m_{h}$, the effective
interaction can be for example expressed as (see Appendix~\ref{AppG} for additional analytical expressions):
\begin{widetext}
\begin{eqnarray}\label{eq70}
{\widetilde{V}}_{{\rm latt},m_{e} = m_{h}}^{\text{ABS}}\left( \overrightarrow{r},\overrightarrow{0}\  \right) = \frac{V_{{\rm latt},m_{e} = m_{h}}^{\text{ABS}}\left( \overrightarrow{r},\overrightarrow{0}\  \right)}{\frac{e^{2}}{4\pi\varepsilon_{0}\varepsilon^{*}R_{\text{pol}}}} = \frac{1}{\rho}\left( \frac{\left( 2\lambda\rho - \lambda^{2}\rho^{2} \right)\left( 1 - e^{- \widetilde{r}/\rho} \right)}{\widetilde{r}/\rho} + \frac{\lambda^{2}\rho^{2}e^{- \widetilde{r}/\rho}}{2} \right).
\end{eqnarray} 
\end{widetext}

Considering $\lambda \rightarrow 1,\rho \rightarrow 1$, one retrieves
the weak coupling limit, the free polaron self-energies as well as the
approximate PB effective interaction Eq.~\eqref{eq66} (vide infra). 
The ABS expressions for
$f_{e\left( h \right),\overrightarrow{k}}^{\min}\left( \overrightarrow{0} \right)$
do not exhibit explicitly the influence of e-h correlations. However,
the influence of the lattice distortions is introduced in the medium
coupling regime, by a reduction of the effective polaron radii
$\left( 0 < \rho < 1 \right)$ and the amplitudes of distortions
$\left( 0 < \lambda < 1 \right)$ (Appendix~\ref{AppG}, Fig.~\ref{figApp2}). 
The relative error on the excitonic polaron binding energy versus PB's model is less than
1.5\% (Appendix~\ref{AppG}, Fig.~\ref{figApp3}).

The ABS variational approach is numerically less efficient than the
direct computation of the effective interaction, self-energies and total
energies within the PB model when one uses the analytical solutions
derived in the present work [Eq.~\eqref{eq52}-\eqref{eq55}]. However, it provides an
accurate approach for excited states and exciton complexes, that are
very hard to tackle with extensions of the PB model. Fig.~\ref{figApp2} in Appendix~\ref{AppG} shows that the variations of the two parameters of the ABS approach 
are very similar. 

We propose here (contribution n° 5) to further reduce the numerical cost of the
variational calculation by considering a single parameter, assuming
$\lambda = \rho$:
\begin{widetext}
\begin{eqnarray}\label{eq71}
{\widetilde{V}}_{latt,m_{e} = m_{h}}^{\text{ABS}_{\text{app}}}\left( \overrightarrow{r},\overrightarrow{0}\  \right) = \frac{V_{latt,m_{e} = m_{h}}^{\text{ABS}_{\text{app}}}\left( \overrightarrow{r},\overrightarrow{0}\  \right)}{\frac{e^{2}}{4\pi\varepsilon_{0}\varepsilon^{*}R_{\text{pol}}}} = \frac{1}{\rho}\left( \frac{\left( 2\rho^{2} - \rho^{4} \right)\left( 1 - e^{- \widetilde{r}/\rho} \right)}{\widetilde{r}/\rho} + \frac{\rho^{4}e^{- \widetilde{r}/\rho}}{2} \right).
\end{eqnarray}
\end{widetext}

Despite this additional simplification, the relative error on the
excitonic polaron binding energy versus PB's model remains moderate and
less than 2.5\% (Appendix~\ref{AppG}, Fig.~\ref{figApp3}). The ABS and simplified ABS 
(ABS$_{\text{app}}$) approaches are the most attractive solutions in the intermediate coupling
regime, compared to the full PB and PBK models (Fig.~\ref{fig8}).

\section{Effective interaction potentials for the weak coupling regime (Hak, PB$_{\text{app}}$, PBK$_{\text{app}}$, Baj)}  \label{SecV}

Haken's historical model (abbreviated as Hak,~\cite{Haken_1956}) is based on approximate
forms of the coefficients $\ f_{e\left( h \right),\overrightarrow{k}}^{\min}$ that do not 
depend on $a_{B}^{\text{eff}}$:

\begin{eqnarray}\label{eq62}
f_{e\left( h \right),\overrightarrow{k}}^{\min}\left( \overrightarrow{0} \right) \approx \frac{1}{\left( 1 + R_{e(h)}^{2}k^{2} \right)}.
\end{eqnarray}

Among approximate models, this model proposed by Haken for the weak coupling 
regime~\cite{Haken_1956} is still very popular. We stress here that its validity is limited to the very weak coupling regime, specifically $\alpha \ll 1$. This is illustrated in Fig.~\ref{fig8}
for TlCl and MAPbI\textsubscript{3} with the comparison to the more
complete PB/PBK approaches.
Within the model, self-energies are those obtained by LLP for free polarons $\sigma_{e(h),1S}\left( \overrightarrow{0} \right) \approx - \alpha_{e(h)}\hbar \omega_{\text{LO}}$. In addition, Haken also considered a reduced mass consistent
with the LLP model.  Haken's approximate expression for the
effective interaction reads:
\begin{widetext}
\begin{eqnarray}\label{eq64}
V_{\text{latt}}^{\text{Hak}}\left( \overrightarrow{r,}f_{e,\overrightarrow{k}}^{\min}\left( \overrightarrow{0} \right),f_{h,\overrightarrow{k}}^{\min}\left( \overrightarrow{0} \right) \right)\approx \sum_{\overrightarrow{k}}^{}{\frac{\left| g_{\overrightarrow{k}} \right|^{2}}{\hbar \omega_{\text{LO}}}\left( f_{e,\overrightarrow{k}}^{\min}\left( \overrightarrow{0} \right)e^{- i\overrightarrow{k}.\overrightarrow{r}} + f_{h,\overrightarrow{k}}^{\min}\left( \overrightarrow{0} \right)e^{i\overrightarrow{k}.\overrightarrow{r}} \right)},
\end{eqnarray} 
\end{widetext}
where the coefficients
$f_{e\left( h \right),\overrightarrow{k}}^{\min}\left( \overrightarrow{0} \right)$
[Eq.~\eqref{eq62}] are consistent with the LLP model for free polarons and do
not depend self-consistently on the exciton
properties.~\cite{Haken_1956} After analytical integration over
$\overrightarrow{k}$, Eq.~\eqref{eq64} leads to an effective interaction term
expressed in real space:
\begin{eqnarray}\label{eq65}
V_{\text{latt}}^{\text{Hak}}\left( \overrightarrow{r}\ ,\overrightarrow{0} \right) \approx \frac{e^{2}}{4\pi\varepsilon_{0}\varepsilon^{*}r}\left( 1 - \frac{1}{2}e^{- r/R_{e}} - \frac{1}{2}e^{-r / R_{h}} \right). \nonumber \\
\end{eqnarray} 
Such a real space expression for the effective interaction is attractive
for a numerical implementation in a Wannier-like equation and is often
referred to as the ``Haken potential''.

PB proposed in 1977 an alternative approximate effective interaction for the weak coupling regime,~\cite{Pollmann_1977} by inserting in the expression of
$V_{\text{latt}}^{\text{PB}}\left( \overrightarrow{r}\ ,\overrightarrow{0} \right)$
Eq.~\eqref{eq49}, the approximate forms of the coefficients 
$\ f_{e\left( h \right),\overrightarrow{k}}^{\min}$ used by Haken Eq.~\eqref{eq62}.
After analytical integration over $\overrightarrow{k}$,
PB obtained a real space expression for the case where
$m_{e} \neq m_{h}$ that reads:

\begin{widetext}
\begin{eqnarray}\label{eq63}
V_{latt,m_{e} \neq m_{h}}^{\text{PB}_{\text{app}}}\left( \overrightarrow{r,}\overrightarrow{0}\  \right) \approx \frac{e^{2}}{4\pi\varepsilon_{0}\varepsilon^{*}r}\Big( 1 + \frac{m_{e}}{\mathrm{\Delta}m}e^{- r / R_{e}} - \frac{m_{h}}{\mathrm{\Delta}m}e^{- r /R_{h}} \Big), \nonumber \\
\end{eqnarray} 
\end{widetext}
with $\mathrm{\Delta}m = m_{h} - m_{e}$. 

Self-energy corrections of the PB model Eq.~\eqref{eq50} are identical to those 
obtained by LLP for free polarons (see Appendix~\ref{AppF}).
It shows that PB approximate effective interaction for the weak coupling  is also 
convenient for a numerical implementation into a Wannier-like equation. This potential 
is often improperly referred to as the ``PB potential'', whereas the
genuine expression of the PB potential
$V_{\text{latt}}^{\text{PB}}\left( \overrightarrow{r}\ ,\overrightarrow{0} \right)$
is Eq.~\eqref{eq49}.

For the simulation of materials with roughly equal masses,
the presence of $\mathrm{\Delta}m$ in the denominators of
$V_{latt,m_{e} \neq m_{h}}^{\text{PB}_{\text{app}}}\left( \overrightarrow{r},\overrightarrow{0}\  \right)$
may be problematic. In this work, we propose a suitable expression for
this specific case (vide infra). Finally, as pointed out by
Kane,~\cite{Kane_1978} expressions for
$f_{e\left( h \right),\overrightarrow{k}}^{\min}\left( \overrightarrow{0} \right)$
such as Eq.~\eqref{eq62} are not sufficient to derive effective masses consistent
with the LLP model in the weak coupling regime.

\begin{figure}[]
\includegraphics[width=0.48\textwidth]{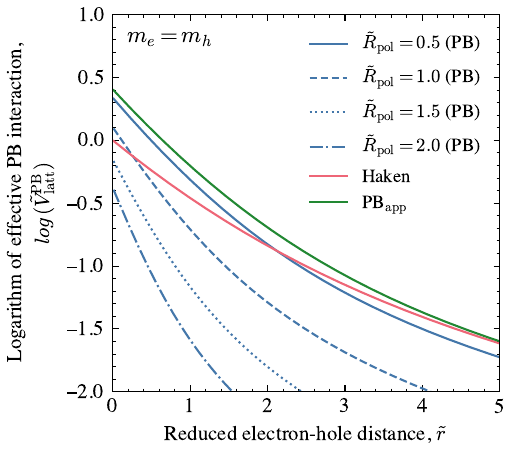}
\caption{
$\log\left( {\widetilde{V}}_{\text{latt}}^{\text{PB}} \right)$
plotted for $m_{e} = m_{h}\ $ against the reduced e-h distance,
for various expressions of the effective potential. The full PB
effective interaction
${\widetilde{V}}_{latt,m_{e} = m_{h}}^{\text{PB}}\left( \overrightarrow{r}\  \right)$
(Eq.~\eqref{eq52}, blue lines) represented for various values of
${\widetilde{R}}_{\text{pol}}$ is compared with approximate
expressions relevant for the weak coupling regime
${\widetilde{R}}_{\text{pol}} \rightarrow 0$. 
${\widetilde{V}}_{latt,m_{e} = m_{h}}^{\text{PB}_{\text{app}}}\left( \overrightarrow{r}\  \right)$
(Eq.~\eqref{eq66}) corresponds to the green line and overlaps the full PB for 
${\widetilde{R}}_{\text{pol}} = 0$. The red line shows
${\widetilde{V}}_{latt,m_{e} = m_{h}}^{\text{Hak}}\left( \overrightarrow{r}\  \right)$
(Eq.~\eqref{eq67}). The e-h distance is
expressed in reduced units:
$\widetilde{r} = \frac{r}{R_{\text{pol}}}$ and ${\widetilde{R}}_{\text{pol}} = \frac{R_{\text{pol}}}{a_{B}}$.} \label{fig9} 
\end{figure}

To understand the differences between Hak and PB approximate
expressions, we provide (contribution n° 6) analytical expressions 
derived for the case where $m_{e} = m_{h}$ ($R_{e} = R_{h} = R_{\text{pol}}$):

\begin{eqnarray}\label{eq66}
{\widetilde{V}}_{latt,m_{e} = m_{h}}^{\text{PB}_{\text{app}}}\left( \overrightarrow{r},\overrightarrow{0}\  \right) &=& \frac{V_{latt,m_{e} = m_{h}}^{\text{PB}_{\text{app}}}\left( \overrightarrow{r},\overrightarrow{0}\  \right)}{\frac{e^{2}}{4\pi\varepsilon_{0}\varepsilon^{*}R_{\text{pol}}}} \nonumber \\ &\approx&  \left( \frac{1 - e^{- \widetilde{r}}}{\widetilde{r}} + \frac{e^{- \widetilde{r}}}{2} \right),
\end{eqnarray} 
with $\widetilde{r} = \frac{r}{R_{\text{pol}}}$. Considering
$R_{e} = R_{h} = R_{\text{pol}}$  :
\begin{eqnarray}\label{eq67}
{\widetilde{V}}_{latt,m_{e} = m_{h}}^{\text{Hak}}\left( \overrightarrow{r},\overrightarrow{0}\  \right) &=& \frac{V_{latt,m_{e} = m_{h}}^{\text{Hak}}\left( \overrightarrow{r},\overrightarrow{0}\  \right)}{\frac{e^{2}}{4\pi\varepsilon_{0}\varepsilon^{*}R_{\text{pol}}}} \nonumber \\ &\approx& \left( \frac{1 - e^{- \widetilde{r}}}{\widetilde{r}} \right),
\end{eqnarray} 
and comparing this expression to the one derived by Haken (Eq.~\eqref{eq65}) evidences that 
the Haken potential is missing the contribution
$\frac{e^{- \widetilde{r}}}{2}$ compared to Eq.~~\eqref{eq66}. This
term is expected to vanish only in the very weak coupling limit where
$a_{B}^{\text{eff}} \gg R_{\text{pol}}$ (Fig.~\ref{fig8}, $\alpha \ll 1$).
For larger coupling regimes, Hak approach significantly deviates from
the complete PBK model (Tab.~\ref{tab:Table3}, Fig.~\ref{fig8}). In many reports, Haken's
potential [Eq.~\eqref{eq65}] is combined in an ad hoc manner with a kinetic energy
term based on LLP effective masses as proposed by Haken
himself.~\cite{Haken1978} On the contrary, PB approximated potential
$V_{\text{latt}}^{\text{PB}_{\text{app}}}\left( \overrightarrow{r}\ ,\overrightarrow{0} \right)$
(Eq.~\eqref{eq63}) is usually implemented~\cite{Men_ndez_Proupin_2015} with bare 
effective masses, in line with PB's initial paper (red line in Fig.~\ref{fig8}).~\cite{Pollmann_1975} As quoted by PB,~\cite{Pollmann_1977} we stress
here again that bare effective masses are not justified in the weak
coupling regime. Consistently with the results of the PBK refined
approach (vide infra), we propose instead to use
$\mathbf{V}_{\mathbf{\text{latt}}}^{\mathbf{\text{PB}}_{\mathbf{\text{app}}}}\left( \overrightarrow{\mathbf{r}}\mathbf{,}\overrightarrow{\mathbf{0}}\mathbf{\ } \right)$
together with the LLP effective masses, in the spirit of
Haken's suggestion, herin quoted PBK$_\text{app}$ (contribution n° 7). 
This new approach, which is mathematically justified
in the present work, appears to be the best approximation to the
full PBK model for the weak coupling regime. It can be used for a
numerical implementation into a Wannier-like equation neglecting
self-consistency necessary for full PB and PBK approaches (green
line in Fig.~\ref{fig8}).~\cite{Kane_1978} As shown in Fig.~\ref{fig9}, the
approximate expressions
${\widetilde{V}}_{latt,m_{e} = m_{h}}^{\text{PB}_{\text{app}}}\left( \overrightarrow{r}\ ,\overrightarrow{0} \right)$
[Eq.~\eqref{eq66}] and
${\widetilde{V}}_{latt,m_{e} = m_{h}}^{\text{Hak}}\left( \overrightarrow{r}\  \right)$
[Eq.~\eqref{eq67}] deviate strongly from the exact expression of the effective
interaction
${\widetilde{V}}_{latt,m_{e} = m_{h}}^{\text{PB}}\left( \overrightarrow{r}\  \right)$
[Eq.~\eqref{eq52}] in the intermediate excitonic polaron coupling regime
(${\widetilde{R}}_{\text{pol}} \geq \frac{1}{4}$, $a_{B,\ eff} < 2\left( R_{e} + R_{h} \right)$).

Bajaj (abbreviated as Baj) proposed~\cite{Bajaj_1974} an ad hoc
modification of Haken's potential to avoid the overestimation of the
exciton binding energy that it entails:
\begin{widetext}
\begin{eqnarray}\label{eq68}
V_{\text{latt}}^{\text{Baj}}\left( \overrightarrow{r}\ ,\overrightarrow{0} \right) \approx \frac{e^{2}}{4\pi\varepsilon_{0}\varepsilon^{*}r} - \frac{e^{2}}{8\pi\varepsilon_{0}\varepsilon^{*}r}\left( \frac{\varepsilon_{\infty}}{\varepsilon_{s}} \right)^{\gamma}\left( e^{-r / R_{e}} + e^{-r / R_{h}} \right).
\end{eqnarray}
\end{widetext}
In this modified potential $\gamma = 3/5$, whereas $\gamma = 0$ in
the Haken potential. But this expression still deviates significantly
from the PB and PBK potentials (Fig.~\ref{fig8}) and is invalid beyond the
weak coupling regime, because self-energies and effective masses of free
polarons are used. Another attempt was made later on by the same author
to extend perturbatively the Haken's approach beyond the free polaron
limit,~\cite{Aldrich_1977} but this extension is again limited, compared 
to the complete PB and PBK approaches.

\section{Extension to multiple polar phonon modes (MLLP, MPB, MPBK, MIad)}  \label{SecVII}
  
When applied to excitonic polarons in halide perovskites and more
generally to polar lattices with atomic motifs containing more than two
atoms, empirical approaches treat the exciton-lattice coupling by means
of a single optical polar mode. For 3D halide perovskites it is known
both theoretically~\cite{Even_2015} and
experimentally~\cite{Ferreira_2020,Sendner_2016} that multiple polar modes play a
role. This is one of the specific features of this class of polar
semiconductors (Tab.~\ref{tab:Table1}). The single mode approximation is related to
the use of the original expression of the Fr\"ohlich interaction. It can
be partially overcome by treating this mode as an effective one, but
this effective frequency is not directly related to an experimental one,
except that they fall roughly in the low-energy range of the lattice
modes (Tab.~\ref{tab:Table1}).  The couplings of
excitons to multiple phonons  have been experimentally observed in a systematic manner. ~\cite{Fu_2017,Fu_2018} To include these effects, one may either rely on an atomistic approach or make approximations in order to turn the multimode polaronic calculation into a tractable problem. The later approach, for example, has been implemented in the past for free
polarons to compute carrier mobilities.~\cite{Sendner_2016,Yu_2016,Verbist_1992,Hellwarth_1999}. 

We propose here a multimode extension of PB, PBK and Iad frameworks (quoted as 
MPB, MPBK, MIad models hereafter) for excitonic polaron (contribution n° 8).
Let's start first by describing how the empirical approaches for the
dielectric properties and the Fr\"ohlich interaction have been originally
extended to multiple polar modes in literature. As discussed in earlier
reports~\cite{Sendner_2016,Zelezn__2023}, polar modes and lattice anharmonicity
are included using the semi-empirical expression of Gervais and
coworkers:~\cite{Gervais_1974}
\begin{eqnarray}\label{eq76}
\varepsilon\left( \omega \right) &=& \varepsilon'\left( \omega \right) + i\varepsilon''\left( \omega \right) \\ &=& \varepsilon_{\infty} + \sum_{\nu}^{}\frac{\omega_{{\rm TO},\nu}^{2}\mathrm{\Delta}\varepsilon_{\nu}}{\omega_{{\rm TO},\nu}^{2} - \omega^{2} - i \omega \gamma_{{\rm TO},\nu}}  \nonumber
\end{eqnarray} 
where $\omega_{{\rm TO},\nu}$ and $\gamma_{{\rm TO},\nu}$ are the frequency of
the TO-mode $\nu$ and the corresponding damping parameter,
respectively. With negligible anharmonicity, this equation is reduced to
Toyozawa's multimode expression:~\cite{Devreese1972}
$\varepsilon\left( \omega \right) = \varepsilon_{\infty} + \sum_{\nu}^{}\frac{\omega_{{\rm TO},\nu}^{2}\mathrm{\Delta}\varepsilon_{\nu}}{\omega_{{\rm TO},\nu}^{2} - \omega^{2}}$.
For a single polar mode:
$\varepsilon\left( \omega \right) = \varepsilon_{\infty} + \frac{\omega_{\text{TO}}^{2}\left( \varepsilon_{s} - \varepsilon_{\infty} \right)}{\omega_{\text{TO}}^{2} - \omega^{2}}$.
A generalization of the LST (Eqs.~\eqref{eq13}) to multimodes can also be
derived:~\cite{Gervais_1974}
\begin{eqnarray}\label{eq77LST}
\frac{\varepsilon\left( \omega \right)}{\varepsilon_{\infty}} = \prod_{\nu}^{}\left( \frac{\omega_{{\rm LO},\nu}^{2} - \omega^{2} + i\omega\gamma_{{\rm LO},\nu}}{\omega_{{\rm TO},\nu}^{2} - \omega^{2} + i\omega\gamma_{{\rm TO},\nu}} \right)
\end{eqnarray} 
where $\omega_{{\rm LO},\nu}$ and $\gamma_{{\rm LO},\nu}$ are the frequency of
the LO mode $\nu$ and the corresponding damping parameter,
respectively. Longitudinal optical modes appear as poles of
$\frac{1}{\varepsilon\left( \omega \right)}$, while transverse optical
modes are poles of $\varepsilon\left( \omega \right)$. In a cubic
crystal, the longitudinal component of the dielectric tensor can be
combined with the general expression of
$\varepsilon\left( \omega \right)$ and the LST equation because
${\ \varepsilon}_{//}\left( \omega \right) = \varepsilon\left( \omega \right)$.
More importantly, as detailed by Toyozawa, a semi-empirical calculation
based on Poisson equation allows computing independently the
longitudinal component of the dielectric tensor as a function of various
contributions to the Fourier transform of the
polarization:~\cite{Devreese1972}
\begin{eqnarray}\label{eq77}
\frac{1}{\varepsilon_{\infty}} - \ \frac{1}{{\ \varepsilon}_{//}\left( \omega \right)} = \sum_{\nu}^{}\frac{2\omega_{{\rm LO},\nu}\varepsilon_{0}{k^{2}\left| g_{\overrightarrow{k},\nu} \right|}^{2}}{\hslash e^{2}\left( \omega_{{\rm LO},\nu}^{2} - \omega^{2} \right)}.
\end{eqnarray} 
Toyozawa pushed forward the analysis in the multimode case with
negligible phonon damping (harmonic approximation) by expanding
$\frac{1}{{\ \varepsilon}_{//}\left( \omega \right)}$ close to
$\omega = \omega_{{\rm LO},\nu}$ (see Appendix~\ref{AppG}), in order
to get the k-dependent e-ph matrix element for each optical phonon
${k^{2}\left| g_{\overrightarrow{k},\nu} \right|}^{2} = \frac{\hslash e^{2}}{\varepsilon_{0}\left. \ \frac{\partial\varepsilon}{\partial\omega} \right|_{\omega_{{\rm LO},\nu}}}$.~\cite{Devreese1972}
This multimode extension of the Fr\"ohlich e-ph coupling matrix element
(Eq.~\eqref{eq12}) was used for alloys of conventional
semiconductors~\cite{Swierkowski_1978,Nash_1987} and, more recently, to investigate free carrier mobility in
MAPbI\textsubscript{3}.~\cite{Yu_2016} Multimode coupling strength
$\alpha_{e,\nu}$ can be introduced with respect to the matrix element
by analogy with the standard theory:
\begin{eqnarray}\label{eq79}
k\left| g_{\overrightarrow{k},\nu} \right| = \hbar \omega_{{\rm LO},\nu}\left( \frac{4\pi\alpha_{e,\nu}}{V} \right)^{1/2}\left( \frac{\hbar}{2m_{e}\omega_{{\rm LO},\nu}} \right)^{1/4}.
\end{eqnarray} 

In the harmonic approximation, the strength of the coupling of an
electron to each polar optical mode of the lattice can then be obtained
as:~\cite{Devreese1972,Swierkowski_1978,Yu_2016}
\begin{eqnarray}\label{eq80}
\alpha_{e,\nu} &=& \frac{e^{2}}{4\pi\varepsilon_{0}\hslash}\left( \frac{m_{e}}{2\hbar \omega_{{\rm LO},\nu}} \right)^{1/2}\frac{1}{\varepsilon_{\infty}}\left( 1 - \frac{\omega_{{\rm TO},\nu}^{2}}{\omega_{{\rm LO},\nu}^{2}} \right) \nonumber \\ &\times& \prod_{\mu \neq \nu}^{}\left( \frac{\omega_{{\rm LO},\nu}^{2} - \omega_{{\rm TO},\mu}^{2}}{\omega_{{\rm LO},\nu}^{2} - \omega_{{\rm LO},\mu}^{2}} \right).
\end{eqnarray} 
By expanding the general expression of
$\varepsilon\left( \omega \right)$ close to
$\omega = \omega_{{\rm TO},\nu}$ we can
demonstrate that (see Appendix~\ref{AppH}):
\begin{eqnarray}\label{eq81}
\mathrm{\Delta}\varepsilon_{\nu} = \varepsilon_{\infty}\left( \frac{\omega_{{\rm LO},\nu}^{2}}{\omega_{{\rm TO},\nu}^{2}} - 1 \right)\prod_{\mu \neq \nu}^{}\left( \frac{\omega_{{\rm TO},\nu}^{2} - \omega_{{\rm LO},\mu}^{2}}{\omega_{{\rm TO},\nu}^{2} - \omega_{{\rm TO},\mu}^{2}} \right).
\end{eqnarray} 

For the treatment of the excitons in the MPB/MPBK/MIad models, we extend
the PB approach (Eq.~\eqref{eq40}) to multiple branches introducing a generalized unitary
transformation:
\begin{eqnarray}\label{eq82}
\widehat{W}\left( \overrightarrow{r} \right) = e^{\sum_{\overrightarrow{k},\nu}^{}\left( {F_{\overrightarrow{k},\nu}^{*}\left( \overrightarrow{Q},\overrightarrow{r} \right)\widehat{a}}_{\overrightarrow{k},\nu} - F_{\overrightarrow{k},\nu}\left( \overrightarrow{Q},\overrightarrow{r} \right){\widehat{a}}_{\overrightarrow{k},\nu}^{+} \right)},
\end{eqnarray} 
assuming independent lattice distortions produced by the various polar
optical modes:
\begin{eqnarray}\label{eq83}
F_{\overrightarrow{k},\nu}\left( \overrightarrow{Q},\overrightarrow{r} \right) \approx \frac{g_{\overrightarrow{k},\nu}^{*}}{\hbar \omega_{\text{LO}}}\left( {f_{e,\overrightarrow{k},\nu}e}^{- is_{h}\overrightarrow{k}.\overrightarrow{r}} - f_{h,\overrightarrow{k},\nu}e^{is_{e}\overrightarrow{k}.\overrightarrow{r}} \right). \nonumber \\
\end{eqnarray} 
The e-h effective interaction potential includes a summation over the
mode contributions:
\begin{widetext}
\begin{eqnarray}\label{eq84}
V_{\text{latt}}^{\text{MPB}}\left( \overrightarrow{r}\  \right) = \text{2}\sum_{\overrightarrow{k},\nu}^{}{\frac{\left| g_{\overrightarrow{k},\nu} \right|^{2}}{\hbar \omega_{{\rm LO},\nu}}\left( f_{e,\overrightarrow{k},\nu}^{\min} + f_{h,\overrightarrow{k},\nu}^{\min} - f_{e,\overrightarrow{k},\nu}^{\min}f_{h,\overrightarrow{k},\nu}^{\min} \right)\cos\left( \overrightarrow{k}.\overrightarrow{r} \right)}\text{~}
\end{eqnarray} 
and the total self-energy corrections read:
\begin{eqnarray}\label{eq85}
\sigma_{e(h),1S} = \text{-}\sum_{\overrightarrow{k},\nu}^{}{\frac{\left| g_{\overrightarrow{k},\nu} \right|^{2}}{\hbar \omega_{\text{LO}},\nu}\left( 2f_{e(h),\overrightarrow{k},\nu}^{\min} - \left( 1 + R_{e\left( h \right),\ \nu}^{2}k^{2} \right){f_{e(h),\overrightarrow{k},\nu}^{\min}}^{2} \right)}
\end{eqnarray} 
\end{widetext}
with mode dependent free polaron radii
${R_{e\left( h \right),\nu\ } = \left( \frac{\hslash}{{2m_{e(h)}\omega}_{{\rm LO},\nu}} \right)}^{1/2}$.

To provide a reference to computing the binding energies, we extend the
LLP empirical approach for free polarons (Eq.~\eqref{eq19}) to multiple phonons
(MLLP), thanks to a unitary transformation:
\begin{eqnarray}\label{eq86}
\widehat{W} = e^{\sum_{\overrightarrow{k},\nu}^{}\left( {F_{\overrightarrow{k},\nu}^{*}\widehat{a}}_{\overrightarrow{k},\nu} - F_{\overrightarrow{k},\nu}{\widehat{a}}_{\overrightarrow{k},\nu}^{+} \right)}
\end{eqnarray} 

At this stage, it is interesting to notice that one of the limitations
of the present multiple phonons implementation is the use of a
generalized LST equation connected to independent harmonic oscillators
[Eq.~\eqref{eq77LST}]. The simplified expression of the coupling strength 
[Eq.~\eqref{eq80}] is therefore derived assuming that the LO/TO resonances 
in the dielectric response are not overlapping, and the mode damping parameters 
are small.

Now, let us consider the specific case of metal halide perovskites. Considering the above mentioned approximation, the present
MLLP/MPB/MPBK/MIad models are expected to be valid solely for the low 
temperature (LT) regime where lattice anharmonicity is reduced (see Appendix \ref{AppA}). However, while experimental data on exciton
binding energies, reduced effective masses and effective dielectric
constants are usually obtained at LT,~\cite{Miyata_2015,Galkowski_2016,Yang_2017,Baranowski_2020,Yamada_2021,Baranowski_2024}
vibrational optical spectroscopy on lattice polar modes is essentially
reported at room temperature (RT).~\cite{Wakamura_2001,Sendner_2016,Miyata_2017,Zhao_2017,Nagai_2018,Lan_2019,Wang_2019,Boldyrev_2020,Maeng_2021}
Experiments at LT performed so far on MA-based 3D perovskites, are leading to more complex results, because LT structural phase transitions
are leading to a large number of closely spaced polar optical resonances in the dielectric
responses.~\cite{P_rez_Osorio_2015,La_o_vorakiat_2015,La_o_vorakiat_2015b,Nagai_2018,Boldyrev_2020,Anikeeva_2023,Zelezn__2023,Frenzel_2023} For the present work, we will consider 
at the end of this section, one of the very few THz experiment performed over a large temperature range.

To gauge the differences between the original versions of PB/PBK models based on a single effective mode and the proposed MPB/MPBK extensions,
let's first consider the fully inorganic 3D halide perovskite
CsPbCl\textsubscript{3}. Cs-based perovskites are less prone to phonon
resonance splitting at LT than MA-based compounds.~\cite{Nagai_2018}
Wakamura's paper on CsPbCl\textsubscript{3} is, to our knowledge, the
first detailed report on multiple mode resonances and LO/TO splitting in
inorganic 3D perovskites especially at LT.~\cite{Wakamura_2001}

\begin{table*}[]
\begin{tabular}{c|c|c|c|c|c|c|c}

\hline \hline 
\multicolumn{2}{c|}{CsPbCl$_3$ modes} & $\hbar \omega_{\text{LO}} / \hbar \omega_{\text{TO}} $ & Number of modes & $\mu^{\text{LLP}}$ & $\mu^{\text{PBK}}$ & $E_{b,1S}$ (meV) & $\varepsilon_{\text{eff}}$ \\ \hline \hline 

PB model         & effective         & 25.6 / 9.6                                             & 1 effective     & 0.372              & 0.205              & 64 (exp. 2K)     & 6.6 (exp. 2K)              \\ 
exp. 40K         & mode 1            & 27.5 / 13.6                                            & 1               & 0.299              & 0.204              & 76               & 6.0                        \\ 
exp. 40K         & mode 2            & 13.3 / 11.5                                            & 2               & 0.308              & 0.205              & 67               & 6.4                        \\
exp. 40K         & mode 3            & 9.9 / 9.8                                              & 3               & 0.309              & 0.205              & 66               & 6.5                        \\
exp. 40K         & mode 4            & 5.5 / 4.8                                              & 4               & 0.318              & 0.205              & 62               & 6.7                \\ \hline \hline 
       
\end{tabular}
\caption{Influence of the presence of multiple polar phonon
modes on excitonic polarons in CsPbCl\textsubscript{3} at LT computed
within the MPB model. Experimental values of the exciton binding
energy, effective dielectric constants and bare effective mass
($\mu = 0.404$) at T=2K are taken from ref.~\cite{Baranowski_2020} and the experimental values of the LO/TO optical mode frequencies at T=40K
from ref.~\cite{Wakamura_2001}. The experimental
LO/TO optical energies are given in the second column. For each line the number of phonon modes is
indicated. In the second line, a single effective LO mode is tuned
within the PB model to match the experimental exciton binding energy.
For the MPB calculations (lines 3-6), the experimental LO/TO optical
mode frequencies are used and included progressively. Computed binding
energies are given in the column 6.
}
\label{tab:Table5}
\end{table*}

Interestingly as shown in Tab.~\ref{tab:Table5} for CsPbCl\textsubscript{3}, both
experimental LO/TO optical frequencies and exciton characteristics are
available from the literature at LT. The binding energy can be evaluated
within PB's model by adjusting the energy of a single effective phonon
(Tab.~\ref{tab:Table1},\ref{tab:Table3}). Within the MPB model, similar values for the exciton binding
energy and effective dielectric constant values are retrieved, provided
that all 4 experimental phonon resonances are considered Tab.~\ref{tab:Table5}. 
We may notice
that, in such a case, the free polaron effective mass
$\mu^{\text{LLP}}$ is smaller than the one derived with a single
effective phonon. It means that the PB approximation of a single
effective phonon tends to overestimate the matrix element of the e-ph
interaction in a multimode situation, by attributing to this mode all
the polarizability responsible for the difference between
$\varepsilon_{\infty}$ and $\varepsilon_{s}$. A more physical
description such as the MPB model, accounts for the various
contributions of the experimental phonon lines to the total
polarizability. It is also clear from Tab.~\ref{tab:Table5} that the various modes do
not equally contribute to the physical properties of the excitonic
polarons in CsPbCl\textsubscript{3}. Mode 1, located at high energy, is
the dominant one. The mode-dependent $\alpha_{(e,\nu)}$ correspond respectively to
0.269, 0.010, 0.025, 3.14, while for a single effective mode, a value of
3.4 is needed.

\begin{table*}[]
\begin{tabular}{c|c|c|c|c|c|c|c}
\hline \hline 
\multicolumn{2}{c|}{MAPbBr$_3$ modes} & $\hbar \omega_{\text{LO}} / \hbar \omega_{\text{TO}} $ & Number of modes & $\mu^{\text{LLP}}$ & $\mu^{\text{PBK}}$ & $E_{b,1S}$ (meV) & $\varepsilon_{\text{eff}}$ \\ \hline \hline 

PB model         & effective         & 13.1 / 5.4                                             & 1 effective     & 0.200              & 0.119              & 25 (exp. 2K)     & 7.5 (exp. 2K)              \\ 
exp. 300K         & mode 1            & 21.3 / 14.6                                            & 1               & 0.146              & 0.118              & 40               & 6.3                        \\ 
exp. 300K         & mode 2            & 13.8 / 9.1                                             & 2               & 0.162              & 0.121              & 18               & 9.4                        \\ 
exp. 300K         & mode 3            & 6.2 / 5.6                                              & 3               & 0.165              & 0.122              & 16               & 10.0       \\ \hline \hline 
               
\end{tabular}
\caption{Influence of the presence of multiple polar phonon
modes on the excitonic polarons in MAPbBr$_3$.
Experimental values of the exciton binding energy, effective dielectric
constants and bare effective mass ($\mu = 0.202$) at T=2K are taken
from ref.~\cite{Galkowski_2016} and the experimental values of the LO/TO
optical mode frequencies at T=300K from ref.~\cite{Zelezn__2023}. The experimental LO/TO 
optical energies are given in the second column. For
each line the number of phonon modes is indicated. In the second line, a
single effective LO mode is tuned within the PB model to match the
experimental exciton binding energy. For the MPB calculations (lines
3-5), the experimental LO/TO optical mode frequencies are used and
included progressively. Computed binding energies are given in the
column 6.}
\label{tab:Table6}
\end{table*}

Tab.~\ref{tab:Table6} describes another example of a 3D halide perovskite,
MAPbBr\textsubscript{3}, for which data on phonons have been collected at
RT,~\cite{Zelezn__2023}, while experimental exciton characteristics
are available from the literature at LT.~\cite{Galkowski_2016} The binding
energy is first computed within PB's model by adjusting the energy of a
single effective phonon (see Tab.~\ref{tab:Table1},\ref{tab:Table3}). 
Using PB's model, it is possible
to almost perfectly match the LT exciton characteristics, however the
model provides little insight into the significance of the extracted
effective phonon energy (13.1 meV) compared to known spectroscopic data.
When the contributions of the phonon resonances at RT are progressively
included within the MPB model, the corresponding binding energy
decreases as the effective dielectric constant increases. Mode 1,
located at high energy is again the dominant one. The mode-dependent
$\alpha_{e,\nu}$ correspond respectively to 0.11, 0.18, 2.2, while for
a single effective mode, a value of 3.0 is needed. Within MPB, the computed
binding energy (16meV) finally ends up being smaller than the LT
experimental one (25meV). This is not a surprise because a temperature
induced dielectric screening effect was predicted early for MA-based 3D
perovskites.~\cite{Even_2014}

\begin{table*}[]
\begin{tabular}{c|c|c|c|c|c|c|c}
\hline \hline 
\multicolumn{2}{c|}{MAPbI$_3$ modes} & $\hbar \omega_{\text{LO}} / \hbar \omega_{\text{TO}} $ & Number of modes & $\mu^{\text{LLP}}$ & $\mu^{\text{PBK}}$ & $E_{b,1S}$ (meV)   & $\varepsilon_{\text{eff}}$ \\ \hline \hline 

PB model         & effective        & 8.2 / 3.1                                              & 1 effective     & 0.184              & 0.106              & 16 (exp. 2K)       & 9.4 (exp. 2K)              \\
PB model         & effective        & 10.8 / 4.1                                             & 1 effective     & 0.171              & 0.107              & 12 (exp. 155-190K) & 10.9 (exp. 155-190K)       \\
exp. 300K         & mode 2           & 16.6 / 7.8                                             & 1               & 0.139              & 0.109              & 11                 & 11.6                       \\
exp. 300K         & mode 3           & 5.0 / 4.0                                              & 2               & 0.144              & 0.111              & 7                  & 14.1                     \\ \hline \hline 
 
\end{tabular}
\caption{Influence of the presence of multiple polar phonon
modes on the excitonic polarons in MAPbI\textsubscript{3}. Experimental
values of the exciton binding energy, effective dielectric constants and
bare effective mass ($\mu = 0.202$) at T=2K and 155-190K are taken
from ref.~\cite{Galkowski_2016} and the experimental values of the LO/TO
optical mode frequencies at T=300K from ref.~\cite{Sendner_2016}. The experimental LO/TO
optical energies are given in the second column. For
each line the number of phonon modes is indicated. For
the MPB calculations (lines 4-5), the experimental LO/TO optical mode
frequencies are used and included progressively. Computed binding
energies are given in the column 6.}
\label{tab:Table7}
\end{table*}

Tab.~\ref{tab:Table7} reports the results computed for MAPbI\textsubscript{3} from
RT phonon spectroscopic data~\cite{Sendner_2016} with experimental
exciton characteristics from the literature reported at 2K and
155-190K.~\cite{Miyata_2015} The binding energy is first computed from
the PB model by adjusting the energy of a single effective phonon for
each temperature. For the MPB model we used several phonon resonances,
but the computed RT binding energy (7meV) is smaller than the one at
very LT (16meV), but closer to the one in the intermediate temperature
range (12meV). Mode 1, located at high energy is again the dominant one.

\begin{figure}[htb]
\includegraphics[width=0.45\textwidth]{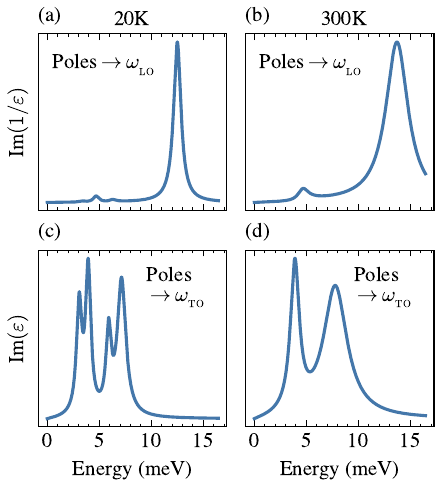}
\caption{Im$\left( 1/\varepsilon \right)$ and
Im$\left( \varepsilon \right)$ of MAPbI\textsubscript{3} obtained from
THz time-domain spectroscopy and plotted as a function of energy for the
lowest (20K) and the highest (300K) temperatures considered in
ref.~\cite{La_o_vorakiat_2015b}. The LO modes frequencies
$\omega_{{\rm LO},\nu}\ $correspond to the poles of $1/\varepsilon$, while
the TO modes frequencies $\omega_{{\rm TO},\nu}$ correspond to the poles of
$\varepsilon$.}\label{fig10}
\end{figure}

To verify whether temperature effects are correctly captured by the MPB model, 
one needs experimental spectroscopic reference data for a wide range of 
temperature, from RT down to LT. Detailed knowledge about the temperature and 
frequency dependence of the dielectric properties of the lattice would also be 
highly desirable. To our knowledge, the paper of La-o-vorakiat and coworkers on 
MAPbI\textsubscript{3}~\cite{La_o_vorakiat_2015b} is the only one providing an 
extended set of spectroscopic data for phonons. The authors performed complete THz
time-domain investigations
tracking the effect of the orthorhombic to tetragonal structural phase
transition in MAPbI\textsubscript{3} observed at LT (about 160K in their
study) on the polar optical modes at the $\Gamma$ point. As expected from the
symmetry of the two phases, the splitting of the two phonon lines
measured at RT is observed, leading to four modes at LT. This transition
is associated with a redistribution of
$\mathrm{\Delta}\varepsilon_{\nu}$ between the modes, when the
mode-specific contributions to the dielectric constant are considered
[Eq.~\eqref{eq81}]. In the original work by La-o-vorakiat and coworkers~\cite{La_o_vorakiat_2015b} the frequencies of the TO modes
$\omega_{{\rm TO},\nu}$ are provided as a function of temperature together
with $\mathrm{\Delta}\varepsilon_{\nu}$ and the TO mode damping
$\gamma_{{\rm TO},\nu}$. Herin we extract the frequencies of the LO modes
$\omega_{{\rm LO},\nu}$ and the e-ph coupling matrix elements
$g_{\overrightarrow{k},\nu}$, which are required for the
implementation of the MPB model, by analysing the
frequency-dependent dielectric constant. We assume
that the structural deviations from the cubic symmetry
(MAPbI\textsubscript{3} at the continuous phase transition from the
I4mcm tetragonal space group to the Pm$\overline{3}$m cubic one
slightly above RT at 327K), are not significant to invalidate 
Toyozawa's analysis as well as the MPB model. Based on the experimental
data of Ref.~\cite{La_o_vorakiat_2015b} and the frequency dependent expression
for $\epsilon (\omega)$ [Eq.~\eqref{eq76}], Fig.~\ref{fig10} shows Im$\left( 1/\varepsilon \right)$ and
Im$\left( \varepsilon \right)$ as a function of energy for the lowest
(20~K) and the highest (300~K) temperatures reported in 
Ref.~\cite{La_o_vorakiat_2015b}. The only piece of experimental information 
that is missing is the temperature-dependent value of
$\varepsilon_{\infty}$, which we approximate to a temperature
independent value of 5. In Fig.~\ref{fig10}, the LO modes frequencies
$\omega_{{\rm LO},\nu}\ $correspond to the poles of $1/\varepsilon$, while
the TO mode frequencies $\omega_{{\rm TO},\nu}$ correspond to the poles of
$\varepsilon$. Both the influence of mode damping at RT and the
splitting of the modes at LT are evident from the plots. Fig.~\ref{fig10} shows 
that the highest frequency TO mode at about 13-14 meV undergoes the largest LO-TO 
splitting and its LO component 
yields the largest contribution to $\frac{1}{\varepsilon\left( \omega \right)}$.

\begin{figure}[]
\includegraphics[width=0.48\textwidth]{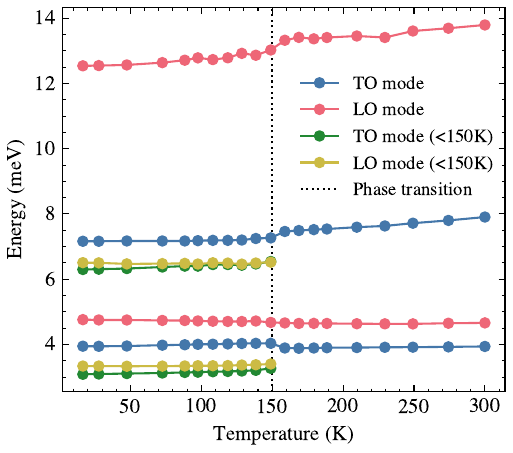}
\caption{Energies of the LO modes as a function of
temperature of MAPbI\textsubscript{3} obtained by a numerical analysis
of experimental data~\cite{La_o_vorakiat_2015b} using Toyozawa's approximate
expression for $\varepsilon\left( \omega \right)\text{.\ }$
Experimental energies of TO modes~\cite{La_o_vorakiat_2015b} are also shown.\label{fig11}}
\end{figure}

Starting from the experimental data, the first step within Toyozawa's
approach which neglects phonon damping, consists in finding numerically
the values LO modes frequencies $\omega_{{\rm LO},\nu}$ leading to a good
match with the experimentally observed dielectric jump
$\mathrm{\Delta}\varepsilon_{\nu}$ (Eq.~\eqref{eq81}). The extracted LO modes
energies together with the experimental TO modes energies are
represented as a function of temperature in Fig.~\ref{fig11}. Despite neglecting
the experimental damping, the LO modes frequencies extracted from this
procedure coincide well at the maxima of
$\text{Im}\left( 1/\varepsilon \right)$ (Fig.~\ref{fig10}). The second step of
Toyozawa's analysis consists in using the LO and TO frequencies to
evaluate the e-ph coupling matrix elements
$g_{\overrightarrow{k},\nu}$ using Eqs.~\eqref{eq79} and \eqref{eq80}. As expected from 
Fig.~\ref{fig10} and ~\ref{fig11}, the highest frequency LO mode at about 13-14 meV 
arises from the largest LO-TO splitting and corresponds to the largest e-ph coupling
matrix elements $g_{\overrightarrow{k},\nu}$ over the entire
temperature range.

\begin{figure}[]
\includegraphics[]{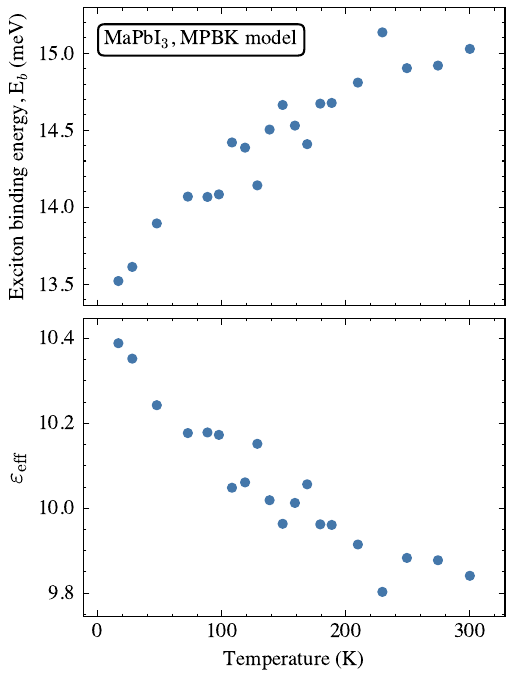}
\caption{Effective exciton binding energy (top) and
dielectric constant (bottom) computed as a function of temperature for
MAPbI\textsubscript{3} using the MPB model and input data from ref.~\cite{La_o_vorakiat_2015b}.\label{fig12}}
\end{figure}

Using the values of the e-ph coupling matrix elements
$g_{\overrightarrow{k},\nu}$, we evaluate the exciton binding energy
and effective dielectric constant as a function of temperature over the
entire temperature range based on the MPB model (Fig.~\ref{fig12}). Both the
calculated quantities undergo a continuous variation across the
temperature range without exhibiting a discontinuity at the first order
phase transition around 160K. It indicates that oscillator strengths are
smoothly redistributed between the vibrational modes, when the number of
phonon modes increases from 2 in the tetragonal phase to 4 in the
orthorhombic phase because of the folding of the BZ.~\cite{Zacharias_2023b} However, the experimentally observed
temperature evolution of both these quantities is not well described
within the MPB model. The model predicts a slight increase of the
exciton binding energy from LT to RT, whereas a sizeable reduction has
been experimentally observed between 2K and the intermediate temperature
range of 155-190K.~\cite{Miyata_2015} This reduction has to be
related to the lattice disorder and to the strong anharmonicity at higher temperatures that affect the phonon spectral properties, but also the electronic states.~\cite{Zacharias_2023b} These effects are not
accounted for within the present PB and the MPB frameworks, which remain
basically 0K theories. Further theoretical developments are needed to incorporate temperature effects in the description of excitonic polarons. Nevertheless, these models can be exploited to predict the Huang-Rhys factors for side bands, evidenced at LT in photoluminescence experiments, related to virtual phonon populations (vide infra). 

\section{Connection with first principles approaches for excitonic polarons}  \label{SecVI}

\subsection{Parameters from first principles approaches for excitonic polaron Hamiltonians}  \label{SecVIA}
Today, first-principles approaches, whose workhorse is DFT, have proven particularly useful in predicting many materials properties related to their electronic structure, including e-ph interactions.~\cite{Giustino_2017} They have also proved effective in the context of semi-empirical methods, supplying parameters especially when experimental data are scarce or lacking. For excitonic polarons, this encompasses carrier effective masses, phonon frequencies, LO-TO splittings, dielectric constants and e-ph coupling strengths. Meanwhile, it is well documented that the favour of DFT  significantly impacts the value of the computed quantities and, in turn, the excitonic polaron properties derived within the semi-empirical model. To illustrate this, we consider the prototypical case of TlCl (see Appendix~\ref{AppCompDetails}). As mentioned earlier, this ionic semiconductor has e-ph scattering dominated by the Fr\"ohlich interaction, with e-ph coupling strength in the intermediate regime, similar to that of halide perovskites, but with a simpler vibrational density of states. 

\begin{figure}[htb]
\includegraphics[]{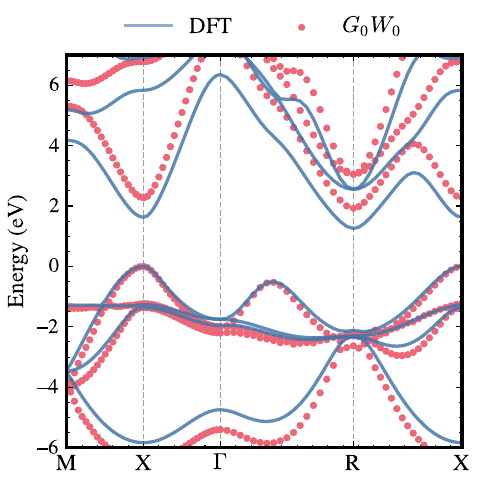}
\caption{Electronic band structures of TlCl computed using semilocal DFT (blue line) and one-shot GW (red circles) methods. The spin-orbit coupling (SOC) effects have been accounted for in these calculations. The band structures are aligned along energy axis by setting the energy of valence band maximum at zero. }\label{TlClelecbands}
\end{figure}

Initial observations can be made on the electronic band structure of TlCl. Without spin-orbit coupling (SOC), the electronic band gap is found direct at the  X point ($1/2,0,1/2$) in reciprocal space of the BZ, while it becomes slightly indirect when SOC is included (band structure with SOC shown in Fig.~\ref{TlClelecbands}). The same conclusion holds when many body effects are included at the GW level based on computed self-energy corrections. In addition, both with and without SOC, the effective masses are anisotropic.  The situation is thus more complex than the simple case of direct band gap with isotropic effective masses underlying the PBK framework. 

To proceed with the semi-empirical modelling based on first-principles parameters, we consider average effective masses at the X point of the BZ (Tab.~\ref{tab:TlClDFT}). The numerical values of the effective masses computed at the different levels of theory show sizeable differences, which will directly reflect in the computed quasiparticule binding energies. The same holds for the other physical observables including static and high frequency dielectric constants, phonon frequencies as well as LO-TO splittings. Noteworthy, among all parameters, the high frequency dielectric constant $\varepsilon_{\infty}$ has a dramatic impact on the predicted excitonic polaron characteristics, mirroring its well known effect on exciton binding energies as obtained in various frameworks including within the BSE in the absence of exciton-phonon coupling.

\begin{table*}[]
\def\arraystretch{1.5}
\centering
\begin{tabular}{c|c|c|c|c|c||c|c|c|c|c|c|c|c|c|c|c}
\hline \hline 
\multicolumn{6}{c||}{\textbf{ab initio}} &\multicolumn{11}{c}{\textbf{empirical models}} \\
\hline \hline 
method & $m_e$ & $m_h$ &  $\varepsilon_{\infty}$ & $\varepsilon_{s}$ & $\hbar \omega_{\rm LO}$ 
& $R_e$ & $R_h$ & $\alpha_e$ & $\alpha_h$ & $\mu$ & $\mu^{\text{PBK}}$ & $\mu^{\text{LLP}}$ & $E_{b}^{\text{PBK}}$ & $E_{b}^{\text{PB}}$ & $\varepsilon^{\text{PBK}}$ & $a_{B}^{\text{PBK}}$\\ 
 &  &  &  & & (meV) & (nm) & (nm) & & & & & & (meV) & (meV) &  & (nm) \\ 
\hline \hline 
DFT & 0.25 & 0.48 & 5.8 & 53.7 & 19.8 & 2.8 & 2.0 & 1.99 & 2.77 & 0.162 & 0.199 & 0.242 
& 3.3 & 4.1 & 26 & 2.8 \\
\hline 
$G_0W_0$ & 0.19 & 0.38 & 5.8 & 53.7 & 19.8 & 3.2 & 2.5 & 1.75 & 2.23 & 0.127 & 0.225 & 0.180 
& 0.7 & 1.3 & 48 & 5.7 \\
\hline 
DFT+SOC & 0.24 & 0.45 & 6.1 & 78.1 & 19.6 & 2.9 & 2.1 & 1.94 & 2.66 & 0.157 & 0.371 & 0.231 
& 0.4 & 1.0 & 73 & 3.9 \\
\hline 
$G_0W_0$+SOC & 0.18 & 0.35 & 6.1 & 78.1 & 19.6 & 3.3 & 2.5 & 1.68 & 2.21 & 0.112 & 0.170 & 0.157 
& 0.3 & 0.4 & 75 & 20.4 \\
\hline \hline                            
\end{tabular}
\caption{Parameters for the implementation of excitonic polaron empirical models as derived from various ab initio approaches: carrier effective masses, dielectric constants and phonon frequencies are computed for TlCl. These parameters are then used to compute polaron and excitonic polaron properties within the PB and PBK empirical models.}
\label{tab:TlClDFT}
\end{table*}

\begin{figure}[]
\begin{center}
    
\includegraphics[]{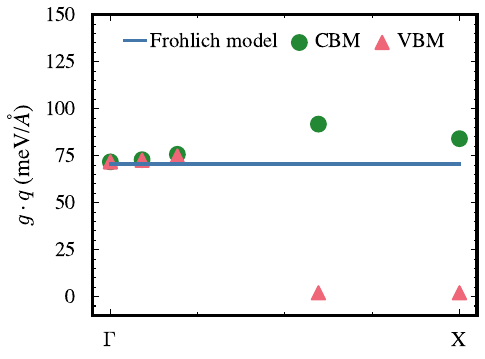}
\caption{Comparison of the dot product between electron-phonon coupling ($g$) and phonon momentum ($q$) computed using empirical Fr\"{o}hlich model and density functional perturbation theory (DFPT), for TlCl. The latter method allows calculations of state-specific electron-phonon interaction and includes both short-range and long-range (Fr\"{o}hlich) contributions. The diagonal elements of the DFPT-computed electron-phonon matrices corresponding to valence band maximum (VBM) and conduction band minimum (CBM) are plotted here. In the limit $q \rightarrow 0$, Fr\"{o}hlich contributions dominate bringing both methods in quantitative agreement.  }\label{TlClFroelich}
\end{center}
\end{figure}

The approximate Fr\"ohlich expression that accounts for the coupling between electrons and polar optical phonons can also be assessed based on DFT calculations. In fact, density functional perturbation theory (DFPT) can be implemented to compute e-ph matrix elements accross the BZ as illustrated for TlCl in Fig.~\ref{TlClFroelich}. Comparison to the Fr\"ohlich model reveals its validity over a significant $q$-range. Overall, this section briefly highlights the pros and cons of using first-principles calculations in connection with empirical models when more advanced developments, such as the full DFT one which will be discussed next, are unusable either for numerical or fundamental reasons. 

\subsection{Full DFT framework for excitonic polarons}  \label{SecVIB}
Progress has been made in recent years on the development of exciton
coupling theory from first principles~\cite{Christiansen_2019,Chen_2020,Antonius_2022},
including approaches to excitonic polarons.~\cite{Dai_2024,Dai_2024b,Bai_2024,Jin_2024}
Some last developments are connected to first-principles approaches for
free polarons.~\cite{Sio_2019} It is thus interesting to make a
comparison between the empirical PBK model detailed in the present work and
recent state-of-the-art atomistic modelling approaches to Fr\"ohlich
excitonic polarons. In their introduction, based on the papers of
Iadonisi and coworkers the authors of refs.~\cite{Dai_2024,Dai_2024b} state
that the empirical formalism is biased to find only localized solutions,
since the dispersions of the exciton bands are not considered. If indeed
the excitonic polaron dispersion is not accounted by the Iad analytical
solution for the ground state,~\cite{Iadonisi_1983} this is generally
not true for the PB formalism, which is precisely based on a unitary
transformation that introduces the excitonic polaron total momentum (Eqs.~\eqref{eq38}). 
Besides, as shown in the present work, it is possible to analyze
the change of the excitonic polaron total mass from PB's formalism,
something that has not yet been performed in detail from first
principles (Fig.~\ref{fig6}, Eqs.~\eqref{eq57}-\eqref{eq59bis}). Indeed, first-principles approaches
allow assessing the exciton~\cite{Cudazzo_2016} or the excitonic
polaron dispersion,~\cite{Dai_2024} but the calculation needs to be
performed for each value of the exciton momentum, and the computational
cost scales linearly with the size of the momentum grid used for
sampling.

For free polarons, first principles approaches give access to envelop
functions for the atomic displacements of the phonon mode $\nu$:
\begin{eqnarray}\label{eq72}
B_{\overrightarrow{q},\nu} = \frac{1}{N_{p}}\sum_{m,n,\overrightarrow{k}}^{}{A_{m,\overrightarrow{k} + \overrightarrow{q}}^{*}\frac{g_{\text{mn},\nu}\left( \overrightarrow{k},\overrightarrow{q} \right)}{\hbar \omega_{\overrightarrow{q},\nu}}}A_{n,\overrightarrow{k},\nu} \,\,
\end{eqnarray} 
where the
$g_{\text{mn},\nu}\left( \overrightarrow{k},\overrightarrow{q} \right)$
are the e-ph coupling matrix elements. Using these envelop functions and
eigenvectors for the phonon modes, the atomic displacements in the
polaron cloud can be computed.~\cite{Sio_2019,Britt_2024} This envelop
function is connected to
$F_{\overrightarrow{k},\nu}^{\min}\left( \overrightarrow{Q} \right)$
in the LLP empirical approach to free polarons extended to multiple
phonons (MLLP model, vide infra, Eq.~\eqref{eq86}). 
We may note that self-consistency is important for the
simulation of the free carrier polaron from first
principles~\cite{Lafuente_Bartolome_2022} on par with the empirical calculations
within LLP's formalism (Eqs.~\eqref{eq28} and \eqref{eq29}).

For excitonic polarons, the recent DFT-based formalism  starts by
combining the electronic ground state and BSE excitation energies with a
quadratic expression for the elastic energy associated with the lattice
distortion and atomic displacements. It involves a BSE Hamiltonian which
depends on the atomic displacements and thus allows to account for a
distorted structure. This Hamiltonian is then expanded linearly as a
function of the displacements, and the total energy is minimized with
respect to the exciton wavefunction
$\Psi\left( \overrightarrow{\mathbf{r}_{\mathbf{e}}}\mathbf{,}\overrightarrow{\mathbf{r}_{\mathbf{h}}} \right)$
and atomic displacements. This is formally equivalent to the functional
and wavefunction minimization at the heart of PB's formalism (Eqs.~\eqref{eq42}-\eqref{eq43}). 

To reduce the computational cost of the DFT-based approach, the
excitonic polaron Hamiltonian is expressed~\cite{Dai_2024} in an
exciton (e-h pair) basis
$\Psi\left( {\overrightarrow{r}}_{e},{\overrightarrow{r}}_{h} \right) = \sum_{s\overrightarrow{Q}}^{}{A_{s,\overrightarrow{Q}}\sum_{\text{vc}\overrightarrow{k}}^{}{a_{\text{vc}\overrightarrow{k}}^{s,\overrightarrow{Q}}\varphi_{c,\overrightarrow{k} + \overrightarrow{Q}}\left( {\overrightarrow{r}}_{e} \right)\varphi_{v,\overrightarrow{k}}^{*}\left( {\overrightarrow{r}}_{h} \right)}}$.
This basis choice brings the first principles formalism close to the
empirical one~\cite{Dai_2024}:
\begin{widetext}
\begin{eqnarray}\label{eq73}
\sum_{s'\overrightarrow{Q}'}^{}\left\lbrack E_{s\overrightarrow{Q}}^{0}\delta_{ss'}\delta_{\overrightarrow{Q}\overrightarrow{Q}'} - \frac{2}{N_{p}}\sum_{\nu}^{}{B_{\overrightarrow{Q} - {\overrightarrow{Q}}^{'},\nu}\mathcal{G}_{ss^{'},\nu}\left( \overrightarrow{Q}',\overrightarrow{Q} - {\overrightarrow{Q}}^{'} \right)} \right\rbrack A_{s^{'},\overrightarrow{Q}'} = EA_{s,\overrightarrow{Q}}
\end{eqnarray} 
where the envelop function is given by
\begin{eqnarray}\label{eq74}
B_{\overrightarrow{Q},\nu} = \frac{1}{N_{p}}\sum_{ss'{\overrightarrow{Q}}^{'}}^{}{A_{s^{'},\overrightarrow{Q}'}^{*}\frac{\mathcal{G}_{ss^{'},\nu}^{*}\left( \overrightarrow{Q}',\overrightarrow{Q} \right)}{\hbar \omega_{\overrightarrow{Q},\nu}}A_{s,\overrightarrow{Q} + \overrightarrow{Q}'}},
\end{eqnarray}
and $E_{s\overrightarrow{Q}}^{0}$ are the eigenvalues of the undistorted
structure. $\mathcal{G}_{ss^{'},\nu}$ are the exciton-phonon matrix
elements given by:
\begin{eqnarray}\label{eq75}
\mathcal{G}_{ss^{'},\nu}\left( \overrightarrow{Q},\overrightarrow{q} \right) =\sum_{\text{vc}\overrightarrow{k}}^{}a_{\text{vc}\overrightarrow{k}}^{s,\overrightarrow{Q} + \overrightarrow{q}*} 
\left[ 
\sum_{\text{c'}}{g_\text{cc'},\nu}\left( \overrightarrow{k} + \overrightarrow{Q},\overrightarrow{q} \right)a_{\text{vc'}\overrightarrow{k}}^{s',\overrightarrow{Q}} - \sum_{\text{v'}}{g_{\text{v'v},\nu}\left( \overrightarrow{k},\overrightarrow{q} \right)a_{\text{v'c}\overrightarrow{k} + \overrightarrow{q}}^{s'\overrightarrow{Q}}} 
\right].
\end{eqnarray} 

\end{widetext}

As shown by the authors of ref.~\cite{Dai_2024}, in the framework
of the Wannier exciton and Fr\"ohlich interaction models, also considered
in the present work, the e-ph coupling matrix elements finally
correspond to Eq.~\eqref{eq11} and a hydrogenic wavefunction for the 1S ground
state can be considered
$\Psi_{1S}\left( \overrightarrow{R_{e}},\overrightarrow{R_{h}} \right) \approx \frac{e^{i\overrightarrow{Q}.\overrightarrow{R}}e^{- \frac{r}{a_{B}^{\text{eff}}}}}{{a_{B}^{\text{eff}}}^{3/2}\pi^{1/2}}$.
Here, $\left( \overrightarrow{R_{e}},\overrightarrow{R_{h}} \right)$
refer to various unit cell positions whereas
$\left( \overrightarrow{R},\overrightarrow{r} \right)$ are the center
of mass and relative e-h positions. This envelop function is connected
to
$F_{\overrightarrow{k},\nu}^{\min}\left( \overrightarrow{Q},\overrightarrow{r} \right)$
in the empirical approach to excitonic polarons extended to multiple
phonons (MPB/MPBK models, Eq.~\eqref{eq86}). These last considerations bridge the
DFT-based approach of Refs.~\cite{Dai_2024,Dai_2024b} with the PB or MPB
(vide infra) formalisms.

In ref.~\cite{Dai_2024} the authors used the case of LiF to
illustrate their novel DFT-based approach. The polaron radii and
$\alpha$ values of LiF reported in Tab.~\ref{tab:Table1} show that it corresponds to
the case of a strongly bound exciton, which is clearly beyond the
validity limit of the LLP model suitable for free polarons. This is
especially true for the hole polaron as pointed out by the authors. LiF hosts
strongly bound Holstein hole polarons, with a hole polaron radius smaller than 
the lattice constant, and Fr\"ohlich electron polarons. 
It is nevertheless interesting to compare the excitonic
polaron properties of LiF computed using the present PBK model and
results obtained using the atomistic approach of Dai and coworkers
(Tab.~\ref{tab:Table4}). In such a case where the interaction with the lattice is very
strong, the self-energies computed by the LLP model deviate from the DFT
ones, as expected, especially for the Holstein hole polaron. However,
the self-energy computed for the excitonic polaron shows a better
agreement between DFT and PB. This results from the strong reduction of the polaronic
distortions due to e-h correlations beyond the weak coupling regime, an
effect which is captured within PB's formalism (Fig.~\ref{fig4}e,f). The exciton
binding energy is significantly underestimated in PB formalism.
This is partly due to the inadequacies of the LLP
approach for free polarons in the strong coupling regime.
Future studies deserve to further compare results from atomistic
approaches with those derived within empirical models, for materials
more adapted to their respective limitations.

\begin{table*}[]
\def\arraystretch{1.5}
\centering
\begin{tabular}{c||c|c|c|c|c|c|c}
\hline \hline 
LiF        & $\mu$ & $\mu^{\text{PBK}}$ & $\mu^{\text{LLP}}$ & $E_{b,1S}$ & $\sigma_e$ (eV) & $\sigma_h$ (eV) & $\sigma_{e,1S} + \sigma_{h,1S}$ (eV) \\ \hline \hline 
Dai et al. &       &                    &                    & 1.88       & 0.2             & 1.98            & 0.46                                 \\
PBK        & 0.158 & 0.161              & 0.180              & 1.35       & 0.38            & 0.85            & 0.42    \\ \hline \hline                            
\end{tabular}
\caption{Excitonic polaron properties for LiF. The first line
reports the results obtained with the first principles approach by Dai
and coworkers.~\cite{Dai_2024} The second line shows results
obtained for LiF within the LLP free polaron and PBK excitonic polaron
empirical models considered in this work. Exciton binding energies
$E_{b,1S}$, electron ($\sigma_{e}$) and hole ($\sigma_{h}$) free
polarons, and excitonic polaron $\sigma_{e,1S} + \sigma_{h,1S}\ $ self-energies
are given in eV.}
\label{tab:Table4}
\end{table*}

\section{Physical observables for excitonic polarons}  \label{SecVIII}

High resolution and time resolved diffraction experiments such as
ultrafast electron diffraction (UED) can be used in principle to extract
the details of the envelop functions of the distortion fields
represented in MLLP/MPB/MPBK/MIad empirical models by
$F_{\overrightarrow{k},\nu}^{\min}\left( \overrightarrow{Q} \right)$
for free polarons and
$F_{\overrightarrow{k},\nu}^{\min}\left( \overrightarrow{r,}\overrightarrow{Q} \right)$
for excitonic polarons.~\cite{Britt_2024} But the practical task of
analyzing these quantities is not easy in complex materials such as
halide perovskites since this information is mainly related to diffuse
scattering, which is a weak component in many
experiments.~\cite{Wu_2017,Cuthriell_2022,Zhang_2023,Seiler_2023,Yazdani_2023} To simply gauge the importance of
the lattice distortion associated to free polarons or excitonic polarons,
the virtual phonon cloud population provides a simple dimensionless
quantity related to the entire distortion field:
\begin{eqnarray}\label{eq87}
N_{1S,\nu}\left( \overrightarrow{0} \right) = \left\langle \varphi_{1S}\left( \overrightarrow{r} \right) \middle| \sum_{\overrightarrow{k}}^{}\left| F_{\overrightarrow{k},\nu}^{\min}\left( \overrightarrow{r,}\overrightarrow{0} \right) \right|^{2} \middle| \varphi_{1S}\left( \overrightarrow{r} \right) \right\rangle. \nonumber  \\
\end{eqnarray} 

\begin{figure}[htb]
\includegraphics[width=0.48\textwidth]{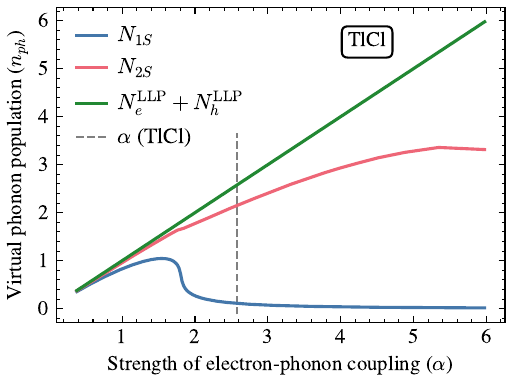}
\caption{Virtual phonon populations of excitonic polarons in
TlCl as a function of $\alpha$, for both 1S (blue) and 2S (red) 
excitons, compared to those obtained for free polarons (green). 
The variation of $\alpha$ is obtained by tuning the phonon
energy. The vertical dashed line indicates the value of $\alpha$ corresponding to the experimental phonon frequency measured for TlCl (Tab.~\ref{tab:Table1}).}\label{fig13}
\end{figure}

A partial cancellation of lattice distortions induced by the
correlations between electron and hole
$N_{1S,\nu}\left( \overrightarrow{0} \right) \neq N_{e,\nu}^{\text{LLP}}\left( \overrightarrow{0} \right) + N_{h,\nu}^{\text{LLP}}\left( \overrightarrow{0} \right)$
is predicted for excitonic polarons~\cite{Meyer_1956} (Fig.~\ref{fig4}b,f).
This is one important difference between excitonic polarons and free (or
non-interacting) polarons. Similar considerations hold for bipolaronic
states (bound states formed from pairs of particles with the same
charge, see Appendix \ref{AppI}).~\cite{Bassani_1991,Devreese2010_arxiv} The PB or MPBs 
approaches further provides convenient ways to quantify the lattice distortion for both the
ground state and the excited states of excitonic polarons. In Fig.~\ref{fig13},
the virtual phonon populations are reported as a function of $\alpha$,
for both 1S and 2S excitons in TlCl and are compared with virtual phonon populations 
of free polarons. The case of a 2S exciton is studied by replacing the function
$G_{1S}\left( \overrightarrow{k},a_{B}^{\text{eff}} \right)$ [Eq.~\eqref{eq47}]
for a 1S exciton by a similar expression for the 2S
exciton.~\cite{Matsuura_1980} As expected from the qualitative picture
in Fig.~\ref{fig4}d, the e-h correlations in a 1S exciton 
impact more importantly the distortions than in case of a 2S exciton. Significant
deviations from the hydrogen-like Rydberg series for the nS excitonic
state energies are thus expected in the intermediate coupling
regime.~\cite{Matsuura_1980} In this coupling regime, the e-h
correlations begin to have an impact on the interaction between the charge carriers and the lattice, in the case of a 2S exciton.

\begin{figure}[htb]
\includegraphics[width=0.48\textwidth]{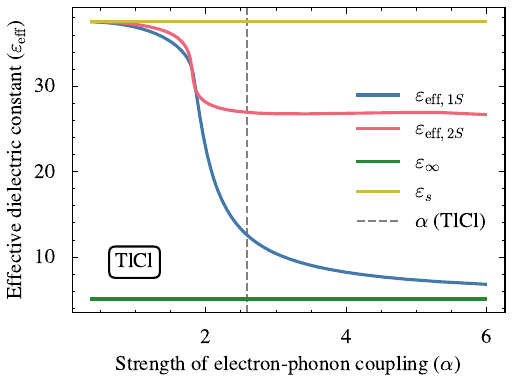}
\caption{Effective dielectric constants for 1S (blue) and 2S
(red) excitons as a function of $\alpha$ for TlCl. The variation of $\alpha$ is obtained by tuning the phonon energy. The vertical dashed line indicates the value of $\alpha$ corresponding to the experimental phonon frequency measured for TlCl (Tab.~\ref{tab:Table1}).}\label{fig14}
\end{figure}

To further illustrate that the interplay between e-h correlated motions
and the polaronic coupling depends on the extension of the exciton
wavefunction, we plot for TlCl the effective dielectric constants
computed for both 1S and 2S excitons as a function of
$\alpha$, keeping the same parameters (Fig.~\ref{fig14}). As expected, both
1S and 2S effective dielectric constants tend toward
$\varepsilon_{\infty}$ in the strong coupling limit, i.e.
$\alpha \rightarrow + \infty$ . Interestingly, within this model a
plateau is predicted for$\ \varepsilon_{eff,2S}$ in the intermediate
coupling regime. Such a plateau is related to the first lobe in the radial
probability density of a 2S exciton wavefunction in the range of small
$\ \widetilde{r}$ values (Fig.~\ref{fig4}d).

\begin{figure}[htb]
\includegraphics[width=0.48\textwidth]{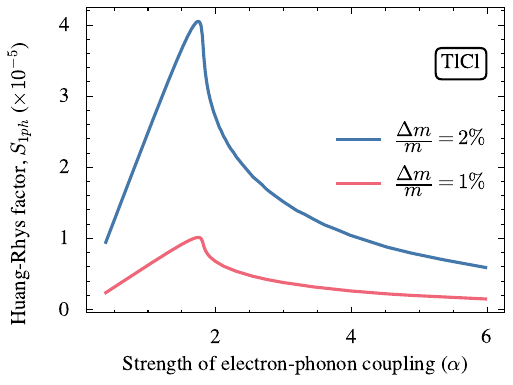}
\caption{ Variation of the Huang-Rhys factor
($S_{1_{\text{ph}}}$) as a function of $\alpha$ for the 1S exciton
for TlCl considering few cases with different electron and hole
effective masses (the relative mass ratio is given by
$\frac{\Delta m}{m}$). The variation of $\alpha$ is obtained by
tuning the phonon energy.}\label{fig15}
\end{figure}

A physical observable, very often used to gauge the importance of the
coupling between an exciton and the lattice, is the strength of the
phonon side bands referred to as the Huang-Rhys factor ($S$). This
factor corresponds to the oscillator strength of the optical transition
from the crystal ground state to an excitonic polaron plus one (real)
phonon ($1_{\text{ph}}$), divided by the oscillator strength of the
excitonic polaron ground state related to the zero-phonon
($0_{\text{ph}}$) line. The latter is usually evaluated using photoluminescence spectroscopy. 
At the level of the PB model with one LO mode, this ratio is
given by:~\cite{Matsuura_1980}
\begin{eqnarray}\label{eq88}
S_{1_{\text{ph}}} = S = \sum_{\overrightarrow{k}}^{}{\frac{\left| g_{\overrightarrow{k}} \right|^{2}}{\left( \hbar \omega_{\text{LO}} \right)^{2}}\left( f_{e,\overrightarrow{k}}\left( \overrightarrow{0} \right) - f_{h,\overrightarrow{k}}\left( \overrightarrow{0} \right) \right)^{2}} 
\end{eqnarray} 
\noindent
and the ratios for the n-phonons ($n_{\text{ph}}$) side bands are
given by
\begin{eqnarray}\label{eq89}
S_{\ n_{\text{ph}}} = \left( S \right)^{n}/n!. 
\end{eqnarray} 
\noindent
These expressions show that the ratio is proportional to the strength of
the interaction and depends on the phonon frequency. 
Eq.~\ref{eq88} will be generalized to multiple phonon branches within the
MPB model (vide infra). It is affected by the correlations between the
electron and hole through the factors
$f_{e\left( h \right),\overrightarrow{k}}\left( \overrightarrow{0} \right)$.
But when the two factors are identical, i.e. for equal masses, they
cancel each other. Fig.~\ref{fig15} reports the computed Huang-Rhys factor as a
function of $\alpha$ for the 1S exciton in TlCl for different values
of the relative difference between the electron and hole effective
masses. Evidently, Huang-Rhys factors are very sensitive to the
peculiarities of the electronic structure, and at $\alpha > 1.5$, they
undergo renormalization due to the onset of stronger e-h
correlations at the cross-over between the weak and medium coupling
regimes.

\begin{figure}[htb]
\includegraphics[width=0.48\textwidth]{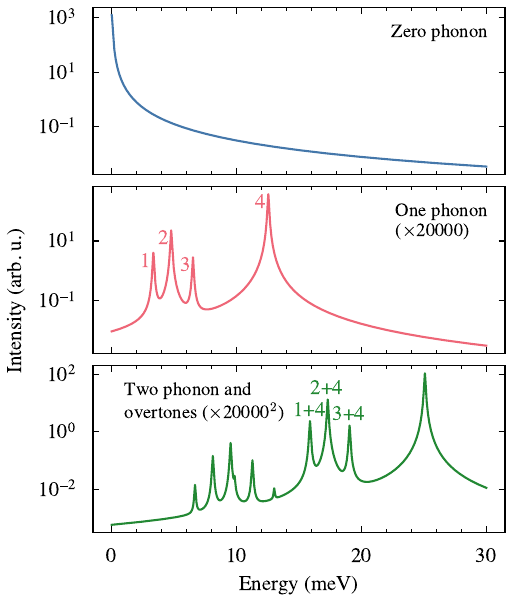}
\caption{Simulation of the intensity as a function of the
energy of the $0_{\text{ph}}$ absorption (top), the first-order
(middle) and second-order (bottom) side band absorptions, including
overtones. The experimental parameters for MAPbI\textsubscript{3} at
T=20K have been used (Figs.~\ref{fig10} and~\ref{fig11}). The theoretical expressions have
been arbitrarily convoluted by Lorentzian function with a broadening of
50 $\mu$eV. The four experimental LO modes  (Figs.~\ref{fig10} and~\ref{fig11}) are indicated on
the first-order side-band absorption curve. Some overtones combining the
LO mode n$^{\circ}$ 4 and the three other modes are indicated by 1+4, 2+4, 3+4. \label{fig16}
}
\end{figure}

We further propose  extending the calculation of Huang-Rhys factors to the case of 
multiple phonon lines within the present MPB model (contribution n° 9). We derive 
a sum rule (see Appendix~\ref{AppK}):
\begin{eqnarray}\label{eq90}
e^{- S}e^{S} &=& e^{- S}\sum_{n = 0}^{+ \infty}\frac{1}{n!}\left( \sum_{\nu}^{}S_{1_{\nu}} \right)^{n} \\ \nonumber &=& e^{- S}\left( 1 + \left( \sum_{\nu}^{}S_{1_{\nu}} \right) + \frac{1}{2}\left( \sum_{\nu}^{}S_{1_{\nu}} \right)^{2} + \ldots \right) 
\end{eqnarray} 
where
$S_{1_{\nu}} = \sum_{\overrightarrow{k}}^{}\left| F_{\overrightarrow{k},\nu}\left( \overrightarrow{0} \right) \right|^{2}$
is the ratio of the one phonon sideband intensity for phonon branch
$\nu$ divided by the $0_{\text{ph}}$ line and
$S = \sum_{\nu}^{}S_{1_{\nu}}$. The last term on the right hand side
of Eq.~\eqref{eq90} corresponds to both two-phonon side band intensities for the
same phonon branch ($\nu_{1} = \nu_{2}$) and overtones involving two
different phonon branches ($\nu_{1} \neq \nu_{2}$). As an example,
Fig.~\ref{fig16} represents the first-order side band absorption (red line) and
second-order side bands and overtones intensity (green line) as a
function of the energy for the case of MAPbI\textsubscript{3} at LT.

\section{Perspectives and conclusion}  \label{SecIX}

\subsection{Extension of the empirical approach for excitonic polarons to quantum nanostructures}
We will briefly discuss how the method of shift operators and the PB
empirical framework can be transferred to analyze excitonic polarons in
quantum nanostructures focusing on 2D quantum-wells and 0D quantum dot
nanostructures. For 2D quantum well structures, the continuous variation
of the $\overrightarrow{k}$ wavevector is lost along the stacking
direction because of quantum confinement and must be replaced by
discrete quantum numbers. This has been initially used to predict 2D
excitons binding energies using effective mass model for the charge
carriers and Wannier Hamiltonian for the excitons, including dielectric
confinement, optical anisotropy, electronic sub-bands and envelope
functions along the stacking axis.~\cite{Miller_1981,Bastard_1982,Matsuura_1984} 
Such approaches, which do not explicitly account for the e-ph coupling, can
also be fruitfully used for 2D layered halide perovskites to study the
dependence of the exciton binding energy on the quantum well thickness
or deviations from the 2D Rydberg series.~\cite{Blancon2018} Extension
of the PB framework to explicitly account for excitonic polarons in 2D
quantum wells was introduced by Matsuura and
coworkers.~\cite{Zheng_1998,Zheng_1998b} Based on unitary transformations related to bulk like phonons and e-ph coupling,~\cite{Matsuura_1988}  the methodology was developed~\cite{Zheng_1997,Zheng_1998,Zheng_1998b}  to include also the quantization of bulk-like optical phonons, and additional interface
phonons.~\cite{Jin_Sheng_1987} In some situations where the weak coupling
regime is justified (Bohr radius larger than the sum of polaron radii),
an approximate bulk-like PB model is appropriate to account for the
exciton-phonon coupling at a reduced computational
cost.~\cite{Senger_2003} 

In general, while for bulk material three characteristic lengths need to be compared, for a quantum well a fourth come into play: the exciton Bohr radius, both polaron radii and in addition
the quantum well thickness. Moreover, compared to conventional semiconductor quantum wells 
studied in the past, 2D halide perovskites show distinctive features: the equivalent quantum 
confinement thickness is extremely small
and the bulk reference materials (3D halide perovskite) belong to the
intermediate regime for the strength of exciton-phonon coupling. Future
theoretical studies based on empirical models for excitonic polarons in
2D halide perovskites would benefit from including a complete PB-like
description combined with quantized bulk-like optical phonons and
interface optical phonons.

We will now describe in detail the extension of the PB framework for the
case of quantum dots (QDs). 0D quantum confinement shall affect all the properties,
including electron, hole and exciton states, as well as the undispersed
bulk optical LO phonons. Surface effects related to the localization of
polarization at the interface between the QD and the outer medium,
potentially add dielectric confinement effects and a coupling with
surface optical (SO) vibrations. For bulk-like quantities, the
continuous variation of the $\overrightarrow{k}$ wavevector is absent
due to quantum confinement and must be replaced by discrete quantum
numbers. A crude approximation consists in neglecting implicitly all the
effects of quantum confinement except for electronic and excitonic
states.~\cite{Senger_2003b} This leads to effective potentials that
include the exciton-polaron coupling but as derived for bulk materials.
Moreover, the effect of quantum confinement on self-energy terms and
reduced mass cannot be included straightforwardly.

At the semi-empirical level the effect of quantum confinement can be
included both for vibrations and e-ph couplings.~\cite{Pan_1988}  In
the case of a spherical QD embedded in a non-polar medium, bulk-like
phonons are replaced by localized phonons with quantum numbers
$\left( n,l,m \right)$ with $n = 1,2,3,\ldots\infty$;
$l = 0,\ 1,2,\ldots\infty\ (S,\ P,\ D,\ldots.)$;
$m = - l,\ldots l\ $. For surface phonons, the spherical symmetry
corresponds to doublets $\left( l,m \right)$ with
$l = 1,2,\ldots\infty\ $; $m = - l,\ldots\ l\ $. For quantum
confined bulk-like polar phonon, the e-ph Hamiltonian is modified and
the Fr\"ohlich expression of the matrix element (Eq.~\eqref{eq12}) transforms to
\begin{eqnarray}\label{eq91}
g_{{\rm LO},\ n,\ l} = \frac{1}{k_{n,l}}\left( \frac{{e^{2}\hbar \omega}_{\text{LO}}}{{j_{l + 1}\left( k_{n,l}R \right)}^{2}R^{3}\varepsilon_{0}} \right)^{1/2}\text{~}\left( \frac{1}{\varepsilon_{\infty}} - \frac{1}{\varepsilon_{s}} \right)^{1/2}. \nonumber \\
\end{eqnarray}
The wavevectors $k_{n,l}$ are now discretized according to the zeros
of the spherical Bessel function ($j_{l}\left( k_{n,l}R \right) = 0$)
where R is the radius of the QD. Noteworthy, the first zeros yield the
most important contributions due to the $\frac{1}{k_{n,l}}$ term.
Effective interaction and self-energies are also affected. The PB
expressions for the LO self-energies obtained for bulk (Eq.~\eqref{eq50}) can also
be reshaped in a similar way for QDs:
\begin{widetext}
\begin{eqnarray}\label{eq92}
\sigma_{e\left( h \right),1S, {\rm LO}} = \text{-}\sum_{n,l,m}^{}{\frac{\left| g_{{\rm LO},n,l} \right|^{2}}{\hbar \omega_{\text{LO}}}\left( 2f_{e\left( h \right),{\rm LO},n,l,m}^{\min}A_{{\rm LO},n,l,m}\left( k_{n,l} \right) - \left( A_{{\rm LO},n,l,m}\left( k_{n,l} \right) + R_{e(h)}^{2}c_{{\rm LO},n,l,m}\left( k_{n,l} \right) \right){f_{e\left( h \right),{\rm LO},n,l,m}^{\min}}^{2} \right)}. \nonumber \\
\end{eqnarray}
\end{widetext}
These self-energies now depend on the characteristics of the exciton
(Bohr radius, \ldots), the polaron radii, but also the geometry (radius $R$)
of the QD through Eq.~\eqref{eq91}. They involve more complex functions such as
$A_{{\rm LO},n,l,m}\left( k_{n,l} \right)$ and
$c_{{\rm LO},n,l,m}\left( k_{n,l} \right)$ in place of 1 and $k^{2}$, and are made up of 
combinations of spherical Harmonics.~\cite{Oshiro_1999}

\subsection{Conclusion and outlook}
The paper reviews empirical approaches to Fr\"ohlich polarons and
excitonic polarons in polar semiconductors based on unitary
transformations. These approaches are relevant for the weak and
intermediate coupling regimes and can be extended from bulk 3D materials
to 2D and 0D nanostructures. Besides extensive review of past achievements, several 
new analytical expressions are derived and discussed  in the context of prototypical ionic semiconductors. More specifically, validity of various popular but approximate approaches for the
simulation of excitonic polarons is assessed in detail. It is found that
these approaches, namely Hak, PB\textsubscript{app}, PBK\textsubscript{app} and Baj, 
are mainly relevant for the weak coupling regime. Among
approximate models, the ABS model and our simplified implementation
(ABS\textsubscript{app})  provide reasonable results also for the intermediate coupling regime and are attractive to further investigate excited states or
exciton complexes. For both weak and intermediate coupling regimes, PB, PBK (including the reduced mass variation) and Iad (full analytic derivation but computationally much more demanding and hardly transferable to lower dimensions) models appear to be the most accurate. Extensions of empirical approaches, earlier limited to single effective polar phonon mode, to cases where multiple polar modes are at play  have been presented, and  demonstrated to be important in more complex materials such as 
the halide perovskite family of semiconductors. Besides bulk materials, the case of quantum wells and QDs is considered, charting the course for future developments.

Empirical approaches  have also been   put into perspective of more recently developed first-principles 
 methods. At first, the pros and cons of using first-principles parameters to calibrate empirical models 
are briefly discussed. Subsequently, a detailed comparison to the state-of-the-art first-principles  approaches
 for polarons and excitonic polarons has been carried out . For now,
 the implementations of new \textit{ab initio} approaches  have been specifically tested on materials belonging to the strong
coupling regime, which hampers their fair comparison with  the various empirical
frameworks considered in this review. Still, these  recent \textit{ab initio}
developments  open  avenues for an
improved calibration of the empirical approaches. So far, work along this
line essentially relied on experimental data that, in the case of ionic semiconductors, 
are often, insufficient or not accurate enough if not conflicting. On the experimental side, 
new experimental protocols based on THz spectroscopy or high-resolution diffraction have
progressed in recent years and shall contribute to this effort. There are still major steps 
 pending before reaching   a consistent understanding and quantitative  comparability between 
experimental and simulated excitonic polaron observables, for instance,  developing models 
to account for temperature effects and lattice disorder.

Last but not least, the methods discussed in this review were historically
introduced for cubic diatomic lattices. When the dimensionality changes and/or the 
composition diversifies, certain approximations must be reevaluated. For instance, ab 
initio calculations on the two-dimensional Fröhlich interaction and analytical models have 
been proposed for 2D systems, including transition metal dichalcogenides.~\cite{Sohier2016}
Akin to metal halide perovskites and TlCl, conventional nitride semiconductors and emerging 
semiconductors such as Ga$_2$O$_3$ are 
also ionic and expected to exhibit a Fröhlich interaction. However, they correspond to a 
more complex case than the compounds considered in this work. Whereas polarons and 
exciton-phonon couplings are mentioned in the literature on nitride 
semiconductors,~\cite{Lee1997,Reimann1998,Zhang2001,Kalliakos2002} 
and polarons in Ga$_2$O$_3$,~\cite{Hao2025} theoretical description of excitonic
polarons in many materials deserves 
further efforts for various reasons: i) the lattice is anisotropic, which requires
an extension of the models discussed in this review, including the Fröhlich
interaction and the PB expression of the excitonic polaron Hamiltonian; ii) the 
coupling of exciton to lattice vibrations is not only related to the Fröhlich 
interaction, which further complicates the picture.

\section{AUTHOR INFORMATION}  \label{SecAI}

\href{mailto:Jacky.even@insa-rennes.fr}{\nolinkurl{Jacky.even@insa-rennes.fr}},
\url{https://cv.hal.science/jacky-even}.

\href{mailto:claudine.katan@univ-rennes.fr}{\nolinkurl{claudine.katan@univ-rennes.fr}} \url{https://cv.hal.science/katan}. 

NOTES

The authors declare no competing financial interest.

ACKNOWLEDGMENT

J.E. acknowledges the financial support from the Institut Universitaire
de France. GENCI A.R.K. acknowledges support from R\'{e}gion Bretagne, France through a SAD PEROPERE grant, and support from Campus France through MOPGA fellowship.

\appendix 

\section{Brief perspective on the physics of halide perovskites}
\label{AppA}
3D halide perovskites are the most promising class of new semiconductors
for solar cell applications.~\cite{Kojima_2009,Im_2011,Lee_2012,Kim_2012}  They are almost
ideal direct band gap semiconductors by comparison to conventional
III-V, II-VI or silicon semiconductors, with low effective masses and
sizeable optical oscillator strengths. However, they exhibit distinct
physical properties owing to their unusual softness of the crystal
lattice and its strong ionicity. The latter characteristic directly
affects the magnitude of the carrier-lattice interactions and the
dielectric properties, including the formation of exciton complexes. 3D
halide perovskites form a novel class of semiconductors with a simple
and original structure of the electronic band edges at the R point of
its cubic BZ, that can be analyzed from an orbital
combination viewpoint by tight binding models coupled with double group
symmetry analyses.~\cite{Boyer_Richard_2016} As a consequence of the perovskite crystallographic structure, the formation of antibonding s-p states at the R point across the electronic band gap is leading to a $S_{z} = \frac{1}{2}$ state at the top of the valence band ($R_{6}^{+}$) and a $J_{z} = \frac{1}{2}$ split-off band ($R_{6}^{-}$) at the bottom of the conduction band due to the presence of a giant spin-orbit coupling~\cite{Even_2013,Even_2014,Even_2015,Even_2016,Even_2018}. It yields
specific selection rules for carrier-phonon scattering mechanisms
including a weak deformation potential mechanism related to acoustic
phonons and stronger polar mechanisms related to optical phonons and low
frequency relaxations, characterized by pseudo-spin
variables.~\cite{Even_2013,Even_2014,Even_2016,Nie_2016}  Density functional theory (DFT)
calculations including spin-orbit coupling predict that 3D halide
perovskites exhibit approximately similar hole ($h$) and electron
effective masses, i.e.
$m_{h} \approx m_{e}.$~\cite{Even_2013,Brivio_2014,Giorgi_2015,Umari_2014,Filip_2015} To the best of our knowledge, the $m_{e} = m_{h}$ approximation, which is quite unusual
among other semiconductors, is consistent with most experimental
observations in case of 3D halide perovskites.~\cite{Miyata_2015,Galkowski_2016,Yang_2017}

Excitonic properties have been at the center of discussions since the
beginning of the hype on perovskite photovoltaics in 2012. Early 2014,
by fitting the line shape of LT absorption, some
of us predicted that $E_{b,1S} \approx 15\ meV$.~\cite{Even_2014}
This motivated the direct measurement of the exciton binding energy in
2015,~\cite{Miyata_2015} which led to a value of $16\ meV$,
significantly smaller than the experimental values reported earlier.
Since then, photovoltaic mechanisms in perovskites are considered 
to be more in line with those of conventional semiconductors, 
than those of Dye Sensitized
Solar Cells, as initially envisioned~\cite{Even2014-vj}. Thus, due to the low exciton
binding energy in 3D halide perovskites, only a negligible proportion of
charge carriers form excitons at the temperatures and carrier densities
corresponding to typical device operating
conditions.~\cite{Fu_2023} Under such conditions, discussions about
the nature of the coupling between charge carriers and the lattice are
essentially related to either electrons or holes. The understanding of
the excitonic properties, especially the exciton fine structure and the
nature of exciton complexes, was further questioned with the advent of
perovskite quantum dots (QD) leading to prospects for LEDs and quantum
light emission.~\cite{Kovalenko_2017} The recent controversy about the
exciton fine structure reflects the importance of a proper description
of perovskite excitonic properties. The dispute between the initial
hypothesis of a fine structure related to a singlet dark ground
state~\cite{Even_2016,Fu_2017} and the alternative proposal of an inverted
dark-bright ordering due to an effective exchange energy stemming from a
Rashba effect,~\cite{Becker_2018,Sercel_2019} was closed only recently in favor
of the initial assertion.~\cite{Tamarat_2019,Tamarat_2020,Tamarat_2023}

Several open questions remain, especially about the nature of the
coupling between the charge carriers and the lattice dynamics. This
coupling is very sensitive to the polar character of the lattice and
exhibits specific selection rules with respect to conventional
semiconductors.~\cite{Even_2016,Yu_2010} Recently developed ab initio
frameworks have contributed to unravelling the peculiar lattice dynamics
in 3D halide perovskites at high temperature, notably identifying that
polymorphous networks correspond to a more realistic description of
these crystals than the highly symmetric reference cubic perovskite
lattice.~\cite{Zhao_2020,Zacharias_2023b} The developed ab initio framework
achieves simultaneous description of vibrational and optoelectronic
properties of highly disordered semiconductors in a strong anharmonic
regime.~\cite{Yaffe_2017,Ferreira_2020,Lanigan_Atkins_2021} Several aspects have to be considered
for the description of these properties, namely the stochastic nature, short range
and anisotropic correlations of atomic motions, among others. This
leads to smearing of the electronic and vibrational densities of states
as well as a smooth change of the electronic band gap at high
temperature. Thus, within the RT regime, conventional
Hamiltonian expressions for e-ph interactions are questionable. In the
low-temperature range, the expression of the polar coupling between
carriers and optical phonons in terms of conventional e-ph
Hamiltonian~\cite{Giustino_2017} is reasonably justified, despite the softness of the lattice. However, the simple e-ph Hamiltonian historically introduced by Fr\"ohlich~\cite{Frohlich1937} is also
inadequate because of the multiplicity of optical vibrational modes that
come into play. We stress that most studies assume that the polar
coupling to optical modes is effectively dominated by a single
one.~\cite{Iaru_2021} This is rigorously incorrect for the cubic
phase of halide perovskites.~\cite{Even_2015} However, as described
early by Toyozawa,~\cite{Toyozawa_1968} Fr\"ohlich's semi-empirical
approach to e-ph coupling can be extended to polar lattices with
multiple optical modes assuming harmonic oscillators with a single
frequency and without dissipation. The polaronic coupling in perovskite
bulk and nanostructures modelled using this approach deserves assessment
and clarification. Existing papers and reviews on polaronic effects in
3D halide perovskites are mainly focused on free polarons and seldom
address specifically excitonic polarons.~\cite{Miyata_2017,Neukirch_2016,Ghosh_2020,Yamada_2022,Zhang_2023} The
effect of e-h correlations is also often not considered in the analysis
of experimental results on ``polaronic''
distortions.~\cite{Guzelturk_2021} There have been a few attempts to
directly perform excitonic polaron calculations with interaction
potentials designed for the weak coupling regime,~\cite{Men_ndez_Proupin_2015}
however the intermediate coupling regime is more relevant to the physics
of 3D halide perovskites.

\section{Unitary transformation introducing a lattice distortion}~\label{AppB}

$\widehat{W} = e^{\sum_{\overrightarrow{k}}^{}\left( {F_{\overrightarrow{k}}^{*}\widehat{a}}_{\overrightarrow{k}} - F_{\overrightarrow{k}}{\widehat{a}}_{\overrightarrow{k}}^{+} \right)}$
includes a shift operator $F_{\overrightarrow{k}}$ which can be
understood as the amplitude of the lattice distortion around the charge
carrier (the sign in the exponential is taken according to the PB convention, which is opposite to the LLP one). Considering the general relation
$e^{- \widehat{B}}\widehat{A}e^{\widehat{B}} = \widehat{A} + \left\lbrack \widehat{A},\widehat{B} \right\rbrack + \frac{1}{2!}\left\lbrack \left\lbrack \widehat{A},\widehat{B} \right\rbrack,\widehat{B} \right\rbrack + \ldots$,
phonon operators are shifted:
${\widehat{W}}^{+}{\widehat{a}}_{\overrightarrow{k}}\widehat{W} = {\widehat{a}}_{\overrightarrow{k}} - F_{\overrightarrow{k}}$
and
${\widehat{W}}^{+}{\widehat{a}}_{\overrightarrow{k}}^{+}\widehat{W} = {\widehat{a}}_{\overrightarrow{k}}^{+} - F_{\overrightarrow{k}}^{*}$

\section{LLP first unitary transformation for the center of mass motion of a free polaron}\label{AppC}

$\hslash\overrightarrow{Q}$ is the eigenvalue of the total momentum of
the polaron
$\widehat{\wp} = {\widehat{p}}_{e} + \sum_{\overrightarrow{k}}^{}{\hslash\overrightarrow{k}{\widehat{a}}_{\overrightarrow{k}}^{+}{\widehat{a}}_{\overrightarrow{k}}}$
. $\widehat{\wp}$ commutes with the e-ph Hamiltonian
($\left\lbrack \widehat{\wp},{\widehat{H}}_{e - ph} \right\rbrack = 0$),
which shows that $\widehat{\wp}$ and ${\widehat{H}}_{e - ph}$ share
the same eigenstates:
${\widehat{H}}_{e - ph}\left| \left. \ \varphi \right\rangle \right.\  = E\left| \left. \ \varphi \right\rangle \right.\ $
and
$\widehat{\wp}\left| \left. \ \varphi \right\rangle \right.\  = \hslash\overrightarrow{Q}\left| \left. \ \varphi \right\rangle \right.\ $.
Next, a unitary transformation is introduced to remove explicitly the
electron position and momentum:
$\widehat{U} = e^{i\left( \overrightarrow{Q} - \sum_{\overrightarrow{k}}^{}{\overrightarrow{k}{\widehat{a}}_{\overrightarrow{k}}^{+}{\widehat{a}}_{\overrightarrow{k}}} \right).{\overrightarrow{r}}_{e}}$,
leading to
${\widehat{U}}^{+}{\widehat{a}}_{\overrightarrow{k}}\widehat{U} = {\widehat{a}}_{\overrightarrow{k}}e^{i\overrightarrow{k}.{\overrightarrow{r}}_{e}}$,
${\widehat{U}}^{+}{\widehat{a}}_{\overrightarrow{k}}^{+}\widehat{U} = {\widehat{a}}_{\overrightarrow{k}}^{+}e^{- i\overrightarrow{k}.{\overrightarrow{r}}_{e}}$
${\widehat{U}}^{+}\widehat{\wp}\widehat{U} = \hslash\overrightarrow{Q} + {\widehat{p}}_{e}$,
${\widehat{U}}^{+}{\widehat{p}}_{e}\widehat{U} = \hslash\overrightarrow{Q} + {\widehat{p}}_{e} - \sum_{\overrightarrow{k}}^{}{\hslash\overrightarrow{k}{\widehat{a}}_{\overrightarrow{k}}^{+}{\widehat{a}}_{\overrightarrow{k}}}$
and to an expression where the electron position is already removed:
\begin{eqnarray} \label{eqc1}
    \ {\widehat{U}}^{+}{\widehat{H}}_{e - ph}\widehat{U} &=& \frac{\left( \hslash\overrightarrow{Q} + {\widehat{p}}_{e} - \sum_{\overrightarrow{k}}^{}{\hslash\overrightarrow{k}{\widehat{a}}_{\overrightarrow{k}}^{+}{\widehat{a}}_{\overrightarrow{k}}} \right)^{2}}{2m_{e}} \\ \nonumber &+& \sum_{\overrightarrow{k}}^{}\left( g_{\overrightarrow{k}}{\widehat{a}}_{\overrightarrow{k}} + g_{\overrightarrow{k}}^{*}{\widehat{a}}_{\overrightarrow{k}}^{+} \right) + \hbar \omega_{\text{LO}}\sum_{\overrightarrow{k}}^{}{{\widehat{a}}_{\overrightarrow{k}}^{+}{\widehat{a}}_{\overrightarrow{k}}}.
    \end{eqnarray}
Next, applying the unitary transformation to
$\widehat{\wp}\left| \left. \ \varphi \right\rangle \right.\  = \hslash\overrightarrow{Q}\left| \left. \ \varphi \right\rangle \right.\ $
and to the eigenstates
$\left| \left. \ \varphi_{\text{CM}} \right\rangle \right.\  = {\widehat{U}}^{+}\left| \left. \ \varphi \right\rangle \right.\ $,
one gets
${\widehat{U}}^{+}\widehat{\wp}\left| \left. \ \varphi \right\rangle \right.\  = {\widehat{U}}^{+}\widehat{\wp}\widehat{U}{\widehat{U}}^{+}\left| \left. \ \varphi \right\rangle \right.\  = \left( \hslash\overrightarrow{Q} + {\widehat{p}}_{e} \right)\left| \left. \ \varphi_{\text{CM}} \right\rangle \right.\ $,
but also
${\widehat{U}}^{+}\hslash\overrightarrow{Q}\left| \left. \ \varphi \right\rangle \right.\  = \hslash\overrightarrow{Q}\left| \left. \ \varphi_{\text{CM}} \right\rangle \right.\ $,
which shows that
${\widehat{p}}_{e}\left| \left. \ \varphi_{\text{CM}} \right\rangle = 0 \right.\ $.
Then, the electron momentum can also be removed from the transformed
Hamiltonian:
\begin{eqnarray} \label{eqc2}
{\widehat{H}}_{\text{CM}}^{\text{LLP}} &=& {\widehat{U}}^{+}{\widehat{H}}_{e - ph}\widehat{U} = \frac{\left( \hslash\overrightarrow{Q} - \sum_{\overrightarrow{k}}^{}{\hslash\overrightarrow{k}{\widehat{a}}_{\overrightarrow{k}}^{+}{\widehat{a}}_{\overrightarrow{k}}} \right)^{2}}{2m_{e}} \\ \nonumber &+& \sum_{\overrightarrow{k}}^{}\left( g_{\overrightarrow{k}}{\widehat{a}}_{\overrightarrow{k}} + g_{\overrightarrow{k}}^{*}{\widehat{a}}_{\overrightarrow{k}}^{+} \right) + \hbar \omega_{\text{LO}}\sum_{\overrightarrow{k}}^{}{{\widehat{a}}_{\overrightarrow{k}}^{+}{\widehat{a}}_{\overrightarrow{k}}}.
\end{eqnarray}
As a result, the description of the system moved from (electron+phonons) to (polaron+phonons) at the expense of additional second and fourth order terms related to the phonon population.

This expression can be alternatively expressed as:
\begin{eqnarray} \label{eqc3}
{\widehat{H}}_{\text{CM}}^{\text{LLP}} &=& \frac{\hslash^{2}Q^{2}}{2m_{e}} + \hslash\Omega_{\text{CM}}^{\text{LLP}}\sum_{\overrightarrow{k}}^{}{{\widehat{a}}_{\overrightarrow{k}}^{+}{\widehat{a}}_{\overrightarrow{k}}} \\ \nonumber &+& \sum_{\overrightarrow{k}}^{}\left\lbrack g_{\overrightarrow{k}}{\widehat{a}}_{\overrightarrow{k}} + c.c \right\rbrack + \sum_{\overrightarrow{k}}^{}{\sum_{\overrightarrow{k'}}^{}{\frac{\hslash^{2}\overrightarrow{k}.\overrightarrow{k'}}{2M}{\widehat{a}}_{\overrightarrow{k}}^{+}{\widehat{a}}_{\overrightarrow{k'}}^{+}{\widehat{a}}_{\overrightarrow{k}}}}{\widehat{a}}_{\overrightarrow{k'}},
\end{eqnarray}
with
$\hslash\Omega_{\text{CM}}^{\text{LLP}} = \hbar \omega_{\text{LO}} - \frac{\hslash^{2}}{M}\overrightarrow{k}.\overrightarrow{Q} + \frac{\hslash^{2}k^{2}}{2m_{e}}$.

\begin{widetext}
\section{LLP second unitary transformation introducing the lattice distortion}\label{AppD}
A second unitary transformation (Eq.~\eqref{eq19}) is used to transform the Hamiltonian obtained in Appendix \ref{AppC}:

\begin{eqnarray} \label{eqd1}
{\widehat{H}}^{\text{LLP}}\left( F_{\overrightarrow{k}}\left( \overrightarrow{Q} \right) \right) &=& {\widehat{W}}^{+}{\widehat{U}}^{+}{\widehat{H}}_{e - ph}\widehat{U}\widehat{W} = {\widehat{W}}^{+}{\widehat{H}}_{\text{CM}}^{\text{LLP}}\widehat{W} =  \frac{\left( \hslash\overrightarrow{Q} - \sum_{\overrightarrow{k}}^{}{\hslash\overrightarrow{k}\left( {\widehat{a}}_{\overrightarrow{k}}^{+} - F_{\overrightarrow{k}}^{*} \right)\left( {\widehat{a}}_{\overrightarrow{k}} - F_{\overrightarrow{k}} \right)} \right)^{2}}{2m_{e}} \\ \nonumber &+& \sum_{\overrightarrow{k}}^{}\left( g_{\overrightarrow{k}}\left( {\widehat{a}}_{\overrightarrow{k}} - F_{\overrightarrow{k}} \right) + g_{\overrightarrow{k}}^{*}\left( {\widehat{a}}_{\overrightarrow{k}}^{+} - F_{\overrightarrow{k}}^{*} \right) \right) + \hbar \omega_{\text{LO}}\sum_{\overrightarrow{k}}^{}{\left( {\widehat{a}}_{\overrightarrow{k}}^{+} - F_{\overrightarrow{k}}^{*} \right)\left( {\widehat{a}}_{\overrightarrow{k}} - F_{\overrightarrow{k}} \right)}.
\end{eqnarray}
The new expression of the eigenstates is related to the initial one by
$\left| \left. \ \psi \right\rangle \right.\  = {\widehat{W}}^{+}\left| \left. \ \varphi_{\text{CM}} \right\rangle \right.\  = {{\widehat{W}}^{+}\widehat{U}}^{+}\left| \left. \ \varphi \right\rangle \right.\ $.
For a `free vacuum' e-ph ground state with zero (real) phonon such as
$\left| \left. \ \psi_{\text{GS}} \right\rangle \right.\  = \phi\left( \overrightarrow{Q} \right)\left| \left. \ 0 \right\rangle \right.\ $,
only
${\widehat{H}}_{0_{\text{ph}}}^{\text{LLP}}\left( F_{\overrightarrow{k}}\left( \overrightarrow{Q} \right) \right)$
needs to be considered:
\begin{eqnarray}\label{eqd2}
E_{\text{GS}}\left( \overrightarrow{Q} \right) = \left\langle \psi_{\text{GS}} \middle| {\widehat{H}}_{0_{\text{ph}}}^{\text{LLP}}\left( F_{\overrightarrow{k}}\left( \overrightarrow{Q} \right) \right)\text{~} \middle| \psi_{\text{GS}} \right\rangle &=& \frac{\hslash^{2}Q^{2} - 2\overrightarrow{Q}.\left( \sum_{\overrightarrow{k}}^{}{\hslash\overrightarrow{k}\left| F_{\overrightarrow{k}}\left( \overrightarrow{Q} \right) \right|^{2}} \right) + \left( \sum_{\overrightarrow{k}}^{}{\hslash\overrightarrow{k}\left| F_{\overrightarrow{k}}\left( \overrightarrow{Q} \right) \right|^{2}} \right)^{2}}{2m_{e}} \nonumber \\ &-& \sum_{\overrightarrow{k}}^{}\left( g_{\overrightarrow{k}}F_{\overrightarrow{k}}\left( \overrightarrow{Q} \right) + c.c \right) + \sum_{\overrightarrow{k}}^{}{\left| F_{\overrightarrow{k}}\left( \overrightarrow{Q} \right) \right|^{2}\left( \hbar \omega_{\text{LO}} + \frac{\hslash^{2}k^{2}}{2m_{e}} \right)}.
\end{eqnarray}
The variational relation between the lattice distortion and
$g_{\overrightarrow{k}}$ is obtained by energy minimization, i.e. solving
$\partial E_{\text{GS}}\left( \overrightarrow{Q} \right) /  F_{\overrightarrow{k}}\left( \overrightarrow{Q} \right) = \ \partial E_{\text{GS}}\left( \overrightarrow{Q} \right)/ \partial F_{\overrightarrow{k}}^{*}\left( \overrightarrow{Q} \right) = 0$:
\begin{eqnarray}\label{eqd3}
g_{\overrightarrow{k}} &=& F_{\overrightarrow{k}}^{*,min}\left( \overrightarrow{Q} \right)\left\lbrack \hbar \omega_{\text{LO}} - \hslash^{2}\frac{\overrightarrow{Q}.\overrightarrow{k}}{m_{e}} + \frac{\hslash^{2}k^{2}}{2m_{e}} + \frac{\hslash^{2}}{m_{e}}\overrightarrow{k}.\sum_{\overrightarrow{k'}}^{}{\overrightarrow{k'}\left| F_{\overrightarrow{k'}}^{\min}\left( \overrightarrow{Q} \right) \right|^{2}} \right\rbrack;
\\
g_{\overrightarrow{k}}^{*} &=& F_{\overrightarrow{k}}^{\min}\left( \overrightarrow{Q} \right)\left\lbrack \hbar \omega_{\text{LO}} - \hslash^{2}\frac{\overrightarrow{Q}.\overrightarrow{k}}{m_{e}} + \frac{\hslash^{2}k^{2}}{2m_{e}} + \frac{\hslash^{2}}{m_{e}}\overrightarrow{k}.\sum_{\overrightarrow{k'}}^{}{\overrightarrow{k'}\left| F_{\overrightarrow{k'}}^{\min}\left( \overrightarrow{Q} \right) \right|^{2}} \right\rbrack.
\end{eqnarray}
\end{widetext}

\subsection{Expressions valid at the bottom of the free polaron dispersion
  ($\overrightarrow{Q} = \overrightarrow{0}$)}

The non-linear term can be removed
$\sum_{\overrightarrow{k}}^{}{\overrightarrow{k}\left| F_{\overrightarrow{k}}^{\min}\left( \overrightarrow{0} \right) \right|^{2}} = \overrightarrow{0}$
because the system does not possess any preferential direction related
to the distortion or the total momentum. In that case,
$F_{\overrightarrow{k}}^{\min}\left( \overrightarrow{0} \right) = \frac{g_{\overrightarrow{k}}^{*}}{{\hslash\omega}_{\text{LO}} + \frac{\hslash^{2}k^{2}}{2m_{e}}}$
and the energy after integration is given by
$E_{\text{GS}}\left( \overrightarrow{0} \right) = \sigma_{e}^{\text{LLP}}\left( \overrightarrow{0} \right) = - \alpha_{e}\hbar \omega_{\text{LO}}$.
The population of virtual phonons for the ground state
$N_{e}^{\text{LLP}}\left( \overrightarrow{Q} \right)$ is directly
connected to the distortion
$F_{\overrightarrow{k}}^{\min}\left( \overrightarrow{Q} \right)$ of
the lattice around the charge carrier through
$N_{e}^{\text{LLP}}\left( \overrightarrow{Q} \right) = \left\langle \varphi_{\text{GS}} \middle| \sum_{\overrightarrow{k}}^{}{{\widehat{a}}_{\overrightarrow{k}}^{+}{\widehat{a}}_{\overrightarrow{k}}} \middle| \varphi_{\text{GS}} \right\rangle = \sum_{\overrightarrow{k}}^{}\left| F_{\overrightarrow{k}}^{\min}\left( \overrightarrow{Q} \right) \right|^{2}$,
leading to a simple expression for
$\overrightarrow{Q} = \overrightarrow{0}$:
$N_{e}^{\text{LLP}}\left( \overrightarrow{0} \right) = \frac{\alpha_{e}}{2}$.

\subsection{Energy dispersion of the free polaron
  ($\overrightarrow{Q} \neq \overrightarrow{0}$)}

The non-linear term is vectorial and is therefore expected to be
parallel to the preferential direction in the system namely
$\overrightarrow{Q}$:
$\sum_{\overrightarrow{k}}^{}{\overrightarrow{k}\left| F_{\overrightarrow{k}}^{\min}\left( \overrightarrow{Q} \right) \right|^{2}} = \eta\overrightarrow{Q}$.
One can integrate over the angle $\theta$ between
$\overrightarrow{k}$ and $\overrightarrow{Q}$, before integration
over the modulus of $\overrightarrow{k}$:
\begin{widetext}
\begin{eqnarray}\label{eqd5}
\eta = \frac{\alpha_{e}\left( 1 - \eta \right)}{\pi q^{3}}\int_{0}^{+ \infty}{\frac{\text{dK}}{2K}\left( \frac{2Kq\left( 1 + K^{2} \right)}{\left( 1 + K^{2} \right)^{2} - 4K^{2}q^{2}} - \frac{1}{2}\log\left( \frac{1 + K^{2} + 2Kq}{1 + K^{2} - 2Kq} \right) \right)} = \frac{\alpha_{e}\left( 1 - \eta \right)}{2q^{3}}\left( \frac{q}{\sqrt{1 - q^{2}}} - {\rm asin}(q) \right), \nonumber \\
\end{eqnarray}
where
$K = \frac{\hbar k}{\left( 2m_{e}\hbar \omega_{\text{LO}} \right)^{1/2}} = R_{e}k$
and
$q = \frac{\hbar Q\left( 1 - \eta \right)}{\left( 2m_{e}\hbar \omega_{\text{LO}} \right)^{1/2}} = R_{e}Q\left( 1 - \eta \right)$.

The expressions of the total energy and virtual phonon population are integrated over wavevector orientation and modulus.
\begin{eqnarray}\label{eqd6}
E_{\text{GS}}\left( \overrightarrow{Q} \right) &=& \frac{\hslash^{2}Q^{2}\left( 1 - \eta \right)^{2}}{2m_{e}} + \frac{2\alpha_{e}\hbar \omega_{\text{LO}}}{\pi q}\int_{0}^{+ \infty}{\frac{\text{dK}}{2K}\left( \frac{2Kq\left( 1 + K^{2} \right)}{\left( 1 + K^{2} \right)^{2} - 4K^{2}q^{2}} - log\left( \frac{1 + K^{2} + 2Kq}{1 + K^{2} - 2Kq} \right) \right)}, \\
\ N_{e}^{\text{LLP}}\left( \overrightarrow{Q} \right) &=& \frac{\alpha_{e}}{2\sqrt{1 - q^{2}}}.
\end{eqnarray}
After simplification the total energy can be recasted in the original
LLP form:
\begin{eqnarray}\label{eqd7a}
E_{\text{GS}}\left( \overrightarrow{Q} \right) = \frac{\hslash^{2}Q^{2}\left( 1 - \eta^{2} \right)}{2m_{e}} - \alpha_{e}\hbar \omega_{\text{LO}}\frac{\text{asin}(q)}{q}.
\end{eqnarray}
The two expressions for $\eta$ and
$E_{\text{GS}}\left( \overrightarrow{Q} \right)$ can be expressed for small q values thanks to
$\text{asin}\left( q \right) \approx q + \frac{q^{3}}{6}$ and
$\eta \approx \frac{\alpha_{e}\left( 1 - \eta \right)}{2q^{3}}\left( \frac{q^{3}}{3} \right) = \frac{\alpha_{e}\left( 1 - \eta \right)}{6}$
yielding the final expression of LLP for the energy dispersion including
a self-energy term and a correction over the charge carrier mass:
\begin{eqnarray}\label{eqd7b}
E_{\text{GS}}\left( \overrightarrow{Q} \right) &\approx& - \alpha_{e}\hbar \omega_{\text{LO}} + \frac{\hslash^{2}Q^{2}}{2m_{e}\left( 1 + \frac{\alpha_{e}}{6} \right)} + \ldots,
\\
N_{e}^{\text{LLP}}\left( \overrightarrow{Q} \right) &\approx& \frac{\alpha_{e}}{2}\left( 1 + \frac{{R_{e}}^{2}Q^{2}}{2\left( 1 + \frac{\alpha_{e}}{6} \right)^{2}} + \ldots \right).
\end{eqnarray}

\section{Energy dispersion of the excitonic polaron ($\overrightarrow{\mathbf{Q}}\mathbf{\neq}\overrightarrow{\mathbf{0}}$\textbf{)}} \label{AppF}
  
\subsection{PB Hamiltonian for a ground state (GS) with zero-phonon}
After two successive unitary transformations the Hamiltonian related to
the interaction between an exciton and an optical polar phonon reads:
\begin{eqnarray}\label{eqf1}
{\widehat{H}}_{\text{PB}}\left( F_{\overrightarrow{k}}\left( \overrightarrow{Q},\overrightarrow{r} \right) \right) = {\widehat{W}\left( \overrightarrow{r} \right)}^{+}{\widehat{U}}^{+}{\widehat{H}}_{X - ph}\widehat{U}\widehat{W}\left( \overrightarrow{r} \right) = {\widehat{H}}_{0_{\text{ph}}}^{\text{PB}} + {\widehat{H}}_{1_{\text{ph}}}^{\text{PB}} + {\widehat{H}}_{2_{\text{ph}}}^{\text{PB}} + {\widehat{H}}_{3_{\text{ph}}}^{\text{PB}} + {\widehat{H}}_{4_{\text{ph}}}^{\text{PB}},
\end{eqnarray}
where
${\widehat{H}}_{n_{\text{ph}}}^{\text{PB}}\left( F_{\overrightarrow{k}}\left( \overrightarrow{Q},\overrightarrow{r} \right) \right)$
are the n-phonon~contributions to the Hamiltonian. The explicit
expression of the $0_{\text{ph}}$ contribution is:
\begin{eqnarray}\label{eq_ajout1a}
{\widehat{H}}_{0_{\text{ph}}}^{\text{PB}}\left( F_{\overrightarrow{k}}\left( \overrightarrow{Q},\overrightarrow{r} \right) \right) &=& E_{g} + \frac{\hslash^{2}Q^{2} - 2\hslash^{2}\overrightarrow{Q}.\sum_{\overrightarrow{k}}^{}{\overrightarrow{k}\left| F_{\overrightarrow{k}}\left( \overrightarrow{Q},\overrightarrow{r} \right) \right|^{2}} + \hslash^{2}\left| \sum_{\overrightarrow{k}}^{}{\overrightarrow{k}\left| F_{\overrightarrow{k}}\left( \overrightarrow{Q},\overrightarrow{r} \right) \right|^{2}} \right|^{2}}{2M} \\ \nonumber +
\frac{\left( \widehat{p} + \mu\overrightarrow{j} \right)^{2}}{2\mu}&+&
\sum_{\overrightarrow{k}}^{}{\frac{\hslash^{2}\left| \nabla F_{\overrightarrow{k}}\left( \overrightarrow{Q},\overrightarrow{r} \right) \right|^{2}}{2\mu}}
- \sum_{\overrightarrow{k}}^{}\left( \rho_{\overrightarrow{k}}g_{\overrightarrow{k}}F_{\overrightarrow{k}}\left( \overrightarrow{Q},\overrightarrow{r} \right) + c.c \right) \\ \nonumber &+& \sum_{\overrightarrow{k}}^{}{\left| F_{\overrightarrow{k}}\left( \overrightarrow{Q},\overrightarrow{r} \right) \right|^{2}\left( \hbar \omega_{\text{LO}} + \frac{\hslash^{2}k^{2}}{2M} \right)},
\end{eqnarray}

with
$\overrightarrow{j} =\sum_{\overrightarrow{k}}^{}{\frac{F_{\overrightarrow{k}}^{*}\left( \overrightarrow{Q},\overrightarrow{r} \right)\left( - i\hslash\nabla F_{\overrightarrow{k}}\left( \overrightarrow{Q},\overrightarrow{r} \right) \right) + c.c}{2\mu}}$.

\subsection{Virtual phonon population: analytical calculation for $m_{e} = m_{h}$ with PB Hamiltonian}

The virtual phonon population for a 1S exciton can be computed thanks to
$N_{1S}\left( \overrightarrow{0} \right) = \left\langle \phi_{1S}\left( \overrightarrow{r} \right) \middle| \sum_{\overrightarrow{k}}^{}\left| F_{\overrightarrow{k}}^{\min}\left( \overrightarrow{r,}\overrightarrow{0} \right) \right|^{2} \middle| \phi_{1S}\left( \overrightarrow{r} \right) \right\rangle$.
When a PB Hamiltonian is used for $m_{e} = m_{h}$ , this expression can
be expressed as a single integral:
$N_{1S}^{\text{PB}}\left( \overrightarrow{0} \right) = N_{1S}^{\text{PBpref}}\int_{0}^{+ \infty}{d\widetilde{k}\frac{\left( 1 - {\widetilde{G}}_{1S} \right)^{3}}{\left( 1 + {{{\widetilde{R}}_{\text{pol}}}^{2}\widetilde{k}}^{2} - {\widetilde{G}}_{1S} \right)}}$
with
${\widetilde{G}}_{1S} = \frac{1}{\left( 1 + \left( \frac{\widetilde{k}}{2} \right)^{2} \right)^{2}}$
and
$N_{1S}^{\text{PBpref}} = \frac{2a_{B}^{\text{vac}}\text{Ry}^{\text{vac}}}{\pi\varepsilon^{*}a_{B}^{\text{eff}}\hbar \omega_{\text{LO}}}$.
After analytical integration over
$\widetilde{k} = ka_{B}^{\text{eff}}$, we find:
\begin{eqnarray}\label{eqf3}
\frac{N_{1S}^{\text{PB}}\left( \overrightarrow{0} \right)}{N_{1S}^{\text{PBpref}}} &=& 2\pi\left( \frac{1}{\sqrt{- \lambda_{+}}} - \frac{1}{\sqrt{- \lambda_{-}}} \right)\left( g_{1} - g_{2}\frac{\lambda_{+} + \lambda_{-} + 2*g3}{\lambda_{+} - \lambda_{-}} \right) \\ \nonumber &+& \pi g_{2}g_{3}\left( \frac{1}{\left( - \lambda_{+} \right)^{3/2}} + \frac{1}{\left( - \lambda_{-} \right)^{3/2}} \right) + \pi g_{2}\left( \frac{1}{\sqrt{- \lambda_{+}}} + \frac{1}{\sqrt{- \lambda_{-}}} \right) - \pi\,
\end{eqnarray}
with $\eta = 4{{\widetilde{R}}_{\text{pol}}}^{2}$\emph{,}
$\lambda_{\pm} = \frac{1}{2\eta}\left( - \left( 1 + 2\eta \right) \pm \sqrt{1 - 4\eta} \right)$,
$g_{1} = \frac{1 + \frac{1}{\eta^{2}}}{\lambda_{+} - \lambda_{-}}$,
$g_{2} = \frac{\eta^{2} + 2\eta - 1}{{\eta^{3}\left( \lambda_{+} - \lambda_{-} \right)}^{2}}$
, $g_{3} = \frac{{2\eta}^{2} + 3\eta - 2}{\eta^{2} + 2\eta - 1}$.

\subsection{Self-energies in the weak coupling regime}.
Within PB's model, the general expression for the single carrier
self-energies is:
\begin{eqnarray} \label{eqf4}
\sigma_{e(h),1S}\left( \overrightarrow{0} \right) = \text{-}\sum_{\overrightarrow{k}}^{}{\frac{\left| g_{\overrightarrow{k}} \right|^{2}}{\hbar \omega_{\text{LO}}}\left( 2f_{e\left( h \right),\overrightarrow{k}}^{\min}\left( \overrightarrow{0} \right) - \left( 1 + R_{e(h)}^{2}k^{2} \right){f_{e\left( h \right),\overrightarrow{k}}^{\min}\left( \overrightarrow{0} \right)}^{2} \right)}.
\end{eqnarray}
In the weak coupling regime,
$f_{e\left( h \right),\overrightarrow{k}}^{\min}\left( \overrightarrow{0} \right) \approx \frac{1}{\left( 1 + R_{e(h)}^{2}k^{2} \right)}$
. The expression of self-energies is simplified to:
\begin{eqnarray}\label{eqf5}
\sigma_{e(h),1S}\left( \overrightarrow{0} \right) = \text{-}\sum_{\overrightarrow{k}}^{}{\frac{\left| g_{\overrightarrow{k}} \right|^{2}}{\hbar \omega_{\text{LO}}}f_{e\left( h \right),\overrightarrow{k}}^{\min}\left( \overrightarrow{0} \right)}.
\end{eqnarray}
Using
$\alpha_{e(h)} = \frac{e^{2}}{8\pi\varepsilon_{0}\hbar \omega_{\text{LO}}R_{e(h)}}\frac{1}{\varepsilon^{*}}$
and
$\int_{0}^{+ \infty}\frac{\text{dK}}{\left( 1 + K^{2} \right)} = \frac{\pi}{2}$,
one finally obtains
$\sigma_{e(h),1S}\left( \overrightarrow{0} \right) = - \alpha_{e(h)}\hbar \omega_{\text{LO}}$.

\subsection{Center of mass motion and internal excitonic polaron energy: 
semi-analytical calculation for $m_{e} = m_{h}$ with PB Hamiltonian.}
  
Analyzing the effects of the excitonic polaron center of mass motion on
the excitonic polaron is more complex than in the case of free
polarons.~\cite{Kane_1978,Behnke_1978} Using PB approach, we propose here a
semi-analytic calculation for $m_{e} = m_{h}$. This assumption, which
is well suited for halide perovskites, leads to several simplifications
in the above expression, including first
$\overrightarrow{j} = \overrightarrow{0}$. Next, the Hamiltonian is
applied to the exciton ground state with zero-phonon:
\begin{eqnarray}\label{eqf6}
E_{\text{GS}}\left( \overrightarrow{Q} \right) = \left\langle \phi_{\ 1S}\left( \overrightarrow{r} \right) \middle| {{\widehat{H}}_{\text{PB}}}^{0}\left( F_{\overrightarrow{k}}\left( \overrightarrow{Q},\overrightarrow{r} \right) \right)\text{~} \middle| \phi_{1S}\left( \overrightarrow{r} \right) \right\rangle.
\end{eqnarray}
According to PB,
$F_{\overrightarrow{k}}\left( \overrightarrow{Q},\overrightarrow{r} \right)$
is parameterized through:
\begin{eqnarray}\label{eqf7}
F_{\overrightarrow{k}}\left( \overrightarrow{Q},\overrightarrow{r} \right) \approx \frac{g_{\overrightarrow{k}}^{*}}{\hbar \omega_{\text{LO}}}\left( {f_{e,\overrightarrow{k}}\left( \overrightarrow{Q} \right)e}^{- is_{h}\overrightarrow{k}.\overrightarrow{r}} - f_{h,\overrightarrow{k}}\left( \overrightarrow{Q} \right)e^{is_{e}\overrightarrow{k}.\overrightarrow{r}} \right)
\end{eqnarray}
and the energy is minimized against
$f_{e,\overrightarrow{k}}\left( \overrightarrow{Q} \right)$ and
$f_{h,\overrightarrow{k}}\left( \overrightarrow{Q} \right)$. For
$m_{e} = m_{h}$, $\alpha_{e} = \alpha_{e} = \alpha$ and
$f_{e,\overrightarrow{k}}^{\min}\left( \overrightarrow{Q} \right) = f_{h,\overrightarrow{k}}^{\min}\left( \overrightarrow{Q} \right) = f_{e\left( h \right),\overrightarrow{k}}^{\min}\left( \overrightarrow{Q} \right)$,
we achieve a simplification of PB parameters:
\begin{eqnarray}\label{eqf8}
f_{e\left( h \right),\overrightarrow{k}}^{\min}\left( \overrightarrow{Q} \right) = \frac{1 - G_{1S}\left( \overrightarrow{k},a_{B}^{\text{eff}} \right)}{1 + K^{2} - \overrightarrow{q}.\overrightarrow{K} - G_{1S}\left( \overrightarrow{k},a_{B}^{\text{eff}} \right)}
\end{eqnarray}
with $\overrightarrow{K} = R_{\text{pol}}\overrightarrow{k}$ ,
$\overrightarrow{q} = R_{\text{pol}}\overrightarrow{Q}\left( 1 - \eta - G_{1S}\left( \overrightarrow{k},a_{B}^{\text{eff}} \right) \right)$, $G_{1S}\left( \overrightarrow{k},a_{B}^{\text{eff}} \right) = \frac{1}{\left( 1 + \left( \frac{ka_{B}^{\text{eff}}}{2} \right)^{2} \right)^{2}}$ and
$\eta\overrightarrow{Q} = \left\langle \phi_{\ 1S}\left( \overrightarrow{r} \right) \middle| \sum_{\overrightarrow{k}}^{}{\overrightarrow{k}\left| F_{\overrightarrow{k}}^{\min}\left( \overrightarrow{Q},\overrightarrow{r} \right) \right|^{2}} \middle| \phi_{1S}\left( \overrightarrow{r} \right) \right\rangle$.

The expression of the energy of the ground state becomes:
\begin{eqnarray}\label{eqf9}
E_{GS}\left( a_{B}^{\text{eff}},\overrightarrow{Q} \right) &=& E_{g} + \frac{\hslash^{2}Q^{2}\left( 1 - \eta \right)^{2}}{2M} + \frac{\hslash^{2}}{2\mu{a_{B}^{\text{eff}}}^{2}} - \frac{e^{2}}{4\pi\varepsilon_{0}\varepsilon_{\infty}a_{B}^{\text{eff}}} \\ \nonumber &+& \frac{\alpha_{e}\hbar \omega_{\text{LO}}}{\pi q}\int_{0}^{+ \infty}{\frac{4dK}{K}\left( 1 - G_{1S} \right)^{2}\left( \frac{\text{Kq}\left( 1 + K^{2} - G_{1S} \right)}{\left( 1 + K^{2} - G_{1S} \right)^{2} - K^{2}q^{2}} - log\left( \frac{1 + K^{2} + Kq - G_{1S}}{1 + K^{2} - Kq - G_{1S}} \right) \right)}
\end{eqnarray}
and 
\begin{eqnarray}\label{eqf10}
\eta = \frac{\alpha_{e}\left( 1 - \eta \right)}{\pi q^{3}}\int_{0}^{+ \infty}{\frac{4dK}{K}\left( 1 - G_{1S} \right)^{3}\left( \frac{\text{Kq}\left( 1 + K^{2} - G_{1S} \right)}{\left( 1 + K^{2} - G_{1S} \right)^{2} - K^{2}q^{2}} - \frac{1}{2}\log\left( \frac{1 + K^{2} + Kq - G_{1S}}{1 + K^{2} - Kq - G_{1S}} \right) \right)}.\end{eqnarray}
For small Q values, the two expressions above can be approximated by:
\begin{eqnarray}\label{eqf11}
E_{GS}\left( a_{B}^{\text{eff}},\overrightarrow{Q} \right) \approx E_{GS}\left( a_{B}^{\text{eff}},\overrightarrow{0} \right) + \frac{\hslash^{2}Q^{2}\left( 1 - \eta \right)^{2}}{2M} + \frac{{\alpha\hslash}^{2}Q^{2}}{2M}\left( I_{3}\left( a_{B}^{\text{eff}} \right) - \eta I_{2}\left( a_{B}^{\text{eff}} \right) \right)
\end{eqnarray}
with the internal excitonic polaron energy
\begin{eqnarray}\label{eqf12}
E_{GS}\left( a_{B}^{\text{eff}},\overrightarrow{0} \right) = E_{g} + \frac{\hslash^{2}}{2\mu{a_{B}^{\text{eff}}}^{2}} - \frac{e^{2}}{4\pi\varepsilon_{0}\varepsilon_{\infty}a_{B}^{\text{eff}}} - \frac{e^{2}}{2\pi^{2}\varepsilon_{0}\varepsilon^{*}R_{\text{pol}}}\int_{0}^{+ \infty}{\text{dK}\frac{\left( 1 - G_{1S} \right)^{2}}{\left( 1 + K^{2} - G_{1S} \right)}}
\end{eqnarray}
and
\begin{eqnarray}\label{eqf13}
\eta \approx \frac{\alpha I_{4}\left( a_{B}^{\text{eff}} \right)}{1 + \alpha I_{3}\left( a_{B}^{\text{eff}} \right)}
\end{eqnarray}
where several integrals have a similar form:
\begin{eqnarray}\label{eqf14}
I_{n}\left( a_{B}^{\text{eff}} \right) = \int_{0}^{+ \infty}{\frac{8K^{2}\text{dK}}{3\pi}\left( \frac{\left( 1 - G_{1S} \right)^{n}}{\left( 1 + K^{2} - G_{1S} \right)^{3}} \right)}
\end{eqnarray}
and
\begin{eqnarray}\label{eqf15}
G_{1S} = \frac{1}{\left( 1 + \left( \frac{K}{2{\widetilde{R}}_{\text{pol}}} \right)^{2} \right)^{2}}. 
\end{eqnarray}
Numerical integrations can be easily performed to get
$M_{\text{PB}}$ for any $\alpha$ value:
\begin{eqnarray}\label{eqf16}
\frac{M^{\text{PB}}}{M} = \frac{1}{\left( 1 - \eta \right)^{2} + \alpha \left( I_{3}\left( a_{B}^{\text{eff}} \right) - \eta I_{2}\left( a_{B}^{\text{eff}} \right) \right)}.
\end{eqnarray}
In the weak coupling limit $G_{1S} \rightarrow 0$ and
$\alpha \rightarrow 0$, using
$\int_{0}^{+ \infty}\frac{K^{2}\text{dK}}{\left( 1 + K^{2} \right)^{3}} = \frac{\pi}{16}$,
$I_{n}\left( a_{B}^{\text{eff}} \right) \rightarrow \frac{1}{6}$. One
can check that the effective mass for the center of mass motion is
consistent with LLP model for free polarons and renormalized to:
\begin{eqnarray}\label{eqf17}
\frac{M^{\text{PB}}}{M} \approx \left( 1 + \frac{\alpha}{6} \right) \approx \frac{M^{\text{LLP}}}{M}
\end{eqnarray}
with $M^{\text{LLP}} = m_{e}^{\text{LLP}} + m_{h}^{\text{LLP}}.$

\section{Kane's models\label{Kane}}
\subsection{Kane's refined variational solution of Pollman-Büttner's model
  for the distortion field (PBK) \label{PBK}}

 For a ground state with zero-phonon and (\(\overrightarrow{Q} = \overrightarrow{0}\)) and a real excitonic wavefunction, the PB Hamiltonian reads:~\cite{Kane_1978,Pollmann_1977}

\begin{eqnarray}
   {\widehat{H}}_{0_{ph}}^{\text{PB}}\left( F_{\overrightarrow{k}}\left( \overrightarrow{0},\overrightarrow{r} \right) \right) = E_{g} + \frac{{\widehat{p}}^{2}}{2\mu} + \sum_{\overrightarrow{k}}^{}\frac{\hslash^{2}\left| \nabla F_{\overrightarrow{k}}\left( \overrightarrow{0},\overrightarrow{r} \right) \right|^{2}}{2\mu} - \sum_{\overrightarrow{k}}^{}\left( \rho_{\overrightarrow{k}}g_{\overrightarrow{k}}F_{\overrightarrow{k}}\left( \overrightarrow{0},\overrightarrow{r} \right) + c.c \right) \nonumber \\ + \sum_{\overrightarrow{k}}^{}{\left| F_{\overrightarrow{k}}\left( \overrightarrow{0},\overrightarrow{r} \right) \right|^{2}\left( {\hslash\omega}_{_\text{LO}} + \frac{\hslash^{2}k^{2}}{2M} \right)}.
\end{eqnarray}

Kane's trial function with additional p-like components for a variational calculation is detailed in the following.

PB initial s-like trial function:

\begin{eqnarray}
F_{\overrightarrow{k}}\left( \overrightarrow{0},\overrightarrow{r} \right) \approx \frac{g_{\overrightarrow{k}}^{*}}{{\hslash\omega}_{_\text{LO}}}\left( {f_{e,\overrightarrow{k}}\left( \overrightarrow{0} \right)}e^{- is_{h}\overrightarrow{k}.\overrightarrow{r}}-f_{h,\overrightarrow{k}}\left( \overrightarrow{0} \right)e^{is_{e}\overrightarrow{k}.\overrightarrow{r}} \right)
\end{eqnarray}

contains only two real parameters
\(f_{e(h),\overrightarrow{k}}\left( \overrightarrow{0} \right)\), 
while the one proposed by Kane contains two more parameters:

\begin{eqnarray}
F_{\overrightarrow{k}}\left( \overrightarrow{0},\overrightarrow{r} \right) \approx \frac{g_{\overrightarrow{k}}^{*}}{{\hslash\omega}_{_\text{LO}}}\left( {f_{e,\overrightarrow{k}} ( \overrightarrow{0})\left( 1 - i\lambda_{e,\overrightarrow{k}}\left( \frac{\overrightarrow{k}.\overrightarrow{r}}{r} \right) \right)}e^{- is_{h}\overrightarrow{k}.\overrightarrow{r}} - f_{h,\overrightarrow{k}} ( \overrightarrow{0} )\left( 1 + i\lambda_{h,\overrightarrow{k}}\left( \frac{\overrightarrow{k}.\overrightarrow{r}}{r} \right) \right)e^{is_{e}\overrightarrow{k}.\overrightarrow{r}} \right)
\end{eqnarray}

The additional refinement parameters \(\lambda_{e(h),\overrightarrow{k}}\) are also real quantities.

Before the minimization step, the expression of the total energy provided by Kane for a ground state
with zero-phonon and (\(\overrightarrow{Q} = \overrightarrow{0}\)),
\(E_{GS}^{\text{PBK}}\left( \overrightarrow{0} \right)\), contains two terms,
the first one \(E_{GS}\left( \overrightarrow{0} \right)\) is the one
already introduced initially by PB and the second
\(E_{GS}^{K}\left( \overrightarrow{0} \right)\) allows a variation of
the reduced mass:

\begin{eqnarray}
E_{GS}^{\text{PBK}}\left( \overrightarrow{0} \right) = \ E_{GS}\left( \overrightarrow{0} \right) + E_{GS}^{K}\left( \overrightarrow{0} \right)
\end{eqnarray}

The first term is given by Eqs.~\eqref{eq48}-\eqref{eq50}) derived by PB:

\begin{eqnarray}
E_{GS}\left( \overrightarrow{0} \right) = \left\langle \phi_{\ 1S}\left( \overrightarrow{r} \right) \middle| E_{g} + \frac{p^{2}}{2\mu} - \frac{e^{2}}{4\pi\varepsilon_{0}\varepsilon_{\infty}r}\text{+}V_{latt}^{\text{PB}}\left( \overrightarrow{r},\ f_{e,\overrightarrow{k}}^{\min}\left( \overrightarrow{0} \right),f_{h,\overrightarrow{k}}^{\min}\ \left( \overrightarrow{0} \right) \right)\text{~} \middle| \phi_{\ 1S}\left( \overrightarrow{r} \right) \right\rangle + \nonumber \\ \sigma_{h,1S}\left( f_{e,\overrightarrow{k}}^{\min}\left( \overrightarrow{0} \right),f_{h,\overrightarrow{k}}^{\min}\ \left( \overrightarrow{0} \right) \right) + \sigma_{e,1S}\left( f_{e,\overrightarrow{k}}^{\min}\left( \overrightarrow{0} \right),f_{h,\overrightarrow{k}}^{\min}\ \left( \overrightarrow{0} \right) \right)
\end{eqnarray}

\begin{eqnarray}
V_{latt}^{\text{PB}}\left( \overrightarrow{r},\ f_{e,\overrightarrow{k}}^{\min}\left( \overrightarrow{0} \right),f_{h,\overrightarrow{k}}^{\min} \right) = \text{2}\sum_{\overrightarrow{k}}^{}{\frac{\left| g_{\overrightarrow{k}} \right|^{2}}{{\hslash\omega}_{_\text{LO}}}\left( f_{e,\overrightarrow{k}}^{\min}\left( \overrightarrow{0} \right) + f_{h,\overrightarrow{k}}^{\min}\left( \overrightarrow{0} \right) - f_{e,\overrightarrow{k}}^{\min}\left( \overrightarrow{0} \right)f_{h,\overrightarrow{k}}^{\min}\left( \overrightarrow{0} \right) \right)\cos\left( \overrightarrow{k}.\overrightarrow{r} \right)}\text{~}
\end{eqnarray}

\begin{eqnarray}
\sigma_{e(h),1S}\left( \overrightarrow{0} \right) = - \sum_{\overrightarrow{k}}^{}{\frac{\left| g_{\overrightarrow{k}} \right|^{2}}{{\hslash\omega}_{_\text{LO}}}\left( 2f_{e(h),\overrightarrow{k}}^{\min}\left( \overrightarrow{0} \right) - \left( 1 + R_{e(h)}^{2}k^{2} \right){f_{e(h),\overrightarrow{k}}^{\min}\left( \overrightarrow{0} \right)}^{2} \right)}
\end{eqnarray}

These equations were rewritten by Kane in the following form:

\begin{eqnarray}
E_{GS}\left( \overrightarrow{0} \right) = \left\langle \phi_{\ 1S}\left( \overrightarrow{r} \right) \middle| E_{g} + \frac{p^{2}}{2\mu} - \frac{e^{2}}{4\pi\varepsilon_{0}\varepsilon_{\infty}r} + \sum_{\overrightarrow{k}}^{}{\frac{\left| g_{\overrightarrow{k}} \right|^{2}}{{\hslash\omega}_{_\text{LO}}}E_{k}^{A}}\text{~} \middle| \phi_{\ 1S}\left( \overrightarrow{r} \right) \right\rangle,
\end{eqnarray}

with $E_{k}^{A} = f_{e,\overrightarrow{k}}\left( \overrightarrow{0} \right)A_{e} + f_{h,\overrightarrow{k}}\left( \overrightarrow{0} \right)A_{h} + {f_{e,\overrightarrow{k}}\left( \overrightarrow{0} \right)}^{2}A_{ee} + {f_{h,\overrightarrow{k}}\left( \overrightarrow{0} \right)}^{2}A_{hh} + {2f}_{e,\overrightarrow{k}}\left( \overrightarrow{0} \right)f_{h,\overrightarrow{k}}\left( \overrightarrow{0} \right)A_{eh},$ 

and 
\begin{itemize}
    \item \(A_{e} = A_{h} = - 2\left( 1 - G_{00} \right)\), 
    \item \(A_{ee} = \left( 1 + R_{e}^{2}k^{2} \right)\), 
    \item \(A_{hh} = \left( 1 + R_{h}^{2}k^{2} \right)\),
    \item \(A_{eh} = - G_{00}\).
\end{itemize}

Notice that the formula for \(A_{e},A_{h}\) reported in the original
paper~\cite{Kane_1978} contains a sign error. Kane further introduces a general integral
form:
\(G_{nm} = \iiint_{}^{}{\frac{\left( i\cos(\theta) \right)^{n}}{r^{m}}e^{i\overrightarrow{k}.\overrightarrow{r}}\left| \phi_{\ 1S}\left( \overrightarrow{r} \right) \right|^{2}d^{3}\overrightarrow{r}}\).
The \(G_{00}\) integral corresponds to the
\(G_{1S}\left( \overrightarrow{k} \right)\) integral initially introduced by PB.~\cite{Pollmann_1977}

The second term is given by the equation~:

\begin{eqnarray}
E_{GS}^{K}\left( \overrightarrow{0} \right) =
\left\langle \phi_{\ 1S}\left( \overrightarrow{r} \right) \middle| \sum_{\overrightarrow{k}}^{}{\frac{\left| g_{\overrightarrow{k}} \right|^{2}}{{\hslash\omega}_{_\text{LO}}}E_{k}^{B}}\text{~} \middle| \phi_{\ 1S}\left( \overrightarrow{r} \right) \right\rangle
\end{eqnarray}

with

\begin{eqnarray}E_{k}^{B} = \lambda_{e,\overrightarrow{k}}\left( \overrightarrow{0} \right)B_{e} + \lambda_{h,\overrightarrow{k}}\left( \overrightarrow{0} \right)B_{h} + {\lambda_{e,\overrightarrow{k}}\left( \overrightarrow{0} \right)}^{2}B_{ee} + {\lambda_{h,\overrightarrow{k}}\left( \overrightarrow{0} \right)}^{2}B_{hh} + {2\lambda}_{e,\overrightarrow{k}}\left( \overrightarrow{0} \right)f_{h,\overrightarrow{k}}\left( \overrightarrow{0} \right)B_{eh}
\end{eqnarray}

and
\begin{eqnarray} 
  B_{e} = - 2f_{e,\overrightarrow{k}}\left( \overrightarrow{0} \right)kG_{10} - \frac{\left( f_{e,\overrightarrow{k}}\left( \overrightarrow{0} \right)\hslash k \right)^{2}}{m_{e}{\hslash\omega}_{_\text{LO}}}\frac{\left( 1 - \cos^{2}(\theta) \right)}{r} \nonumber  \\
  - f_{e,\overrightarrow{k}}\left( \overrightarrow{0} \right)f_{h,\overrightarrow{k}}\left( \overrightarrow{0} \right)\left( \frac{\hslash^{2}k^{2}}{m_{h}{\hslash\omega}_{_\text{LO}}}\left( G_{21} + G_{01} \right) - 2kG_{10} \right) \label{eqBe}
\end{eqnarray}

\begin{eqnarray}\label{eqBh}
  B_{h} = - 2f_{h,\overrightarrow{k}}\left( \overrightarrow{0} \right)kG_{10} - \frac{\left( f_{h,\overrightarrow{k}}\left( \overrightarrow{0} \right)\hslash k \right)^{2}}{m_{h}{\hslash\omega}_{_\text{LO}}}\frac{\left( 1 - \cos^{2}(\theta) \right)}{r}\nonumber  \\
  - f_{e,\overrightarrow{k}}\left( \overrightarrow{0} \right)f_{h,\overrightarrow{k}}\left( \overrightarrow{0} \right)\left( \frac{\hslash^{2}k^{2}}{m_{e}{\hslash\omega}_{_\text{LO}}}\left( G_{21} + G_{01} \right) - 2kG_{10} \right)
\end{eqnarray}
\begin{eqnarray}\label{eqBee}
  B_{ee} = \left( f_{e,\overrightarrow{k}}\left( \overrightarrow{0} \right)k \right)^{2}\left( \cos^{2}(\theta) + \frac{\hslash^{2}k^{2}\cos^{2}(\theta)}{{2m}_{e}{\hslash\omega}_{_\text{LO}}} + \frac{\hslash^{2}\left( 1 - \cos^{2}(\theta) \right)}{2\mu{\hslash\omega}_{_\text{LO}}r^{2}} \right)
\end{eqnarray}
\begin{eqnarray}\label{eqBhh}
  B_{hh} = \left( f_{h,\overrightarrow{k}}\left( \overrightarrow{0} \right)k \right)^{2}\left( \cos^{2}(\theta) + \frac{\hslash^{2}k^{2}\cos^{2}(\theta)}{{2m}_{h}{\hslash\omega}_{_\text{LO}}} + \frac{\hslash^{2}\left( 1 - \cos^{2}(\theta) \right)}{2\mu{\hslash\omega}_{_\text{LO}}r^{2}} \right)
\end{eqnarray}

\begin{eqnarray}\label{eqBeh}
  B_{eh} = \frac{f_{e,\overrightarrow{k}}\left( \overrightarrow{0} \right)f_{h,\overrightarrow{k}}\left( \overrightarrow{0} \right)k^{2}}{2}\left( - G_{20} + \frac{\hslash^{2}}{\mu{\hslash\omega}_{_\text{LO}}}\left( G_{02} + G_{22} + {kG}_{11} + {kG}_{31} \right) \right)
\end{eqnarray}
Notice that the formulas for \(B_{e},B_{h}\) reported in the original
paper~\cite{Kane_1978} contain a sign error for their first term
\(- 2f_{e,\overrightarrow{k}}\left( \overrightarrow{0} \right)kG_{10}\).
The last term of \(B_{eh}\) also contains a corrupted denominator in the
original paper, reporting \(G_{20} / {\mu\hslash\omega_{_\text{LO}}}\)
instead of \(G_{20}\).

The self-energy expressions are thus slightly modified by adding the
first two terms of \(B_{ee(hh)}\):

\begin{eqnarray}
\sigma_{e(h),1S}\left( \overrightarrow{0} \right) = - \sum_{\overrightarrow{k}}^{}{\frac{\left| g_{\overrightarrow{k}} \right|^{2}}{{\hslash\omega}_{_\text{LO}}}\left( 2f_{e(h),\overrightarrow{k}}^{\min}\left( \overrightarrow{0} \right) - \left( 1 + R_{e(h)}^{2}k^{2} \right){f_{e(h),\overrightarrow{k}}^{\min}\left( \overrightarrow{0} \right)}^{2}\left( 1 + \frac{\left( {k\lambda}_{e(h),\overrightarrow{k}} \right)^{2}}{3} \right) \right)}
\end{eqnarray}\

Eqs.~\eqref{eqBe}, \eqref{eqBh}, \eqref{eqBeh} as well as the third terms of \eqref{eqBee} and \eqref{eqBhh} rest of the terms of \(E_{k}^{B}\ \)are new contributions to the
lattice mediated electron-hole interaction $V_{latt}^{\text{PB}}$.

\subsection{Kane's refined variational solution using a hydrogenic trial
  function for the 1S excitonic polaron (PBK)}

\begin{figure}[htb]
\includegraphics[]{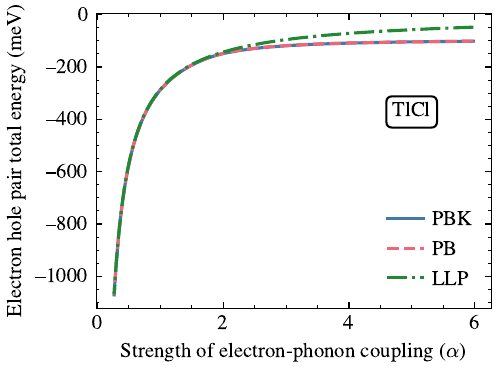}
\caption{\label{figApp4}Variation of the electron hole pair total 
energy as a function of the strength of the electron-phonon coupling $\alpha$ 
computed for TlCl considering parameters reported in Tab.~\ref{tab:Table1}. 
The LLP dashed-dotted green curve corresponds to a pair of free polarons.  
The PB (PBK) exciton binding energies can be estimated by calculating the 
difference between the PB (PBK) and LLP electron hole pair total energies. The 
variation of $\alpha$ is obtained by tuning the phonon energy.}
\end{figure}

For practical variational calculations, Kane also considered a 1S wavefunction that reads:

\begin{eqnarray}
\phi_{\ 1S}\left( \overrightarrow{r} \right) = \frac{e^{- \frac{r}{a_{B}^{\text{eff}}}}}{{a_{B}^{\text{eff}}}^{3/2}\pi^{1/2}}
\end{eqnarray}

Introducing dimensionless quantities \(x = \frac{ka_{B}^{\text{eff}}}{2}\),
\(\eta_{e(h)} = 4\left( \frac{R_{e(h)}}{a_{B}^{\text{eff}}} \right)^{2} = 4{\widetilde{R}}_{e(h)}^{2}\),
\(\delta = 1 - \frac{atan(x)}{x.}\) (Note that the expression of
\(\delta\) given in the initial paper should not be misinterpreted as in Ref.~\cite{Lizaraga2025}, where it was reported to read
\(\delta = 1 - \frac{1}{xtan(x)}\)). A 1S wavefunction allows to rewrite Eqs.~\eqref{eqBh}-\eqref{eqBeh}

\begin{eqnarray}\label{eqBe2}
B_{e} = \frac{4}{a_{B}^{\text{eff}}}\left( f_{e,\overrightarrow{k}}{\left( \overrightarrow{0} \right)x^{2}}G_{00} - \frac{{\eta_{e}\left( f_{e,\overrightarrow{k}}\left( \overrightarrow{0} \right)x \right)}^{2}}{3} - f_{e,\overrightarrow{k}}\left( \overrightarrow{0} \right)f_{h,\overrightarrow{k}}\left( \overrightarrow{0} \right)\left( \eta_{h}\delta + x^{2}G_{00} \right) \right)
\end{eqnarray}
\begin{eqnarray}\label{eqBh2}
  B_{h} = \frac{4}{a_{B}^{\text{eff}}}\left( f_{h,\overrightarrow{k}}{\left( \overrightarrow{0} \right)x^{2}}G_{00} - \frac{{\eta_{h}\left( f_{h,\overrightarrow{k}}\left( \overrightarrow{0} \right)x \right)}^{2}}{3} - f_{e,\overrightarrow{k}}\left( \overrightarrow{0} \right)f_{h,\overrightarrow{k}}\left( \overrightarrow{0} \right)\left( \eta_{e}\delta + x^{2}G_{00} \right) \right)
\end{eqnarray}
\begin{eqnarray}\label{eqBee2}
  B_{ee} = \frac{4}{3}\left( \frac{f_{e,\overrightarrow{k}}\left( \overrightarrow{0} \right)x}{a_{B}^{\text{eff}}} \right)^{2}\left( 1 + \eta_{e}\left( 1 + x^{2} \right) + \eta_{h} \right)
\end{eqnarray}
\begin{eqnarray}\label{eqBhh2}
  B_{hh} = \frac{4}{3}\left( \frac{f_{h,\overrightarrow{k}}\left( \overrightarrow{0} \right)x}{a_{B}^{\text{eff}}} \right)^{2}\left( 1 + \eta_{h}\left( 1 + x^{2} \right) + \eta_{e} \right)
\end{eqnarray}
\begin{eqnarray}\label{eqBeh2}
  B_{eh} = 4{\left( \frac{f_{e,\overrightarrow{k}}\left( \overrightarrow{0} \right)f_{h,\overrightarrow{k}}\left( \overrightarrow{0} \right)}{{a_{B}^{\text{eff}}}^{2}} \right)}\left( \left( \eta_{e} + \eta_{h} \right)\delta + \delta - x^{4}G_{00} \right)
\end{eqnarray}

Compared to Eqs.~\eqref{eqBe}-\eqref{eqBeh} as reported in the original paper, the 
simplified expressions Eqs.~\eqref{eqBe2}-\eqref{eqBeh2} were almost correct in 
Ref.~\cite{Kane_1978} Still, the first term of \(B_{e}\ (B_{h})\) 
\(\frac{4}{a_{B}^{\text{eff}}}f_{e(h),\overrightarrow{k}}{\left( \overrightarrow{0} 
\right)x}^{2}G_{00}\) has the uncorrect sign in the original paper by Kane.~\cite{Kane_1978}
However, in his analysis of the reduced
mass for the the weak coupling regime, Kane showed that the first terms of
\(B_{e}\ (B_{h})\) vanish. Thus, the sign error in \(B_{e}\ (B_{h})\) has 
no impact on this analytical result. 
  
\section{Iadonisi's approach to PB's model for zero phonon and
  $\overrightarrow{\mathbf{Q}}\mathbf{=}\overrightarrow{\mathbf{0}}$: (Iad)} \label{AppE}

To perform a partial summation of the exact solution proposed as a series by Iadonisi and coworkers and expressed in the main text using Whittaker functions, it is interesting to start with Lommel integrals
for the Whittaker functions:~\cite{Martin_1952}
\begin{eqnarray}\label{eqe1}
\begin{pmatrix}
M_{\frac{1}{\widetilde{\zeta}},l + \frac{1}{2}}\left( 2\widetilde{\zeta}\widetilde{r} \right)\int_{\widetilde{r}}^{+ \infty}{W_{\frac{1}{\widetilde{\zeta}},l + \frac{1}{2}}\left( 2\widetilde{\zeta}\widetilde{u} \right)M_{0,l + \frac{1}{2}}\left( {2is}_{e}\widetilde{k}\widetilde{u} \right)d\widetilde{u}} \\
 + W_{\frac{1}{\widetilde{\zeta}},l + \frac{1}{2}}\left( 2\widetilde{\zeta}\widetilde{r} \right)\int_{0}^{\widetilde{r}}{M_{\frac{1}{\widetilde{\zeta}},l + \frac{1}{2}}\left( 2\widetilde{\zeta}\widetilde{u} \right)M_{0,l + \frac{1}{2}}\left( {2is}_{e}\widetilde{k}\widetilde{u} \right)d\widetilde{u}} \\
\end{pmatrix} = \left( \frac{2\widetilde{\zeta}\Gamma\left( 2\left( l + 1 \right) \right)}{\left( {\widetilde{\zeta}}^{2} + {s_{e}}^{2}{\widetilde{k}}^{2} \right)\Gamma\left( l + 1 - \frac{1}{\widetilde{\zeta}} \right)} \right)M_{0,l + \frac{1}{2}}\left( {2is}_{e}\widetilde{k}\widetilde{r} \right) \nonumber 
\\  + 
\frac{1}{\left( {\widetilde{\zeta}}^{2} + {s_{e}}^{2}{\widetilde{k}}^{2} \right)} \begin{pmatrix}
M_{\frac{1}{\widetilde{\zeta}},l + \frac{1}{2}}\left( 2\widetilde{\zeta}\widetilde{r} \right)\int_{\widetilde{r}}^{+ \infty}{W_{\frac{1}{\widetilde{\zeta}},l + \frac{1}{2}}\left( 2\widetilde{\zeta}\widetilde{u} \right)M_{0,l + \frac{1}{2}}\left( {2is}_{e}\widetilde{k}\widetilde{u} \right)\frac{2}{\widetilde{u}}d\widetilde{u}} \\
 + W_{\frac{1}{\widetilde{\zeta}},l + \frac{1}{2}}\left( 2\widetilde{\zeta}\widetilde{r} \right)\int_{0}^{\widetilde{r}}{M_{\frac{1}{\widetilde{\zeta}},l + \frac{1}{2}}\left( 2\widetilde{\zeta}\widetilde{u} \right)M_{0,l + \frac{1}{2}}\left( {2is}_{e}\widetilde{k}\widetilde{u} \right)\frac{2}{\widetilde{u}}d\widetilde{u}} \\
\end{pmatrix},
\end{eqnarray}
where the Jacobian of the $\left( M,W \right)$ Whittaker functions is
introduced in the first term after analytical calculations of the
partial Lommel integrals. Next, it is necessary to connect Whittaker and
Bessel functions:
\begin{eqnarray}\label{eqe2}
    M_{0,l + \frac{1}{2}}\left( {2is}_{e}\widetilde{k}\widetilde{r} \right) = i^{l + 1}2^{2l + \frac{3}{2}}\Gamma\left( l + \frac{3}{2} \right)\left( s_{e}\widetilde{k}\widetilde{r} \right)^{1/2}J_{l + \frac{1}{2}}\left( s_{e}\widetilde{k}\widetilde{r} \right)
\end{eqnarray}
and finally, perform partial infinite summations through:
\begin{eqnarray}\label{eqe3}
e^{is_{e}\widetilde{k}\widetilde{r}\cos\left( \theta \right)} = \sum_{l = 0}^{+ \infty}{i^{l}\sqrt{\frac{\pi}{2}}\left( 2l + 1 \right)J_{l + \frac{1}{2}}\left( s_{e}\widetilde{k}\widetilde{r} \right)\left( s_{e}\widetilde{k}\widetilde{r} \right)^{- 1/2}P_{l}\left( \cos\left( \theta \right) \right)}.
\end{eqnarray}
The leading first term of
$F_{\overrightarrow{k}}\left( \overrightarrow{r} \right)$ has a form
which is now close to the ones obtained by PB and Kane:
\begin{eqnarray}\label{eqe4}
\frac{g_{\overrightarrow{k}}^{*}}{\hbar \omega_{\text{LO}}}\left( {\frac{1}{\left( {\widetilde{R}}_{e}^{2} + {\widetilde{R}}_{h}^{2} + 1 + {\widetilde{R}}_{e}^{2}{\widetilde{k}}^{2} \right)}e}^{- is_{h}\overrightarrow{k}.\overrightarrow{r}} - \frac{1}{\left( {\widetilde{R}}_{e}^{2} + {\widetilde{R}}_{h}^{2} + 1 + {\widetilde{R}}_{h}^{2}{\widetilde{k}}^{2} \right)}e^{is_{e}\overrightarrow{k}.\overrightarrow{r}} \right)
\end{eqnarray}
and the second term leads to a better convergence as a function of $l$ than the initial series:
\begin{eqnarray}\label{eqe5}
\frac{g_{\overrightarrow{k}}^{*}}{\hbar \omega_{\text{LO}}}\sum_{l = 0}^{+ \infty}\begin{matrix}
 \\
\begin{pmatrix}
\left( - 1 \right)^{l + 1}\int_{0}^{+ \infty}{\left( e^{\widetilde{r} - \widetilde{u}} - 1 + \frac{2}{\widetilde{u}\left( {\widetilde{\zeta}}^{2} + {s_{e}}^{2}{\widetilde{k}}^{2} \right)} \right)\left( \frac{\widetilde{u}}{s_{e}\widetilde{k}{\widetilde{r}}^{2}} \right)^{\frac{1}{2}}M_{\frac{1}{\widetilde{\zeta}},l + \frac{1}{2}}^{1}\left( 2\widetilde{\zeta}\widetilde{r} \right)M_{\frac{1}{\widetilde{\zeta}},l + \frac{1}{2}}^{2}\left( 2\widetilde{\zeta}\widetilde{u} \right)J_{l + \frac{1}{2}}\left( s_{e}\widetilde{k}\widetilde{u} \right)d\widetilde{u}} \\
 + \int_{0}^{+ \infty}{\left( e^{\widetilde{r} - \widetilde{u}} - 1 + \frac{2}{\widetilde{u}\left( {\widetilde{\zeta}}^{2} + {s_{h}}^{2}{\widetilde{k}}^{2} \right)} \right)\left( \frac{\widetilde{u}}{s_{h}\widetilde{k}{\widetilde{r}}^{2}} \right)^{\frac{1}{2}}M_{\frac{1}{\widetilde{\zeta}},l + \frac{1}{2}}^{1}\left( 2\widetilde{\zeta}\widetilde{r} \right)M_{\frac{1}{\widetilde{\zeta}},l + \frac{1}{2}}^{2}\left( 2\widetilde{\zeta}\widetilde{u} \right)J_{l + \frac{1}{2}}\left( s_{h}\widetilde{k}\widetilde{u} \right)d\widetilde{u}} \\
\end{pmatrix} \\
 \times \left( \frac{\left( - i \right)^{l}\sqrt{\frac{\pi}{8}}\left( 2l + 1 \right)\Gamma\left( l + 1 - \frac{1}{\widetilde{\zeta}} \right)P_{l}\left( \cos\left( \theta \right) \right)}{\left( {\widetilde{R}}_{e}^{2} + {\widetilde{R}}_{h}^{2} \right)\widetilde{\zeta}\Gamma\left( 2\left( l + 1 \right) \right)} \right) \\
\end{matrix}.
\end{eqnarray}

\section{Adamowski, Bednarek and Suffczynski approximation to  PB's model (ABS) } \label{AppJ}

Within PB's model, the full expression of the effective interaction can be integrated thanks to the ABS approximation. For $m_{e} = m_{h}$, a variational form is assumed for
$f_{e(h),\overrightarrow{k}}\left( \overrightarrow{0} \right)$:
$f_{e(h),\overrightarrow{k}}\left( \overrightarrow{0} \right) \approx \frac{\lambda \rho}{1 + \left( \rho R_{\text{pol}} \right)^{2}k^{2}}$. These coefficients can be introduced into the general expression of the effective interaction leading to:
\begin{eqnarray}\label{eqj1}
V_{latt,m_{e} = m_{h}}^{\text{ABS}}\left( \overrightarrow{r},\overrightarrow{0}\  \right) = \frac{e^{2}}{2\pi^{2}\varepsilon_{0}\varepsilon^{*}}\int_{0}^{+ \infty}{\frac{\sin\left( \text{kr} \right)}{\text{kr}}\text{dk}\left( \frac{2\lambda\rho}{1 + \rho^{2}R^{2}k^{2}} - \frac{\lambda^{2}\rho^{2}}{\left( 1 + \rho^{2}R^{2}k^{2} \right)^{2}} \right)},\ 
\end{eqnarray}
which can be further integrated to yield:
\begin{eqnarray}\label{eqj2}
{\widetilde{V}}_{latt,m_{e} = m_{h}}^{\text{ABS}}\left( \overrightarrow{r},\overrightarrow{0}\  \right) = \frac{V_{latt,m_{e} = m_{h}}^{\text{ABS}}\left( \overrightarrow{r},\overrightarrow{0}\  \right)}{\frac{e^{2}}{4\pi\varepsilon_{0}\varepsilon^{*}R_{\text{pol}}}} = \frac{1}{\rho}\left( \frac{\left( 2\lambda\rho - \lambda^{2}\rho^{2} \right)\left( 1 - e^{- \widetilde{r}/\rho} \right)}{\widetilde{r}/\rho} + \frac{\lambda^{2}\rho^{2}e^{- \widetilde{r}/\rho}}{2} \right).
\end{eqnarray}
The general expression of self-energies can be integrated analytically
as well, leading to:
\begin{eqnarray}\label{eqj3}
\sigma_{e(h)}\left( \overrightarrow{0} \right) = - 2\alpha\hbar \omega_{\text{LO}}\left( \lambda - \frac{\lambda^{2}\left( 1 + \rho^{2} \right)}{4\rho} \right).
\end{eqnarray}
Finally, assuming a 1S trial excitonic polaron wavefunction such as
$\phi_{1S}\left( \overrightarrow{r} \right) = \frac{e^{- \frac{r}{a_{B}^{\text{eff}}}}}{{a_{B}^{\text{eff}}}^{3/2}\pi^{1/2}}$, we obtain:
\begin{eqnarray}\label{eqj4}
\frac{E_{GS}\left( a_{B}^{\text{eff}},\overrightarrow{0} \right)}{\text{Ry}_{\text{vac}}} &=& \frac{{a_{B}^{\text{vac}}}^{2}}{\mu{a_{B}^{\text{eff}}}^{2}} - \frac{2a_{B}^{\text{vac}}}{\varepsilon_{\infty}a_{B}^{\text{eff}}} - \frac{4a_{B}^{\text{vac}}}{\varepsilon^{*}R_{\text{pol}}}\left( \lambda - \frac{\lambda^{2}\left( 1 + \rho^{2} \right)}{4\rho} \right) \\ \nonumber &+& \frac{2a_{B}^{\text{vac}}}{\varepsilon^{*}a_{B}^{\text{eff}}}\left( \left( 2\lambda\rho - \lambda^{2}\rho^{2} \right)\left( 1 - \frac{1}{\left( 1 + \frac{a_{B}^{\text{eff}}}{2\rho R_{\text{pol}}} \right)^{2}} \right) + \frac{\lambda^{2}\rho^{2}a_{B}^{\text{eff}}}{2\rho R_{\text{pol}}\left( 1 + \frac{a_{B}^{\text{eff}}}{2\rho R_{\text{pol}}} \right)^{3}} \right)
\end{eqnarray}

\begin{figure}[htb]
\includegraphics[]{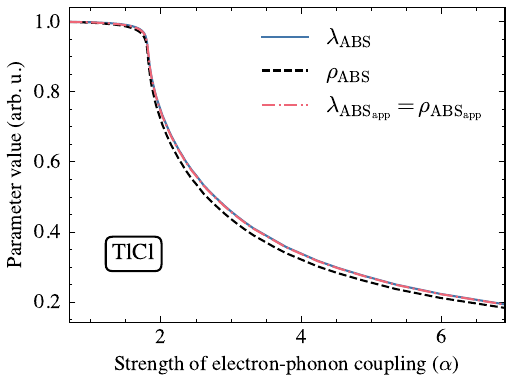}
\caption{Variation of ABS variational parameters $\lambda$ (blue line) and $\rho$ (dashed 
black line) against $\alpha$ for model parameters relevant to TlCl taken from 
Tab.~\ref{tab:Table1}.
The result of a simplified version of the ABS method (ABS$_{\text{app}}$) with a single 
adjustable parameter ($\lambda = \rho)$ is indicated by the red line. The variation of $\alpha$ is obtained by tuning the phonon energy.\label{figApp2}}
\end{figure}

\begin{figure}[htb]
\includegraphics[]{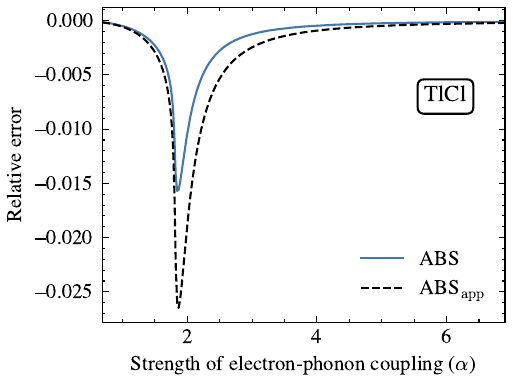}
\caption{ABS relative error (blue line) on the excitonic polaron binding energy by comparison to PB's model as a function of $\alpha$ for model parameters relevant to 
TlCl taken from Tab.~\ref{tab:Table1}.
The relative error obtained by a simplified version of the ABS method 
(ABS$_{\text{app}}$) with a single
adjustable parameter is indicated by the dashed black line. The variation of $\alpha$ is 
obtained by
tuning the phonon energy.\label{figApp3}}
\end{figure}

\section{Toyozawa's analysis of the Fr\"ohlich interaction in cubic,   polar and harmonic lattices with more than one longitudinal optical mode} \label{AppG}

In the harmonic approximation we have:
\begin{eqnarray}\label{eqg1}
\frac{1}{\varepsilon_{\infty}} - \frac{1}{\varepsilon\left( \omega \right)} = \sum_{\nu}^{}\frac{2\omega_{{\rm LO},\nu}\varepsilon_{0}{k^{2}\left| g_{\overrightarrow{k},\nu} \right|}^{2}}{\hslash e^{2}\left( \omega_{{\rm LO},\nu}^{2} - \omega^{2} \right)}.
\end{eqnarray}

Considering the longitudinal modes as poles of the dielectric constant
$\varepsilon\left( \omega_{{\rm LO},\nu} \right) = 0$ and expanding
$\frac{1}{\varepsilon\left( \omega \right)}$ close to
$\omega = \omega_{{\rm LO},\nu}\ $in this expression
($\omega = \omega_{{\rm LO},\nu} + \delta$, $\delta \rightarrow 0$):
\begin{eqnarray}\label{eqg2}
{k^{2}\left| g_{\overrightarrow{k},\nu} \right|}^{2} = \frac{\hslash e^{2}}{\varepsilon_{0}\left. \ \frac{\partial\varepsilon}{\partial\omega} \right|_{\omega_{{\rm LO},\nu}}}.
\end{eqnarray}

Considering the inverse of the generalized LST equation:
$\frac{\varepsilon_{\infty}}{\varepsilon\left( \omega \right)} = \prod_{\nu}^{}\left( \frac{\omega_{{\rm TO},\nu}^{2} - \omega^{2}}{\omega_{{\rm LO},\nu}^{2} - \omega^{2}} \right)$,
and expanding it close to $\omega = \omega_{{\rm LO},\nu}$, we obtain:
\begin{eqnarray}\label{eqg3}
\left. \ \frac{\partial\varepsilon}{\partial\omega} \right|_{\omega_{{\rm LO},\nu}} = \varepsilon_{\infty}\left( \frac{\omega_{{\rm LO},\nu}}{\omega_{{\rm LO},\nu}^{2} - \omega_{{\rm TO},\nu}^{2}} \right)\prod_{\mu \neq \nu}^{}\left( \frac{\omega_{{\rm LO},\nu}^{2} - \omega_{{\rm LO},\mu}^{2}}{\omega_{{\rm LO},\nu}^{2} - \omega_{{\rm TO},\mu}^{2}} \right).
\end{eqnarray}

\section{Analysis of Gervais multimode expression for cubic, polar and   harmonic lattices with more than one longitudinal optical mode} \label{AppH}

Considering the generalized LST equation:
$\frac{\varepsilon\left( \omega \right)}{\varepsilon_{\infty}} = \prod_{\nu}^{}\left( \frac{\omega_{{\rm LO},\nu}^{2} - \omega^{2}}{\omega_{{\rm TO},\nu}^{2} - \omega^{2}} \right)$,
and expanding it close to $\omega = \omega_{{\rm TO},\nu}\ $
($\omega = \omega_{{\rm TO},\nu} + \delta$, $\delta \rightarrow 0$), we have:
\begin{eqnarray}\label{eqh1}
\varepsilon\left( \omega_{{\rm TO},\nu} + \delta \right) \approx \varepsilon_{\infty}\frac{\omega_{{\rm LO},\nu}^{2} - \omega_{{\rm TO},\nu}^{2}}{- 2\omega_{{\rm TO},\nu}\delta}\prod_{\mu \neq \nu}^{}\left( \frac{\omega_{{\rm TO},\nu}^{2} - \omega_{{\rm LO},\mu}^{2}}{\omega_{{\rm TO},\nu}^{2} - \omega_{{\rm TO},\mu}^{2}} \right).
\end{eqnarray}
Starting from
$\varepsilon\left( \omega \right) = \varepsilon_{\infty} + \sum_{\nu}^{}\frac{\omega_{{\rm TO},\nu}^{2}\mathrm{\Delta}\varepsilon_{\nu}}{\omega_{{\rm TO},\nu}^{2} - \omega^{2}}$,~\cite{Gervais_1974}, we 
expand it close to $\omega = \omega_{{\rm TO},\nu}$ to obtain:
\begin{eqnarray}\label{eqh2}
\varepsilon\left( \omega_{{\rm TO},\nu} + \delta \right) \approx \frac{\mathrm{\Delta}\varepsilon_{\nu}\omega_{{\rm TO},\nu}^{2}}{- 2\omega_{{\rm TO},\nu}\delta},
\end{eqnarray}
which shows that
$\mathrm{\Delta}\varepsilon_{\nu} = \varepsilon_{\infty}\left( \frac{\omega_{{\rm LO},\nu}^{2}}{\omega_{{\rm TO},\nu}^{2}} - 1 \right)\prod_{\mu \neq \nu}^{}\left( \frac{\omega_{{\rm TO},\nu}^{2} - \omega_{{\rm LO},\mu}^{2}}{\omega_{{\rm TO},\nu}^{2} - \omega_{{\rm TO},\mu}^{2}} \right)$.
 
\section{A simplified model for a Fr\"ohlich bipolaron} \label{AppI}

A simplified model for a bipolaron (two equivalent charges, i.e. two holes or two electrons
numbered 1 and 2 and abbreviated as
$\text{bip}$) at the same level of theory than PB's approach for
excitonic polaron leads to the effective Hamiltonian:
\begin{eqnarray}\label{eqi1a}
{{\widehat{H}}_{\text{bip}} = E}_{g} + \frac{{\widehat{P}}^{2}}{2M} + \frac{{\widehat{p}}^{2}}{2\mu} + \frac{e^{2}}{4\pi\varepsilon_{0}\varepsilon_{\infty}r} + {\hslash\omega}_{\text{LO}}\sum_{\overrightarrow{k}}^{}{{\widehat{a}}_{\overrightarrow{k}}^{+}{\widehat{a}}_{\overrightarrow{k}}} + \sum_{\overrightarrow{k}}^{}\left\lbrack \rho_{\overrightarrow{k}}{e^{i\overrightarrow{k}.\overrightarrow{R}}g}_{\overrightarrow{k}}{\widehat{a}}_{\overrightarrow{k}} + {\rm c.c} \right\rbrack,
\end{eqnarray}
where
$\rho_{\overrightarrow{k}} = e^{is_{1}\overrightarrow{k}.\overrightarrow{r}} + e^{- is_{2}\overrightarrow{k}.\overrightarrow{r}}$
with $s_{1} = s_{2} = \frac{1}{2}$~\cite{Bassani_1991}.

A simplified variational approach can be based on a PB-like Hamiltonian
(see ~\cite{Bassani_1991} for a complete approach) that yields:
\begin{eqnarray}\label{eqi1b}
F_{\overrightarrow{k}}\left( \overrightarrow{r} \right) &\approx& \frac{g_{\overrightarrow{k}}^{*}}{\hbar \omega_{\text{LO}}}\left( {f_{1,\overrightarrow{k}}^{\min}e}^{- is_{1}\overrightarrow{k}.\overrightarrow{r}} + f_{2,\overrightarrow{k}}e^{is_{2}\overrightarrow{k}.\overrightarrow{r}} \right),
\\
E_{GS, bip}\left( a_{B}^{\text{eff}},\overrightarrow{0} \right) &=& \left\langle \phi_{\text{bip},\ 1S}\left( \overrightarrow{r} \right) \middle| E_{g} + \frac{p^{2}}{2\mu} + \frac{e^{2}}{4\pi\varepsilon_{0}\varepsilon_{\infty}r}\text{+}V_{\text{latt}}\left( \overrightarrow{r,}f_{1,\overrightarrow{k}}^{\min}\left( \overrightarrow{0} \right),f_{2,\overrightarrow{k}}^{\min}\left( \overrightarrow{0} \right) \right)\text{~} \middle| \phi_{{\rm bip},\ 1S}\left( \overrightarrow{r} \right) \right\rangle \\ \nonumber
&+& \sigma_{1,1S}\left( f_{1,\overrightarrow{k}}^{\min}\left( \overrightarrow{0} \right),f_{2,\overrightarrow{k}}^{\min}\left( \overrightarrow{0} \right) \right) + \sigma_{2,1S}\left( f_{1,\overrightarrow{k}}^{\min}\left( \overrightarrow{0} \right),f_{2,\overrightarrow{k}}^{\min}\left( \overrightarrow{0} \right) \right),
\end{eqnarray}
including a lattice mediated attractive interaction to the
electron-electron (or hole-hole) effective repulsion given by:
\begin{eqnarray}\label{eqi4}
V_{\text{latt}}^{\text{PB}}\left( \overrightarrow{r},f_{1,\overrightarrow{k}}^{\min}\left( \overrightarrow{0} \right),f_{2,\overrightarrow{k}}^{\min}\left( \overrightarrow{0} \right) \right) = \text{2}\sum_{\overrightarrow{k}}^{}{\frac{\left| g_{\overrightarrow{k}} \right|^{2}}{\hbar \omega_{\text{LO}}}\left( - f_{1,\overrightarrow{k}}^{\min}\left( \overrightarrow{0} \right) - f_{2,\overrightarrow{k}}^{\min}\left( \overrightarrow{0} \right) + f_{1,\overrightarrow{k}}^{\min}\left( \overrightarrow{0} \right)f_{2,\overrightarrow{k}}^{\min}\left( \overrightarrow{0} \right) \right)\cos\left( \overrightarrow{k}.\overrightarrow{r} \right)}
\end{eqnarray}
and a sum of self-energy corrections
$\sigma_{1,1S}\left( \overrightarrow{0} \right) + \sigma_{2,1S}\left( \overrightarrow{0} \right)$
in the presence of the bipolaron, with:
\begin{eqnarray}\label{eqi5}
\sigma_{1(2),1S}\left( \overrightarrow{0} \right) = \text{-}\sum_{\overrightarrow{k}}^{}{\frac{\left| g_{\overrightarrow{k}} \right|^{2}}{\hbar \omega_{\text{LO}}}\left( 2f_{1(2),\overrightarrow{k}}^{\min}\left( \overrightarrow{0} \right) - \left( 1 + R_{1(2)}^{2}k^{2} \right){f_{1(2),\overrightarrow{k}}^{\min}\left( \overrightarrow{0} \right)}^{2} \right)}.
\end{eqnarray}
For a 1S trial bipolaron wavefunction such as
$\phi_{1S}\left( \overrightarrow{r} \right) = \frac{e^{- \frac{r}{a_{B}^{\text{eff}}}}}{{a_{B}^{\text{eff}}}^{3/2}\pi^{1/2}}$
(see~\cite{Bassani_1991} for a complete approach), the
expressions of
$f_{1(2),\overrightarrow{k},1S}\left( \overrightarrow{0} \right)$ can
be simplified to:
\begin{eqnarray}\label{eqi6}
f_{e(h),\overrightarrow{k},1S}\left( \overrightarrow{0} \right) = \frac{1 + G_{1S}\left( \overrightarrow{k},a_{B}^{\text{eff}} \right)}{1 + {R_{\text{pol}}}^{2}k^{2} + G_{1S}\left( \overrightarrow{k},a_{B}^{\text{eff}} \right)}
\end{eqnarray}
and the total energy :
\begin{eqnarray}\label{eqi7}
E_{GS, bip}\left( a_{B}^{\text{eff}},\overrightarrow{0} \right) = E_{g} + \frac{\hslash^{2}}{2\mu{a_{B}^{\text{eff}}}^{2}} + \frac{e^{2}}{4\pi\varepsilon_{0}\varepsilon_{\infty}a_{B}^{\text{eff}}} - \frac{e^{2}}{2\pi^{2}\varepsilon_{0}\varepsilon^{*}R_{\text{pol}}}\int_{0}^{+ \infty}{\text{dK}\frac{\left( 1 + G_{1S} \right)^{2}}{\left( 1 + K^{2} + G_{1S} \right)}}
\label{bipolaron}
\end{eqnarray}
with $\overrightarrow{K} = R_{\text{pol}}\overrightarrow{k}$.
Effective interaction and self-energies have been merged in a single integral.

\begin{figure}[htb]
\includegraphics[]{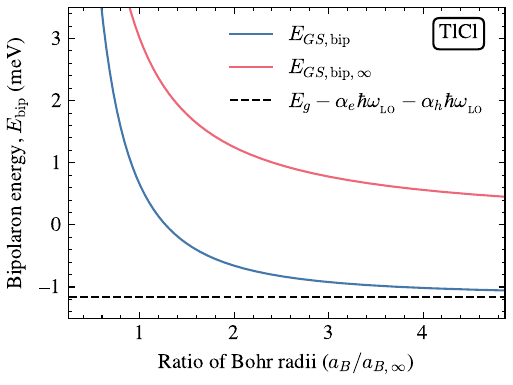}
\caption{Variational calculation for the 1S bipolaron
energy illustrated for TlCl. The internal bipolaron energy is computed as a function of the
Bohr radius within the present PB framework (blue, Eqs.~\eqref{bipolaron}) and for a simple interaction model 
considering $\varepsilon_{\text{eff}} = \varepsilon_{\infty}$ (red). The Bohr radius is 
scaled by its value at the minimum energy for a Wannier exciton with 
$\varepsilon_{\text{eff}} = \varepsilon_{\infty}.$ The energies are
scaled with the Rydberg energy of a Wannier exciton with 
$\varepsilon_{\text{eff}} = \varepsilon_{\infty}.$ The minimum energy for a pair of free 
carriers (band gap) is taken as a reference ($E_g=0$) to define the vertical axis. The 
minimum energy for a pair of free polarons is lowered by self-energy terms computed within 
LLP theory (black dashed line).}\label{figApp1}
\end{figure}

\section{Computation of multiband Huang-Rhys factors ($\mathbf{S}$) within the MPB model} \label{AppK}

In the dipolar electric approximation within PB's framework, the matrix
element for a direct optical transition from the crystal ground state
$\left| \left. \ \text{GS} \right\rangle \right.\ $ to an excitonic
polaron state
$\left| \left. \ \phi_{n}\left( \overrightarrow{r} \right),n_{\overrightarrow{k}} \right\rangle \right.\ $
characterized by a wavefunction
$\phi_{n}\left( \overrightarrow{r} \right)$ (n=1S, 2S, \ldots) for the
relative electron motion and the creation of $n_{\overrightarrow{k}}$
phonons at position $\overrightarrow{k}$ in the dispersion of the LO
optical phonon branch is given by:~\cite{Iadonisi_1983,Matsuura_1980}
\begin{eqnarray}\label{eqak1}    
\left\langle \text{GS} \middle| {\widehat{H}}_{\text{D.E}} \middle| \phi_{n}\left( \overrightarrow{r} \right),n_{\overrightarrow{k}} \right\rangle \approx \overrightarrow{e}.{\overrightarrow{M}}_{c,v}\psi_{n}\left( \overrightarrow{0} \right)\left\langle 0_{\text{ph}} \middle| \ \widehat{W}\left( \overrightarrow{0} \right) \middle| n_{\overrightarrow{k}} \right\rangle\delta\left( E\left( \phi\left( \overrightarrow{r} \right),n_{\overrightarrow{k}} \right) + \frac{\hslash^{2}k_{\text{opt}}^{2}}{2M} - \hslash\omega \right),
\end{eqnarray}
where $\hbar \omega$ and $\overrightarrow{e}$ are the energy and
polarization of the incoming photon, ${\overrightarrow{M}}_{c,v}$ is
the dipolar matrix element between the monoelectronic Bloch functions in
the conduction and valence bands, $0_{\text{ph}}$ stands for the
absence of thermal phonon population and
$\widehat{W}\left( \overrightarrow{r} \right)$ is the unitary
transformation of the PB model. The light dispersion can be neglected
$\left( k_{\text{opt}} \approx 0 \right)$ and thus we obtain~\cite{Iadonisi_1983}:
\begin{eqnarray}\label{eqak2}    
E\left( \phi_{n}\left( \overrightarrow{r} \right),n_{\overrightarrow{k}} \right){\approx E}_{n}\left( a_{B}^{\text{eff}},\overrightarrow{0} \right) + n_{\overrightarrow{k}}\hbar \omega_{\overrightarrow{k}}. 
\end{eqnarray}
For the MPB model, we introduced a generalized unitary transformation
$\widehat{S}\left( \overrightarrow{r} \right) = e^{\sum_{\overrightarrow{k},\nu}^{}\left( {F_{\overrightarrow{k},\nu}^{*}\left( \overrightarrow{r} \right)\widehat{a}}_{\overrightarrow{k},\nu} - F_{\overrightarrow{k},\nu}\left( \overrightarrow{r} \right){\widehat{a}}_{\overrightarrow{k},\nu}^{+} \right)}$
assuming independent lattice distortions produced by the various polar
optical modes $\nu$.

\subsection{Zero-phonon line}
To estimate the amplitude of the zero-phonon line of the 1S excitonic polaron within the MPB model we use:
\begin{eqnarray}    \label{eqak3}
I_{1S,0_{\text{ph}}} \propto \left| M_{1S,0_{\text{ph}}} \right|^{2}\delta\left( E_{GS}\left( a_{B}^{\text{eff}},\overrightarrow{0} \right) - \hslash\omega \right),
\end{eqnarray}
where
\begin{eqnarray}    \label{eqak4}
M_{1S,0_{\text{ph}}} &=& \left\langle \text{GS} \middle| {\widehat{H}}_{\text{D.E}} \middle| \phi_{1S}\left( \overrightarrow{r} \right),0_{\text{ph}} \right\rangle \approx \overrightarrow{e}.{\overrightarrow{M}}_{c,v}\phi_{1S}\left( \overrightarrow{0} \right)\left\langle 0_{\text{ph}} \middle| \ \widehat{W}\left( \overrightarrow{0} \right) \middle| 0_{\text{ph}} \right\rangle,
\\
\left\langle 0_{\text{ph}} \middle| \ \widehat{W}\left( \overrightarrow{0} \right) \middle| 0_{\text{ph}} \right\rangle &=& \left\langle 0_{\text{ph}} \middle| \ e^{\sum_{\overrightarrow{k},\nu}^{}\left( {F_{\overrightarrow{k},\nu}^{*}\left( \overrightarrow{0} \right)\widehat{a}}_{\overrightarrow{k},\nu} - F_{\overrightarrow{k},\nu}\left( \overrightarrow{0} \right){\widehat{a}}_{\overrightarrow{k},\nu}^{+} \right)} \middle| 0_{\text{ph}} \right\rangle.
\end{eqnarray}

Introducing
$\widehat{A} = \sum_{\overrightarrow{k},\nu}^{}{F_{\overrightarrow{k},\nu}^{*}\left( \overrightarrow{0} \right)\widehat{a}}_{\overrightarrow{k},\nu}$
and
$\widehat{B} = - \sum_{\overrightarrow{k},\nu}^{}{F_{\overrightarrow{k},\nu}\left( \overrightarrow{0} \right){\widehat{a}}_{\overrightarrow{k},\nu}^{+}}$,
we find $\left\lbrack \widehat{A},\widehat{B} \right\rbrack = - S$
with
$S = \sum_{\nu}^{}S_{\nu} = \sum_{\overrightarrow{k},\nu}^{}\left| F_{\overrightarrow{k},\nu}\left( \overrightarrow{0} \right) \right|^{2}$.
Using the relations
$e^{\widehat{A} + \widehat{B}} = e^{\widehat{A}}e^{\widehat{B}}e^{- \frac{1}{2}\left\lbrack \widehat{A},\widehat{B} \right\rbrack}$
and
$e^{\widehat{A}}e^{\widehat{B}} = e^{\widehat{B}}e^{\widehat{A}}e^{- \left\lbrack \widehat{B},\widehat{A} \right\rbrack}$,
which are valid since
$\left\lbrack \widehat{A},\widehat{B} \right\rbrack$ commutes with
$\widehat{A}$ and $\widehat{B}$, we obtain:
\begin{eqnarray}    \label{eqak6}
\left\langle 0_{\text{ph}} \middle| \ \widehat{S} \middle| 0_{\text{ph}} \right\rangle = e^{- \frac{S}{2}}
\,{\rm and}\, \left| M_{1S,0_{\text{ph}}} \right|^{2} \propto e^{- S}
\end{eqnarray}
after using the relation
\begin{eqnarray}    \label{eqak7}
e^{\widehat{A}}\left| \left. \ 0_{\text{ph}} \right\rangle \right.\ \  = \sum_{n = 0}^{+ \infty}\frac{1}{n!}\left( \sum_{\overrightarrow{k},\nu}^{}{F_{\overrightarrow{k},\nu}^{*}\left( \overrightarrow{0} \right)\widehat{a}}_{\overrightarrow{k},\nu} \right)^{n}\left| \left. \ 0_{\text{ph}} \right\rangle \right.\  = \left|  \ 0_{\text{ph}} \right\rangle .
\end{eqnarray}

\subsection{One-phonon replica}
The amplitude of an one phonon replica, phonon branch $\nu$ at
wavevector $\overrightarrow{k}$ of energy
$\hbar \omega_{\overrightarrow{k},\nu}$, for the 1S excitonic polaron is:
\begin{eqnarray}    \label{eqak8}
I_{1S,1_{\overrightarrow{k},\nu}} \propto \left| M_{1S,1_{\overrightarrow{k},\nu}} \right|^{2}\delta\left( E_{GS}\left( a_{B}^{\text{eff}},\overrightarrow{0} \right) + \hbar \omega_{\overrightarrow{k},\nu} - \hslash\omega \right),
\end{eqnarray}
where:
\begin{eqnarray}    \label{eqak9}
M_{1S,1_{\overrightarrow{k},\nu}} \approx \overrightarrow{e}.{\overrightarrow{M}}_{c,v}\phi_{1S}\left( \overrightarrow{0} \right)\left\langle 0_{\text{ph}} \middle| \ \widehat{W}\left( 0 \right) \middle| 1_{\overrightarrow{k},\nu} \right\rangle.
\end{eqnarray}
Using the same transformation we have:
\begin{eqnarray}    \label{eqak11}
e^{\widehat{A}}\left| \left. \ 1_{{\overrightarrow{k}}_{1},\nu} \right\rangle \right.\ \  &=& \sum_{n = 0}^{+ \infty}\frac{1}{n!}\left( \sum_{\overrightarrow{k},\nu}^{}{F_{\overrightarrow{k},\nu}^{*}\left( \overrightarrow{0} \right)\widehat{a}}_{\overrightarrow{k},\nu} \right)^{n}{\widehat{a}}_{{\overrightarrow{k}}_{1},\nu}^{+}\left| \left. \ 0_{\text{ph}} \right\rangle \right.\  = F_{{\overrightarrow{k}}_{1},\nu}^{*}\left( \overrightarrow{0} \right)\left|  \ 0_{\text{ph}} \right\rangle, \ 
\\
\left\langle 0_{\text{ph}} \middle| \ \widehat{S} \middle| 1_{\overrightarrow{k},\nu} \right\rangle &=& e^{- \frac{S}{2}}\left\langle 0_{\text{ph}} \middle| \ e^{\widehat{B}}e^{\widehat{A}} \middle| 1_{\overrightarrow{k},\nu} \right\rangle= e^{- \frac{S}{2}}F_{\overrightarrow{k},\nu}^{*}\left( \overrightarrow{0} \right),
\end{eqnarray}
and thus the ratio of $I_{1S,1_{\overrightarrow{k},\nu}}$ and
$I_{1S,0_{\text{ph}}}$ is given by:
\begin{eqnarray}    \label{eqak12}
\frac{\left| M_{1S,1_{\overrightarrow{k},\nu}} \right|^{2}}{\left| M_{1S,0_{\text{ph}}} \right|^{2}} = \left| F_{\overrightarrow{k},\nu}\left( \overrightarrow{0} \right) \right|^{2}.
\end{eqnarray}

For a phonon branch $\nu$ without dispersion
($\hbar \omega_{\overrightarrow{k},\nu} = \hbar \omega_{\nu}$), it is
possible to sum over $\overrightarrow{k}\ $the one phonon replica
$I_{1S,1_{\nu}}$ contributions observed at
$\hslash\omega \approx E_{GS}\left( a_{B}^{\text{eff}},\overrightarrow{0} \right) + \hbar \omega_{\nu}$. This yields:
\begin{eqnarray}    \label{eqak13}
\frac{I_{1S,1_{\nu}}}{I_{1S,0_{\text{ph}}}} = \frac{\sum_{\overrightarrow{k}}^{}\left| M_{1S,1_{\overrightarrow{k},\nu}} \right|^{2}}{\left| M_{1S,0_{\text{ph}}} \right|^{2}} = S_{\nu} = \sum_{\overrightarrow{k}}^{}\left| F_{\overrightarrow{k},\nu}\left( \overrightarrow{0} \right) \right|^{2}.
\end{eqnarray}

\subsection{Two-phonons replica}

To compute the two-phonon replica for the same phonon branch $\nu$, it
is necessary to consider two contributions ($\frac{1}{2}\ $in the
following expression to avoid double-counting):
\begin{eqnarray}    \label{eqak14}
I_{1S,2_{\nu}} &\propto& \frac{1}{2}\sum_{{\overrightarrow{k}}_{1}}^{}{\sum_{{\overrightarrow{k}}_{2} \neq {\overrightarrow{k}}_{1}}^{}\left| \left\langle 0_{\text{ph}} \middle| \ \widehat{S} \middle| 1_{{\overrightarrow{k}}_{1},\nu};1_{{\overrightarrow{k}}_{2},\nu} \right\rangle \right|^{2}}\delta\left( E_{GS}\left( a_{B}^{\text{eff}},\overrightarrow{0} \right) + \hbar \omega_{{\overrightarrow{k}}_{1}\nu} + \hbar \omega_{{\overrightarrow{k}}_{2},\nu} - \hslash\omega \right)  \nonumber \\ &+& \sum_{\overrightarrow{k_{1}}}^{}\left| \left\langle 0_{\text{ph}} \middle| \ \widehat{S} \middle| 2_{{\overrightarrow{k}}_{1},\nu} \right\rangle \right|^{2}\delta\left( E_{GS}\left( a_{B}^{\text{eff}},\overrightarrow{0} \right) + 2\hbar \omega_{{\overrightarrow{k}}_{1}\nu} - \hslash\omega \right),
\end{eqnarray}
with
$\left| \left. \ 1_{{\overrightarrow{k}}_{1},\nu};1_{{\overrightarrow{k}}_{2},\nu} \right\rangle \right.\  = {\widehat{a}}_{{\overrightarrow{k}}_{1},\nu}^{+}{\widehat{a}}_{{\overrightarrow{k}}_{2},\nu}^{+}\left| \left. \ 0_{\text{ph}} \right\rangle \right.\ $
and
$\left| \left. \ 2_{{\overrightarrow{k}}_{1},\nu} \right\rangle \right.\  = \frac{({{\widehat{a}}_{{\overrightarrow{k}}_{1},\nu}^{+}})^{2}}{\sqrt{2}}\left| \left. \ 0_{\text{ph}} \right\rangle \right.\ $. 
\\
Using 
$$
\left\langle 0_{\text{ph}} \middle| \frac{(\widehat{a}_{{\overrightarrow{k}}_{1},\nu})^2(\widehat{a}^{+}_{{\overrightarrow{k}}_{1},\nu})^2}{2} \middle| 0_{\text{ph}} \right\rangle = 
\left\langle 0_{\text{ph}} \middle| \frac{(\widehat{a}_{{\overrightarrow{k}}_{1},\nu})^2\widehat{a}^{+}_{{\overrightarrow{k}}_{1},\nu}}{2} \middle| 1_{{\overrightarrow{k}}_{1},\nu} \right\rangle = 
\left\langle 0_{\text{ph}} \middle| \frac{(\widehat{a}_{{\overrightarrow{k}}_{1},\nu})^2}{\sqrt{2}} \middle| 2_{{\overrightarrow{k}}_{1},\nu} \right\rangle = 
\left\langle 0_{\text{ph}} \middle| \widehat{a}_{{\overrightarrow{k}}_{1},\nu} \middle| 1_{{\overrightarrow{k}}_{1},\nu} \right\rangle = 1
$$ \\

we also obtain:
\begin{eqnarray}    \label{eqak15}
\left\langle 0_{\text{ph}} \middle| \ \widehat{S} \middle| 1_{{\overrightarrow{k}}_{1},\nu};1_{{\overrightarrow{k}}_{2},\nu} \right\rangle &=& e^{- \frac{S}{2}}\left\langle 0_{\text{ph}} \middle| \ e^{\widehat{B}}e^{\widehat{A}} \middle| 1_{{\overrightarrow{k}}_{1},\nu};1_{{\overrightarrow{k}}_{2},\nu} \right\rangle = e^{- \frac{S}{2}}F_{{\overrightarrow{k}}_{1},\nu}^{*}\left( \overrightarrow{0} \right)F_{{\overrightarrow{k}}_{2},\nu}^{*}\left( \overrightarrow{0} \right),
\\
\left\langle 0_{\text{ph}} \middle| \ \widehat{S} \middle| 2_{{\overrightarrow{k}}_{1},\nu} \right\rangle &=& e^{- \frac{S}{2}}\left\langle 0_{\text{ph}} \middle| \ e^{\widehat{B}}e^{\widehat{A}} \middle| 2_{{\overrightarrow{k}}_{1},\nu} \right\rangle = \frac{e^{- \frac{S}{2}}}{\sqrt{2}}{F_{{\overrightarrow{k}}_{1},\nu}^{*}\left( \overrightarrow{0} \right)}^{2}.
\end{eqnarray}
Considering an optical phonon branch without dispersion, the ratio of
the two-phonon replica observed at
$\hslash\omega \approx E_{GS}\left( a_{B}^{\text{eff}},\overrightarrow{0} \right) + 2\hbar \omega_{\nu}$
to the one of the zero-phonon line observed at
$\hslash\omega \approx E_{GS}\left( a_{B}^{\text{eff}},\overrightarrow{0} \right)$
is:
\begin{eqnarray}    \label{eqak17}
S_{2_{\nu}} = \frac{I_{1S,2_{\nu}}}{I_{1S,0_{\nu}}} = \frac{\left( S_{\nu} \right)^{2}}{2}.
\end{eqnarray}

\subsection{ Two-phonons overtone}
\begin{eqnarray} \label{eqak18}
    I_{1S,1_{\nu_{1}},1_{\nu_{2}}} \propto \sum_{\overrightarrow{k_{1}}}^{}{\sum_{\overrightarrow{k_{2}}}^{}\left| \left\langle 0_{\nu} \middle| \ \widehat{S} \middle| 1_{\overrightarrow{k_{1}},\nu_{1}};1_{\overrightarrow{k_{2}},\nu_{2}} \right\rangle \right|^{2}}\delta\left( E_{GS}\left( a_{B,eff},\overrightarrow{0} \right) + \hbar \omega_{\overrightarrow{k_{1}},\nu_{1}} + \hbar \omega_{\overrightarrow{k_{2}},\nu_{2}} - \hslash\omega \right),
\end{eqnarray}
leading for phonon branches without dispersion to:
\begin{eqnarray}  \label{eqak19}
    \frac{I_{1S,1_{\nu_{1}},1_{\nu_{2}}}}{I_{1S,0_{\text{ph}}}} = S_{\nu_{1}}S_{\nu_{2}}.
\end{eqnarray}

\subsection{Sum rule for the MPB model}

Summing all the contributions of the phonon replica and overtones for
the various phonon branches leads to a sum rule similar to the one
predicted for the PB model:~\cite{Iadonisi_1983,Matsuura_1980}

\begin{eqnarray}  \label{eqak20}
e^{- S}e^{S} = e^{- S}\sum_{n = 0}^{+ \infty}\frac{1}{n!}\left( \sum_{\nu}^{}S_{\nu} \right)^{n} = e^{- S}\left( 1 + \left( \sum_{\nu}^{}S_{\nu} \right) + \frac{1}{2}\left( \sum_{\nu}^{}S_{\nu} \right)^{2} + \ldots \right),
\end{eqnarray}
where the contributions of the zero-phonon line, the one-phonon replica, and the two-phonon lines (replica+overtones) appear progressively.

\section{Details of \textit{ab initio} calculations on TlCl } \label{AppCompDetails}

\subsection{Determination of electronic structure}

All first-principles calculations have been carried out using the {\sc PWscf} code of the {\sc Quantum ESPRESSO} package, utilizing a plane-wave basis formulation of Kohn-Sham DFT implemented therein~\cite{Giannozzi2009-vi,Giannozzi2017-pl}. The interaction between the ionic core and the valence electrons is modelled using optimized norm-conserving Vanderbilt (ONCV) pseudopotentials from the {\sc PseudoDojo} library~\cite{Hamann2013-ne,pseudodojo_2018}. The exchange and correlation interactions among the electrons have been approximated using a Perdew–Burke–Ernzerhof parametrization of exchange–correlation functional revised for description of physical properties of solids (PBEsol)~\cite{Perdew2008-rq}. Truncation of the plane-wave basis is set to a sufficiently large kinetic energy cut-off value of 120 Ry. The BZ sampling is achieved using dense  Monkhorst-Pack k-point grids of size 18\texttimes 18\texttimes 18 along the reciprocal lattice vectors. 

The lattice constant of cubic TlCl was optimized with stringent tolerances for calculated forces on atoms (10\textsuperscript{-6} Ry/Bohr) and residual pressure (0.1 kbar). The resulting equilibrium lattice constant is 3.768\AA.
Spin–orbit coupling (SOC) is included using fully-relativistic pseudopotentials and expanding the electronic states in spinorial basis, to achieve accurate description of the electronic structure. Due to the self-interaction error intrinsic to semilocal DFT, the DFT eigenvalues cannot be used to accurately calculate the band gaps.

We therefore performed a single-shot $G_0W_0$ calculation to perturbatively correct the DFT eigenvalues using diagonal elements of the dynamic electron self-energy, using the {\sc YAMBO} package~\cite{Marini2009-mw,Sangalli2019-yn} in which quasiparticle corrections are evaluated from the many-body self-energy. The screened Coulomb interaction and the correlation part of the electron self-energy are evaluated using a plane-wave basis truncated to an energy cut-off of 4 Ry. The response functions are computed using summations over a total of 150 bands, including 20 occupied bands. We explicitly calculate the quasi-particle energies for the 8 highest occupied and 6 lowest unoccupied bands. The resulting quasiparticle bands exhibit an increased gap and refined dispersions impacting the effective-masses.

By fitting parabolic dispersions to the quasiparticle bands in small neighborhoods of the VBM and CBM, we then extracted the effective-mass tensors:

\begin{eqnarray}
m_{ij}^{-1} = \dfrac{1}{\hbar^2} \dfrac{\partial^2 E(k)}{\partial k_i \partial k_j}\rvert_{k=k_0}.
\end{eqnarray}\label{eff_mass_expr}

\subsection{Calculation of electron-phonon interaction}

The general electron-phonon coupling matrix within the framework of Density functional perturbation theory (DFPT):
\begin{eqnarray}\label{e-ph_expresion}
    g_{mn\nu}(\mathbf{k,q}) = \langle \psi_{m\mathbf{k+q}}\lvert \Delta_{\mathbf{q},\nu}V \rvert \psi_{n,\mathbf{k}} \rangle = \sum_{\kappa \alpha p} \, \sqrt{\dfrac{\hbar}{2M_{\kappa} \omega_{\mathbf{q},\nu}}} e^{\alpha \kappa}_{\mathbf{q},\nu} \, e^{i\mathbf{q.R}_p} \, \langle \psi_{m\mathbf{k+q}}\lvert \partial_{\kappa \alpha p}V \rvert \psi_{n,\mathbf{k}} \rangle,
\end{eqnarray}
where $\psi_{n,\mathbf{k}}$ is the initial Kohn-Sham wave function of the electron of wavevector $\mathbf{k}$ in band $n$ scattered by phonon of branch $\nu$ and wavevector $\mathbf{q}$ into a final state of wavevector $\mathbf{k}+\mathbf{q}$ in band $m$. $\omega_{\mathbf{q},\nu}$ is the phonon angular frequency, $M_{\kappa}$ the mass of atom $\kappa$ in the unit cell, $\mathbf{R}_p$ the lattice vector of the $p$-th unit cell in the Born-von-Karman supercell, and $e^{\alpha \kappa}_{\mathbf{q},\nu}$ the component along direction $\alpha$ of the phonon eigenvector in the unit cell. $\Delta_{\mathbf{q},\nu}V$ stands for the change in the self-consistent potential induced by the phonon and $\partial_{\kappa \alpha p}V$ for the derivative of the potential with respect to a displacement along $\alpha$ of atom $\kappa$ in the $p$-th unit cell.

This coupling matrix element can be split into two parts:
\begin{eqnarray}\label{g_composition}
g_{mn\nu}(\mathbf{k,q}) = g^S_{mn\nu}(\mathbf{k,q}) +g^L_{mn\nu}(\mathbf{k,q}),
\end{eqnarray}
where the superscripts $S$ and $L$ correspond to short-range and long-range contributions, respectively. The long-range contribution arises from the electric field generated by longitudinal optic phonons and diverges as $1/||\mathbf{q}||$ when $\mathbf{q} \rightarrow 0$. C. Verdi et al. developed a method to calculate the Fr\"{o}hlich electron-phonon matrix elements within an \textit{ab initio} framework~\cite{Verdi_2015,Sio_2022}. The formula they obtain is:
\begin{eqnarray}\label{frohlich_expression_a}
g^L_{mn\nu}(\mathbf{k,q}) = i \dfrac{4\pi}{\Omega} \dfrac{e^2}{4\pi\epsilon_0} \sum_{\kappa}  \left( \dfrac{\hbar}{2M_\kappa \omega_{\mathbf{q},\nu}} \right)^{1/2} \times \sum_{G\neq -\mathbf{q}} \dfrac{(\mathbf{q+G}) \cdot \mathbf{Z}^{*}_{\kappa} \cdot \mathbf{e}^\kappa_{\mathbf{q},\nu}}{ (\mathbf{q+G}) \cdot \varepsilon^{\infty} \cdot (\mathbf{q+G})} \times \langle \psi_{m\mathbf{k+q}}\lvert e^{i (\mathbf{q+G})\cdot (\mathbf{r-\tau}_\kappa)} \rvert \psi_{n,\mathbf{k}} \rangle,
\end{eqnarray}
where $\Omega$ is the unit cell volume, $\mathbf{\tau}_K$ is the equilibrium position of atom $\kappa$, $\mathbf{G}$ are the reciprocal lattice vectors, and $\mathbf{Z}^{*}_{\kappa}$ is the dimensionless Born-effective charge tensor associated to atom $\kappa$. An expression corresponding more closely to Fr\"{o}hlich's original work can be obtained by keeping only the $\mathbf{G}=0$ term representing the macroscopic electric field generated by LO phonons and responsible for the $1/||\mathbf{q}||$ divergence when $\mathbf{q} \rightarrow 0$. Then, assuming that $e^{i (\mathbf{q+G})\cdot (\mathbf{r-\tau}_\kappa)}$ varies slowly over a unit cell yields the intraband Fr\"{o}hlich matrix elements:
\begin{eqnarray}\label{frohlich_expression_b}
g^L_{n\nu}(\mathbf{k,q}) = i \dfrac{4\pi}{\Omega} \dfrac{e^2}{4\pi\epsilon_0} \sum_{\kappa}  \left( \dfrac{\hbar}{2M_\kappa \omega_{\mathbf{q},\nu}} \right)^{1/2} \times \dfrac{\mathbf{q} \cdot \mathbf{Z}^{*}_{\kappa} \cdot \mathbf{e}^\kappa_{\mathbf{q},\nu} e^{-i \mathbf{q} \cdot \mathbf{\tau}_\kappa} }{ \mathbf{q} \cdot \varepsilon^{\infty} \cdot \mathbf{q}}. 
\end{eqnarray}
A further simplification can be made by taking the limit $\mathbf{q} \rightarrow 0$ for non-diverging quantities:
\begin{eqnarray}\label{frohlich_expression_d}
g^L_{n\nu}(\mathbf{k,q}) = i \dfrac{4\pi}{\Omega} \dfrac{e^2}{4\pi\epsilon_0} \sum_{\kappa}  \left( \dfrac{\hbar}{2M_\kappa \omega_{\mathbf{q}\rightarrow 0,\nu}} \right)^{1/2} \times \dfrac{\mathbf{q} \cdot \mathbf{Z}^{*}_{\kappa} \cdot \mathbf{e}^\kappa_{\mathbf{q}\rightarrow 0,\nu}  }{ \mathbf{q} \cdot \varepsilon^{\infty} \cdot \mathbf{q}}. 
\end{eqnarray}
Note that this expression is independent of the electronic state $n$ and that the dot product $\mathbf{q} \cdot \mathbf{Z}^{*}_{\kappa} \cdot \mathbf{e}^\kappa_{\mathbf{q}\rightarrow 0,\nu}$ vanishes for TO modes, while retaining only the LO modes. For the case of TlCl, which has cubic symmetry and only two atoms in the unit cell, $\varepsilon^{\infty}$ is a scalar and $\omega_{\mathbf{q}\rightarrow 0,\nu}$ is simply the single LO frequency $\omega_{_{_\text{LO}}}$. Then the above expression reads:
\begin{eqnarray}\label{frohlich_expression_c}
g^L(\mathbf{k,q}) = i \dfrac{4\pi}{\Omega} \dfrac{e^2}{4\pi\epsilon_0\epsilon^\infty} \sum_{\kappa}  \left( \dfrac{\hbar}{2M_\kappa \omega_{_{_\text{LO}}}} \right)^{1/2} \times \dfrac{\mathbf{q} \cdot \mathbf{Z}^{*}_{\kappa} \cdot \mathbf{e}^\kappa_{\mathbf{q}\rightarrow 0,_{_\text{LO}}}  }{ q^2}. 
\end{eqnarray}
This is the expression used to estimate in \ref{SecVIA} the Fr\"{o}hlich matrix element in TlCl and compare it to the full DFPT matrix elements for the VBM and CBM. 

\end{widetext}

\bibliographystyle{apsrev4-2}
\bibliography{references}{}

\end{document}